\documentclass[twocolumn,preprintnumbers,amsmath,amssymb,final]{revtex4}

\usepackage{showkeys}

\usepackage{graphicx}
\usepackage{dcolumn}
\usepackage{bm}
\usepackage{amsmath,amssymb}
\usepackage{upgreek}
\usepackage{color}
\usepackage{pstricks,pst-plot,pst-node} 
\usepackage{infix-RPN}
\usepackage{psfrag}
\newcommand{\delA}{\delta \hspace{-.5ex}A}
\newcommand{\ext}{\text{ext}}
\newcommand{\opt}{\text{opt}}
\newcommand{\wall}{\text{wall}}
\newcommand{\stray}{\text{stray}}
\newcommand{\domain}{\text{domain}}

\newcommand{\dd}{\operatorname{d}\!}
\definecolor{mgray}{gray}{0.6}
\definecolor{mdgray}{gray}{0.5}
\definecolor{mlgray}{gray}{0.7}
\definecolor{ddgray}{gray}{0.4}
\definecolor{llgray}{gray}{0.8}
\definecolor{dddgray}{gray}{0.2}
\definecolor{lllgray}{gray}{0.9}
\definecolor{dgreen}{rgb}{0.06,0.6,0.06}
\begin{document}
%

 \title{
The Formation and Coarsening of the Concertina Pattern}

 \author{Jutta Steiner}
 \affiliation{Institute for Applied Mathematics, University of Bonn, Endenicher Allee 60, 53115 Bonn, Germany
 }%
\author{Holm Wieczoreck}
 \affiliation{
Leibniz Institute for Solid State and Materials Research Dresden (IFW Dresden), Inst. f. Metallic Materials, Helmholtzstr. 20, 01069 Dresden, Germany
 }%
 \author{Rudolf Sch\"afer}
 \affiliation{
Leibniz Institute for Solid State and Materials Research Dresden (IFW Dresden), Inst. f. Metallic Materials, Helmholtzstr. 20, 01069 Dresden, Germany
 }%
\author{Jeffrey McCord}
 \affiliation{
Institute of Ion Beam Physics and Materials Research, Forschungszentrum Dresden-Rossendorf, 01328 Dresden, Germany
 }%
 \author{Felix Otto}
\affiliation{Max Planck Institute for Mathematics in the Sciences, Inselstraße 22, 04103 Leipzig, Germany}

 \date{\today}

 \begin{abstract}
The concertina is a magnetization pattern in
elongated thin-film elements of a soft material.
It is a ubiquitous domain pattern that occurs in the process
of magnetization reversal in direction of the long axis of the 
small element.
\smallskip

Van den Berg argued that this pattern grows
out of the flux closure domains as the external
field is reduced.
Based on experimental observations and theory,
we argue that in sufficiently elongated thin-film
elements, the concertina pattern rather bifurcates 
from an oscillatory buckling mode.
\smallskip

Using a reduced model derived by asymptotic analysis and
investigated by numerical simulation, we quantitatively
predict the average period of the concertina pattern
and qualitatively predict its hysteresis. In particular,
we argue that the experimentally observed coarsening of the concertina
pattern is due to secondary bifurcations related to an Eckhaus
instability.
\smallskip

We also link the concertina pattern to the magnetization ripple
and discuss the effect of a weak (crystalline or induced) anisotropy.

 \end{abstract}

\maketitle
\section{Introduction}
\begin{figure}[htb]
  \begin{center}
  \includegraphics[height=0.06\textwidth]{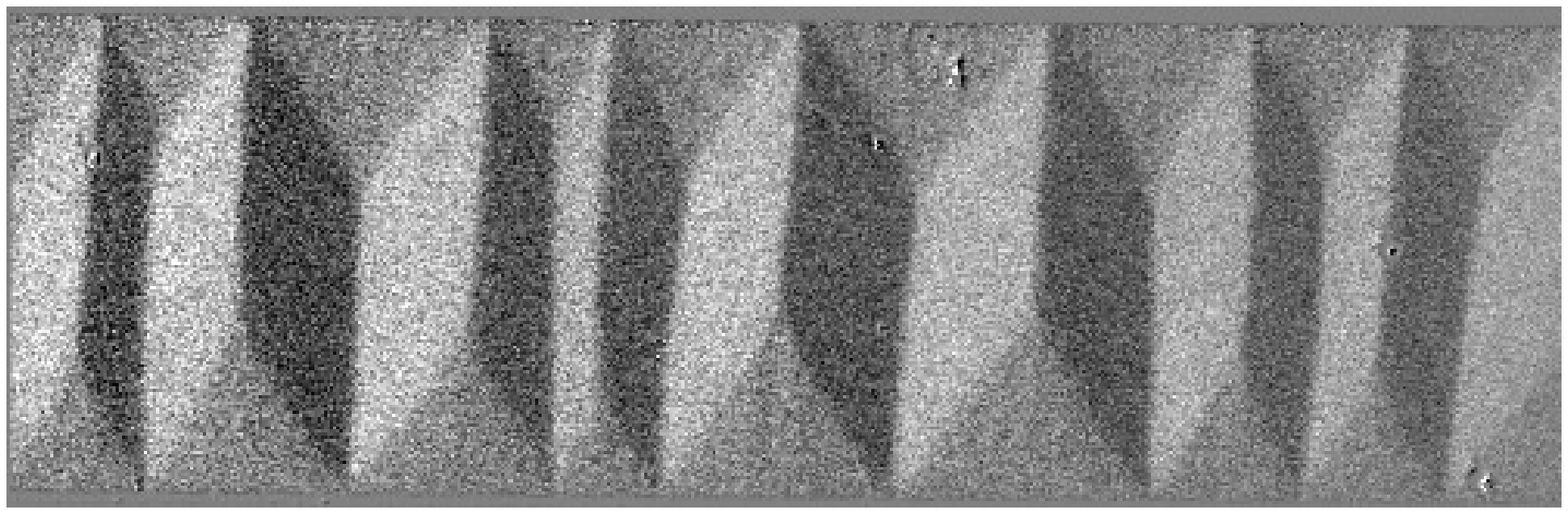}
\psset{unit=0.2}
\begin{pspicture}(3,-2.7)(3,2.7)
\rput[c]{0}(-5.8,-2){\psline[linecolor=blue,linewidth=2pt,arrowinset=0,arrowlength=.3]{->}(-.5,0)(.5,0)}
\rput[c]{50}(-5.4,0){\psline[linecolor=blue,linewidth=2pt,arrowinset=0,arrowlength=.3]{->}(-.5,0)(.5,0)}
\rput[c]{-50}(-6.4,0){\psline[linecolor=blue,linewidth=2pt,arrowinset=0,arrowlength=.3]{->}(-.5,0)(.5,0)}
\rput[c]{0}(-7,2){\psline[linecolor=blue,linewidth=2pt,arrowinset=0,arrowlength=.3]{->}(-.5,0)(.5,0)}
\rput[c]{0}(-4.6,2){\psline[linecolor=blue,linewidth=2pt,arrowinset=0,arrowlength=.3]{->}(-.5,0)(.5,0)}
\end{pspicture}  
\quad
\begin{pspicture}(-2.,-2.7)(3.4,2.7)
\rput[c]{0}(0,0){\psline[linecolor=blue,linewidth=2pt,arrowinset=0,arrowlength=1]{->}(-1.5,0)(1.5,0)}
\pscircle[linewidth=1pt,fillstyle=none,fillcolor=dddgray,linecolor=black](0,0){2}
\pscircle[linewidth=.3pt,fillstyle=solid,fillcolor=mgray,linecolor=black](2.4,0){3pt}
\pscircle[linewidth=.3pt,fillstyle=solid,fillcolor=mgray,linecolor=black](-2.4,0){3pt}
\pscircle[linewidth=.3pt,fillstyle=solid,fillcolor=dddgray,linecolor=black](0,2.4){3pt}
\pscircle[linewidth=.3pt,fillstyle=solid,fillcolor=lllgray,linecolor=black](0,-2.4){3pt}
\end{pspicture}   
  \scalebox{-1}[1]{ \includegraphics[height=0.06\textwidth]{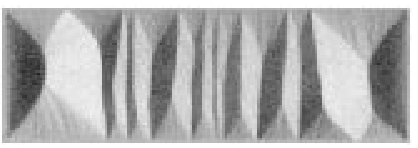}}
\begin{pspicture}(.8,-2.7)(.8,2.7)
\rput[c]{0}(-6.8,-2){\psline[linecolor=blue,linewidth=2pt,arrowinset=0,arrowlength=.3]{->}(-.5,0)(.5,0)}
\rput[c]{50}(-6.4,0){\psline[linecolor=blue,linewidth=2pt,arrowinset=0,arrowlength=.3]{->}(-.5,0)(.5,0)}
\rput[c]{-50}(-7.3,0){\psline[linecolor=blue,linewidth=2pt,arrowinset=0,arrowlength=.3]{->}(-.5,0)(.5,0)}
\rput[c]{0}(-6,2){\psline[linecolor=blue,linewidth=2pt,arrowinset=0,arrowlength=.3]{->}(-.5,0)(.5,0)}
\rput[c]{0}(-7.6,2){\psline[linecolor=blue,linewidth=2pt,arrowinset=0,arrowlength=.3]{->}(-.5,0)(.5,0)}
\end{pspicture}
  \end{center}
  \caption
{Concertina in a very elongated (length $2$~mm) sample of width $50 \,\upmu$m and thickness
    $50$~nm (left) and in a sample of width $35 \,\upmu$m, thickness
    $40$~nm and moderate length $110\,\upmu$m (right). The left image shows
    only the center of the stripe which is less than $10$ percent of
    the whole sample. As indicated by the blue arrows, the gray scales encode
    the transversal component of the magnetization in the domains.}
\label{fig:concertina}
\end{figure}

To our knowledge, the term {\it concertina} was introduced by van den Berg 
et.\ al.\ in \cite[p.880]{vdB}. In that paper he explains the
formation of this domain pattern in Permalloy thin-film elements
that are fairly thick (thickness $t=350$nm), 
with a rectangular cross section (width $\ell=15\upmu$m)
that is {\it not too elongated} (length $60\upmu$m). 
After near-saturation along the long axis, 
a concertina pattern grows out of the flux closure domains 
at the short edges of the cross section during subsequent reduction of
the external field $H_\ext$, until the pattern eventually invades the entire sample,
see Figure \ref{fig:concertina} on the right. 
Our experimental observations 
suggest that in {\it very elongated} samples (length $2$nm, thickness $10$ to $150$nm,
width $10$ to $100\upmu$m) a bifurcation is at the origin of the
concertina. As a consequence, the pattern forms simultaneously   
all over the sample. We will report on
van den Berg's explanation in more detail after introducing the micromagnetic model.

\begin{figure*}[htb]
 \includegraphics[width=0.9\textwidth,height=0.03375\textwidth]{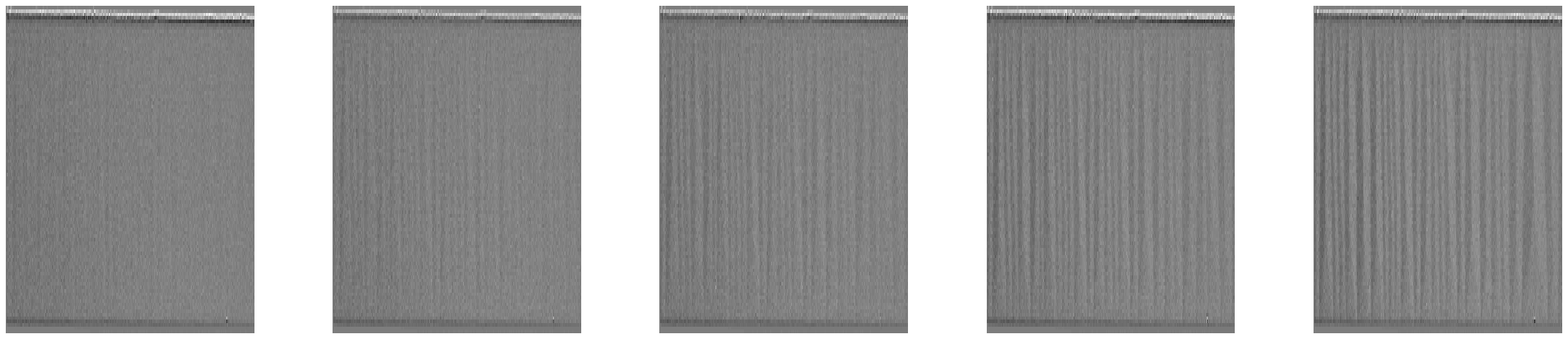}
 \includegraphics[width=0.9\textwidth,height=0.05625\textwidth]{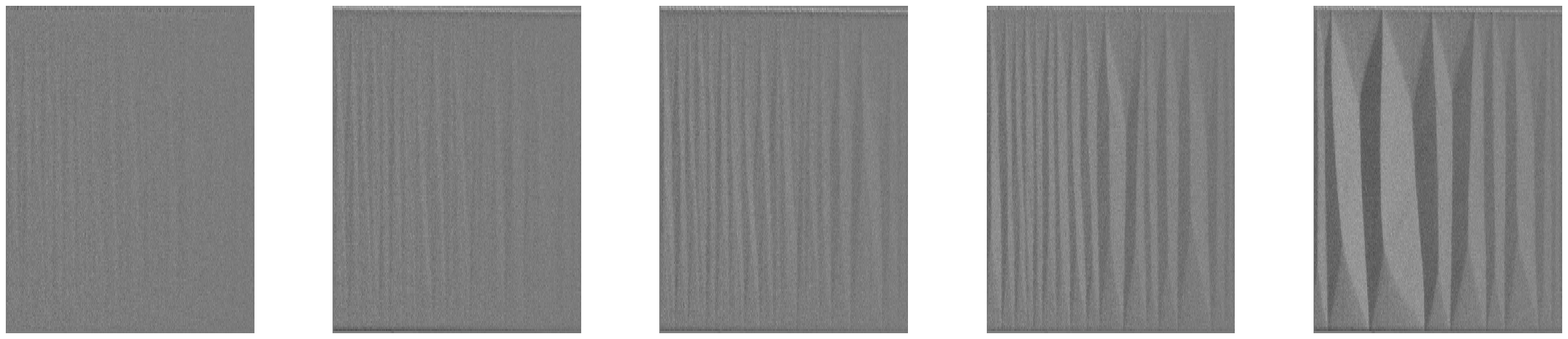}
 \includegraphics[width=0.9\textwidth,height=0.03375\textwidth]{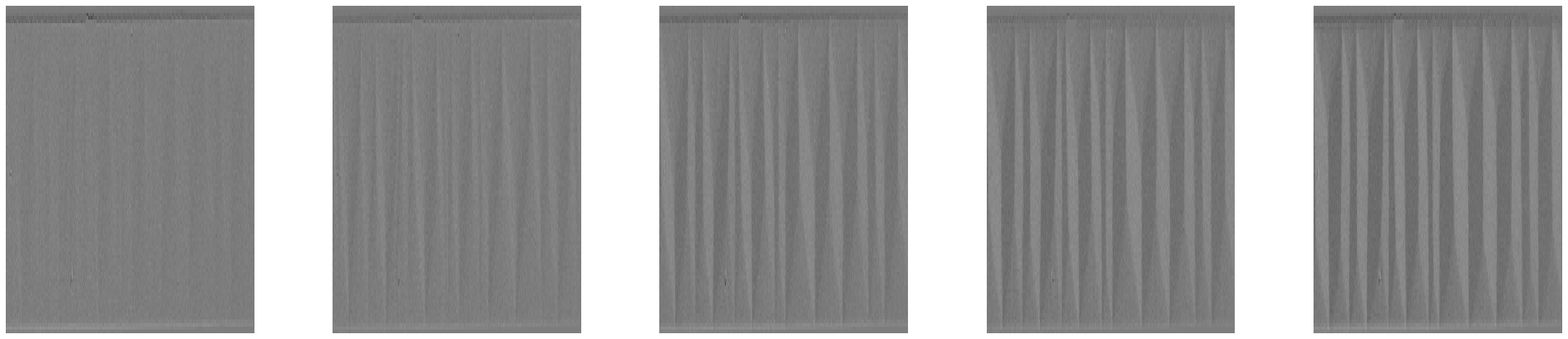}
 \includegraphics[width=0.9\textwidth,height=0.05625\textwidth]{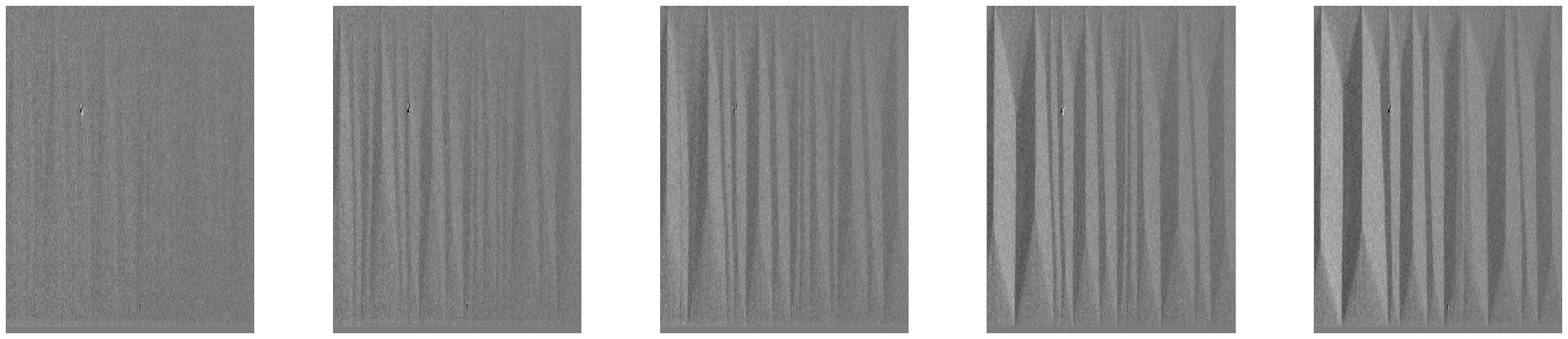}
\caption{Formation of the concertina pattern in the experiment: The
  pictures show a section near the center of the elongated thin-film
  element. The unstable mode grows
  into a domain-wall pattern which coarsens several times. The two
  upper series show a sample of $30$nm thickness of low
  anisotropy. The two lower series show a sample of $30$nm thickness
  of higher anisotropy. The width is $30\upmu$m and $50\upmu$m,
  respectively.}
\label{fig:coarsening}
\end{figure*} 
\subsection{The micromagnetic energy}
The variation of the applied magnetic field in the experiments
is on a very slow time-scale so that the magnetization always relaxes
to equilibrium. Hence our theoretical analysis is based on the
micromagnetic (free) energy, which we introduce now. The magnetization
of a ferromagnetic sample occupying some domain
$\Omega$ is described by a vector field $m=(m_1,m_2,m_3)$.
The micromagnetic energy $E(m)$ is given by 
\begin{align}
\begin{split}
E(m)=\,&d^2\int_{\Omega}|\nabla
m|^2\,\dd
x+\int_\text{all space}|H_\stray|^2\,\dd
x \\&{}-Q\int_{\Omega}(m\cdot e)^2\,\dd
x-2\,\int_{\Omega}H_\ext\cdot m\,\dd x.
\end{split}
\label{B1:Micromag}
\end{align}
The energy \eqref{B1:Micromag} is already
partially, i.e., except for lengths, non-dimensionalized. In
particular the
magnetization has length one. Outside of the sample, it vanishes identically:
\begin{equation}
  \label{constraint}
  \begin{split}
  &|m|^2=1 \quad\text{ in the sample }\Omega,\\
&m=0 \quad\text{ outside of the sample } \Omega.
    \end{split}
\end{equation}
Let us now briefly introduce and discuss the different energy contributions: 

\smallskip
The first contribution in \eqref{B1:Micromag} is the so called exchange energy. (The gradient
acts component wise, i.e., $|\nabla m|^2=\sum_{i=1}^{3}\sum_{j=1}^{3}
(\partial_i m_j)^2$.) This term favors a
uniform magnetization. The material parameter $d$ is called the
exchange length and measures the relative strength of exchange
with respect to stray-field energy, see below. The exchange length is typically of
the order of a few nano-meters.

\smallskip
The second contribution in \eqref{B1:Micromag} is the stray-field energy.  
The static Maxwell equations state that the magnetization $m$ generates a
stray field $H_\stray$ that is described by
\begin{equation}
  \begin{split}
    &\nabla \times H_\stray=0,\\
&\nabla \cdot(H_\stray+m)=0,
  \end{split}
\label{Maxwell}
\end{equation}
where both equations hold in the whole space and $B=H_\stray+m$ is the magnetic induction.
Hence the stray field is generated by the
divergence of the magnetization. Since the magnetization is
discontinuous at the  
boundary of the sample $\partial \Omega$, cf.\ \eqref{constraint}, the 
second equation in \eqref{Maxwell} has to be understood in the following sense:
\begin{align}
\begin{split}
\nabla \cdot H_\stray&=
\begin{cases}
-\nabla \cdot m &\text{ in the sample }\Omega\\
0 &\text{ outside of the sample } \Omega,
\end{cases}\\
[H_\stray\cdot \nu]&=m\cdot \nu \text{ on the boundary
}\partial\Omega,
\end{split}
\label{classical}
\end{align}
where $\nu$ denotes the normal to the boundary of the sample, and
$[H_\stray\cdot \nu]$ denotes the jump that $H_\stray\cdot \nu$
experiences across the boundary $\partial \Omega$.
Hence we distinguish two different types of sources of the
stray field -- in analogy to electrostatics one commonly speaks
of charges -- namely
\begin{align*}
  \text{magnetic volume charges }\quad \nabla &\cdot m \quad\text{ in }\quad\Omega\quad \text{ and }\\
  \text{magnetic surface charges } \quad\nu &\cdot m \quad\text{ on }\quad\partial \Omega.
\end{align*}
Later on we will also use the following equivalent distributional
formulation of \eqref{classical}, which is obtained by testing with smooth functions $\zeta$ vanishing at infinity, namely 
\begin{align}\label{alternative} 
\int_\text{all space} H_\stray\cdot\nabla \zeta\,\dd x=-\int_{\Omega}m \cdot
\nabla \zeta\, \dd x.
\end{align}

The third contribution in \eqref{B1:Micromag} models a uniaxial anisotropy, i.e., the
preference of an easy axis $e=(e_1,e_2,e_3)$ in a material.
The material parameter $Q>0$ is called the quality
factor. It measures the
relative strength of anisotropy with respect to stray-field energy.
A uniaxial anisotropy can for example come in form of crystalline or induced
anisotropy. Notice that the polycrystalline anisotropy in a material
like Permalloy can be described as a position-dependent easy axis $e(x)$. 

\smallskip
The last contribution in \eqref{B1:Micromag} is called Zeeman
energy. This term models the interaction and favors the
alignment of the magnetization with an external magnetic field
$H_\ext$.  
\smallskip

The specific material parameters of our samples are discussed in Section \ref{Sec:Experiments}.
\subsection{Van den Berg's explanation of the concertina}
We now return to van den Berg's paper \cite{vdB}.
Combining the explanation of the formation of the concertina
pattern therein with the insight from \cite{Bryant,DKMO+S},
we can give the following updated version of the explanation presented
in \cite[Sections A \& B]{vdB}: 
It is a theory on the {\it two-dimensional} {\it mesoscopic} magnetization
pattern; by two-dimensional, we understand that the magnetization is in-plane,i.e., $m_3=0$, and independent
of the thickness direction, i.e.,  $m=m(x_1,x_2)$; by mesoscopic, we
understand that the walls are replaced by sharp discontinuity curves
that are charge-free in the sense that the normal component of the
magnetization does not jump; moreover, the magnetization is tangential
to the lateral edges of the sample so that there are no surface charges. 
In sufficiently large thin-film elements and for sufficiently low
external fields, \cite{Bryant} now postulate that the two-dimensional
mesoscopic magnetization pattern arranges
itself in such a way that the corresponding continuous
magnetic charge density $\sigma=-(\partial_1m_1+\partial_2m_2)$ generates
a stray field $H_{stray}$ that expels the external field $H_\ext$ from
inside of the sample (like in electrostatics). 
\smallskip

In \cite{DKMO05}, see \cite{DKMO+S} for an efficient account,
it is shown that in the regime of sufficiently large thin-film
elements (i.e., $t\ll\ell$ and $\ell t\gg d^2\log\frac{\ell}{t}$ with
comparable lateral dimensions of the order $\sim \ell$),
this principle extends to moderately large fields (of the order $\sim\frac{t}{\ell}$):
In this case, the stray field $H_\text{stray}$ in general can no longer expel
the external field $H_\ext$ everywhere in the sample, 
since the (total) charge density 
$\sigma=-(\partial_1m_1+\partial_2m_2)$ is limited by $m_1^2+m_2^2=1$.
The charge density $\sigma$ is uniquely determined by a 
{\it convex} variational problem only involving the stray-field energy
and the Zeeman energy. At least some aspects of the mesoscopic
two-dimensional magnetization pattern $(m_1,m_2)$ can be recovered from $\sigma$:
The characteristics of $(m_1,m_2)$, i.e., the curves along which $(m_1,m_2)$ is normal (called ``trajectories in \cite{vdB}), 
have curvature given by $\sigma$.
However, due to the potential discontinuity curves of
the mesoscopic magnetization $(m_1,m_2)$, this seemingly rigid
condition does not suffice to determine $(m_1,m_2)$ -- even the
fact that the discontinuity curves are charge-free is still not enough. 
Notice that it is easy to construct a particular solution $(m_1,m_2)$
for any charge density $\sigma$ via the maximal solution of
a modified eikonal equation \cite[p.2987]{DKMO+S}).  
On the other hand, in the region
where the external field has penetrated, the magnetization $(m_1,m_2)$ is uniquely
determined, cf.\ \cite[p.2987]{DKMO+S},
and has no discontinuities, cf.\ \cite[p.883]{vdB}.

\smallskip

Van den Berg gives a recipe how to construct a solution that
corresponds to the experimental observation of a concertina pattern
growing out of the flux closure pattern: For sufficiently large
external fields ($H_\ext=(h_\ext,0,0)$, $h_\ext \gg \frac{t}{\ell}$), $H_\ext+H_{stray}$
does not vanish in the sample besides in the vicinity of the
two distant edges; as a consequence walls only occur in the
two flux closure pattern there. 
As the external field is reduced, the penetrated region shrinks
as the walls invade the sample. Each of the two flux closure pattern 
has a ``doublet'', that is, a point on one of the long edges where two wall
segments intersect. Note that the doublets were created as the central
$180^\circ$-wall of
the initial Landau state moved towards the edge where it broke up due to the
application of a large external field at the very beginning. The
inner, i.e., most distant to the short edges, ones of the doublet walls fade out in the
middle (with respect to the long edges) of the cross section. Van den
Berg postulates that the position of the doublets does not change as the
external field decreases further. As a consequence, each of the two inner
walls grows -- necessarily in direction of the characteristic -- till is
hits the opposite edge. There it must generate a ``triplet'' (a point
on the edge where three walls meet); the middle wall must 
coincide with the previous one originating in the doublet. Again, as
the external field is further reduced, the position of each triplet is
supposed to be fixed, the inner of the three walls
grows towards the opposite edge. 
\begin{figure}
  \centering
  \input{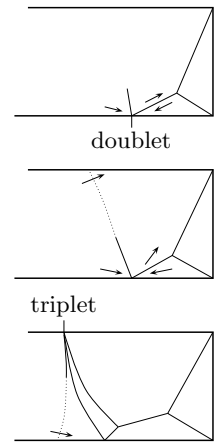}
  \caption{The creation of a triplet out of the initial doublet as described and sketched by van den Berg in \cite{vdB}}
\end{figure}

This process repeats itself
till the two half-concertina structures growing from the short edges
are linked in the middle (with respect to the short edges) of the
cross section. For very elongated samples of length $L\gg \ell$, the
linking is expected at a field strength of order  
$h_\ext\sim t \ell L^{-2} \ln
t\ell^{-1}$ and thus differs from the field at the beginning of the
growth process by a {\it factor} $\ell^2 L^{-2}$ (up to a logarithm). 
Speaking in mathematical terms, van den Berg postulates that the
positions of doublets and triplets remain fixed as the field is
decreased and
appeals to continuity, i.e., the pattern should depend continuously
on the value of the external field, to overcome the non-uniqueness of
$(m_1,m_2)$ mentioned in the previous paragraph. 

\subsection{Van den Berg's vs.\ our explanation}
Our explanation for the formation of the concertina pattern is
very different from the one of van den Berg. Instead of
a {\it successive} outgrowth (along the sample) of the closure domains, we explain the concertina as
a {\it simultaneous} outgrowth (along the sample) of an unstable mode, best captured in very elongated
thin-film elements.  Indeed, our experiments were performed on
thin-film elements of thicknesses $t$ in the range of $10$nm to $150$nm,
widths $\ell$ in the range of $10\upmu$m to $100\upmu$m, 
but lengths in the range
of $2$mm. We recorded the pattern at three different sections,
equidistant and equidistant to the short edges of the cross section,
and observed qualitatively the same pattern at the same values
of the external field. 
\smallskip 

Not surprisingly, our theoretical predictions are quite different 
from those in \cite{vdB} -- already in terms of scaling. Van den
Berg's explanation entails two different scales of the external field

\begin{itemize}
\item $h_\ext^\text{begin}\sim \frac{t}{\ell}$ for the beginning of the build-up
  process and
\item $h_\ext^\text{end}\sim t \ell L^{-2} \ln
t\ell^{-1}$ for the completion when the external field is totally
expelled from the sample
\end{itemize}
whereas in our case there is one critical field $h_\ext^*$ at which
the simultaneous formation of the concertina along the sample --
independently of the specific position -- due to an interior instability begins. This critical field is given by 
$h_\ext^*\sim - d^{2/3}\ell^{-4/3}t^{2/3}$, see Regime III in Subsection
\ref{modes}, for isotropic samples -- thus the instability would only occur
after the field is reversed and thus when the van den Berg concertina
has already invaded the sample. However, as we discuss later, the critical
field is shifted in case of a transversal anisotropy $h_\ext^* \leadsto
h_\ext^*+Q$. It turns out that even for relatively weak transversal anisotropy 
the formation thus starts before the field is reversed, see Section
\ref{anisotropy}, a). For very elongated samples, i.e., $L\gg \ell$, we
have that $h_\ext^\text{begin} \gg h_\ext^\text{end}$ and
$h_\ext^\text{end}=0$ for the limit case of an infinitely extended
sample. The strength of the anisotropy and the geometry of the majority
of the samples that we investigated is such that $h_\ext^\text{begin} \gg
h_\ext^*\gg h_\ext^\text{end}$. We thus expect the following
scenario in very elongated samples: At
$h_\ext^\text{begin}$ the van den Berg build-up process starts at the tips of
the sample. As the field is reduced, the concertina grows slowly into
the sample from the tips. Meanwhile, as $h_\ext^*$ is attained our instability occurs all over the sample --
sufficiently far away from the tips and way before the van den Berg
linking could take place in the center of the sample, see Table \ref{tab:fields}.
\begin{table}[htb]
  \centering
  \begin{tabular}{|l|c|c|c|}
\hline
    &$h_\ext^\text{begin}$&$h_\ext^*$ &$h_\ext^\text{end}$\\\hline\hline
Weak anisotropy &&&\\($Q=1.3 \times 10^{-4}$) &&&\\\hline
$\ell=30\upmu$m&$1\times 10^{-3}$ & {\color{red}$-4.0\times 10^{-5}$} &$1.6\times 10^{-6} $\\\hline
$\ell=50\upmu$m&$6\times 10^{-4}$&$4.2 \times 10^{-5}$&$2.8\times 10^{-6}$\\\hline\hline
Stronger anisotropy &&&\\($Q=5.0 \times 10^{-4}$)&&&\\\hline
$\ell=30\upmu$m&$1\times 10^{-3}$ & $3.4\times 10^{-4} $&$1.6\times 10^{-6} $\\\hline
$\ell=50\upmu$m&$6\times 10^{-4}$&$4.2 \times 10^{-4}$&$2.8\times 10^{-6}$\\\hline
  \end{tabular}
  \caption{Comparison of the characteristic fields in van den Berg's
    theory of the concertina and the critical field in our instability
    for the samples shown in Figure \ref{fig:coarsening} (thickness
    $t=30$nm, length $L=2$mm. Apart from the sample of weak anisotropy and small width, the characteristic fields appear in the expected order.}
\label{tab:fields}
\end{table}

\begin{figure}[htb]
  \centering
\includegraphics[trim = 0mm 8mm 0mm 8mm, clip, width=0.12\textwidth]{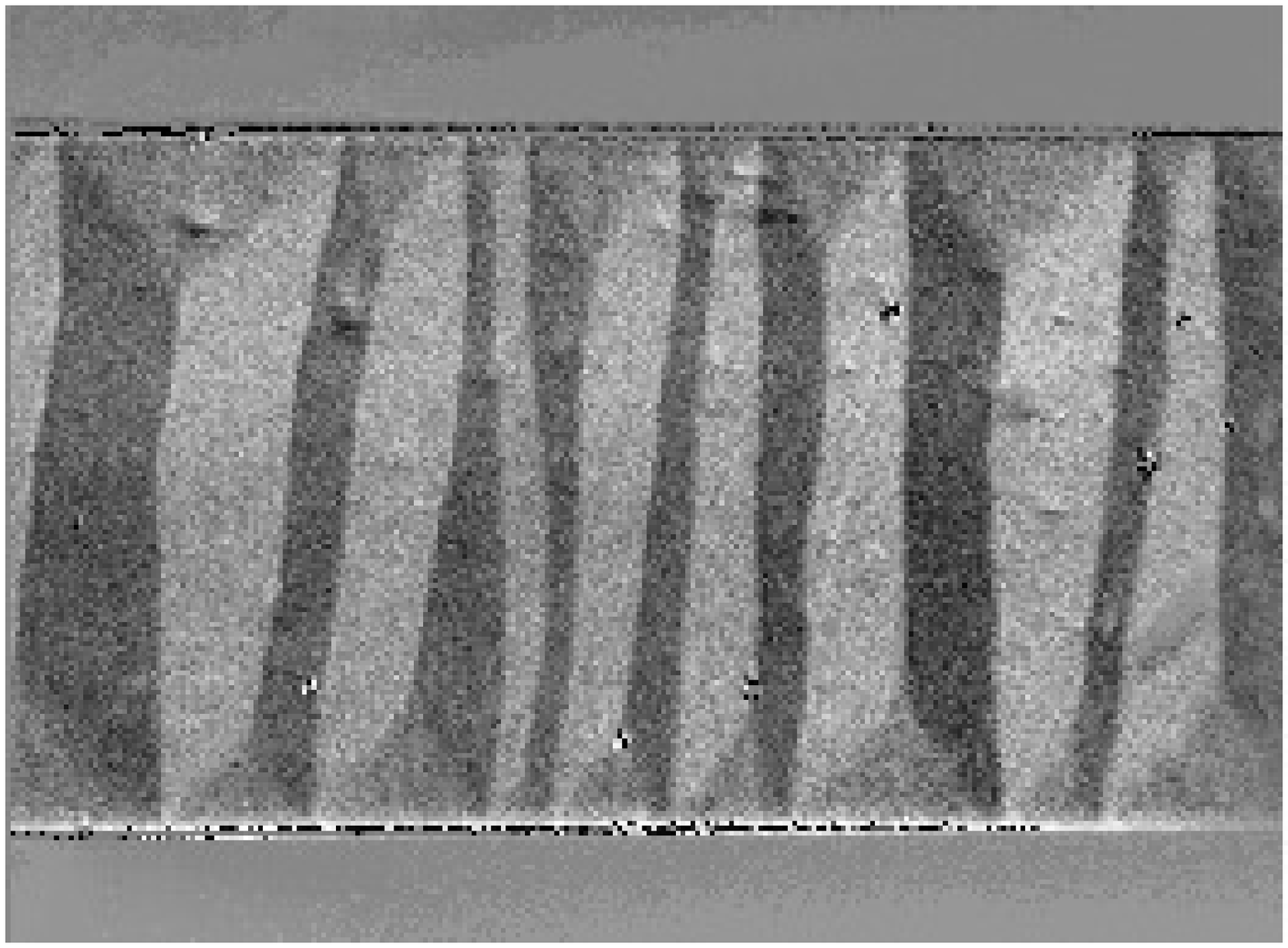}\quad
\includegraphics[trim = 0mm 8mm 0mm 8mm, clip, width=0.12\textwidth]{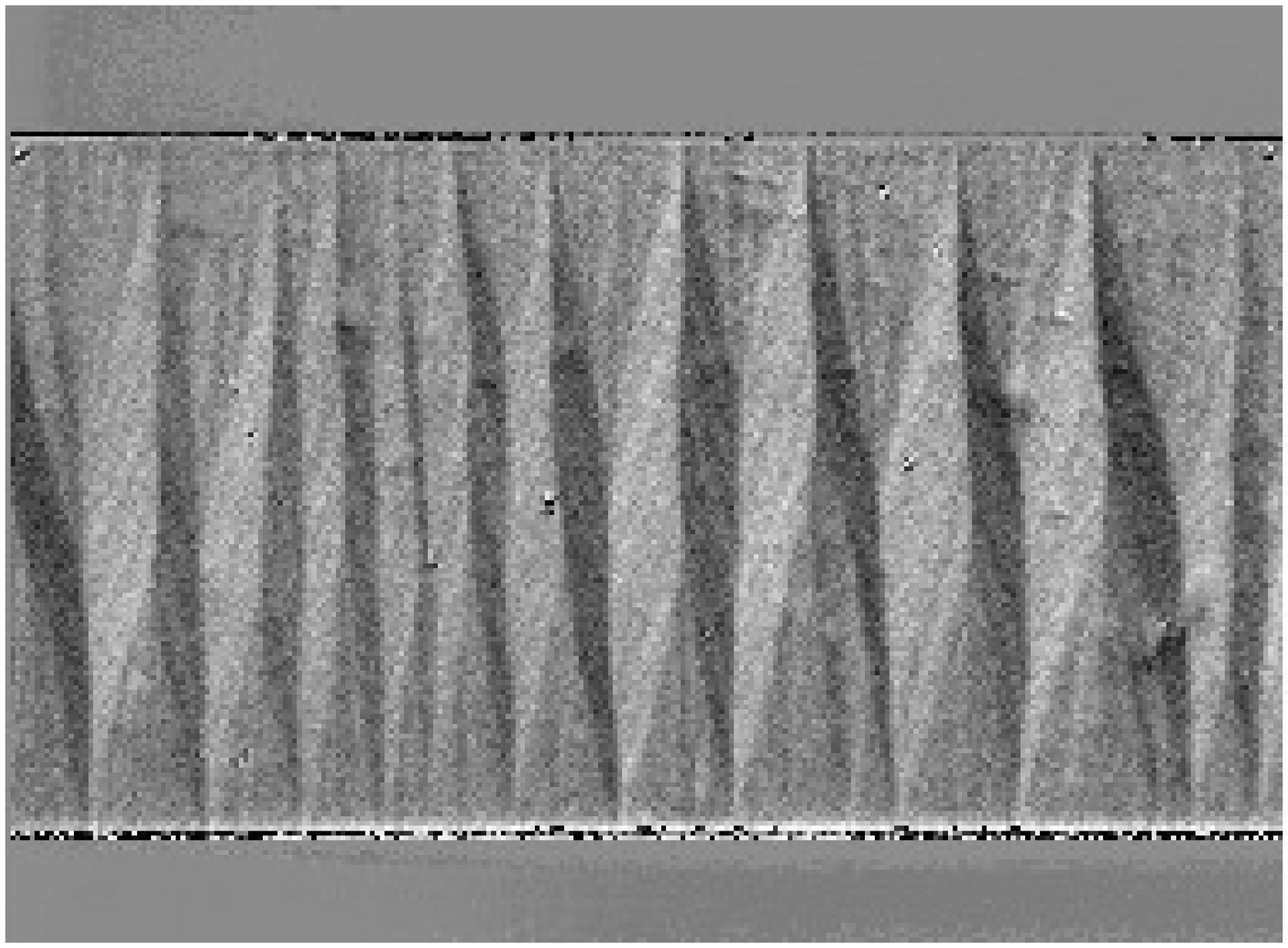}\quad
\includegraphics[trim = 0mm 8mm 0mm 8mm, clip, width=0.12\textwidth]{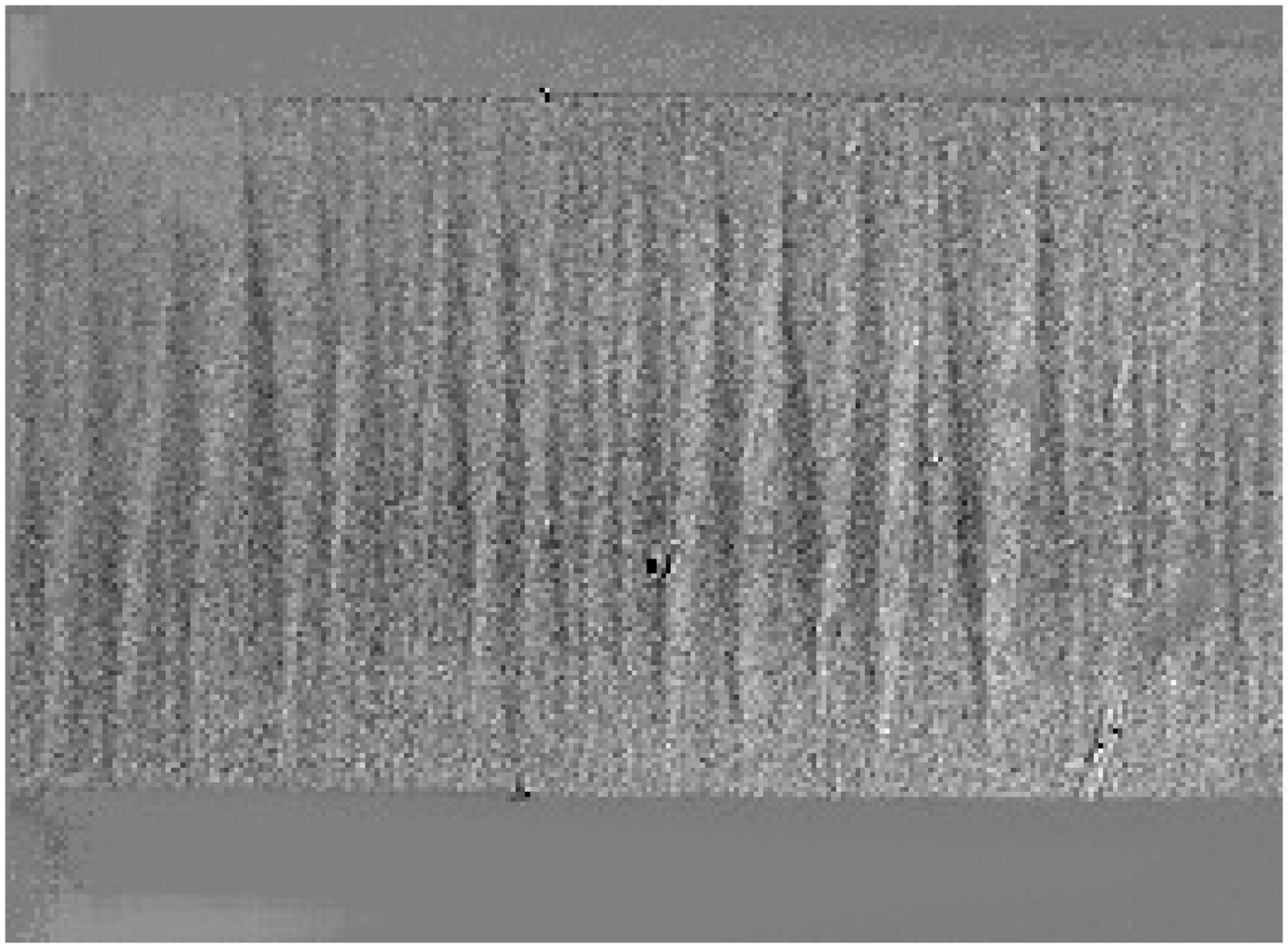}
\caption{Concertina in Permalloy samples of width $\ell=100 \upmu$m
  and thickness $t=30$nm (left), $t=80 $nm (center), and $t=300$nm
  (right). The width average period of the pattern is a decreasing
  function of the thickness $t$.}  
\label{thickness}
\end{figure}

Whereas in \cite{vdB} the appropriate
scale for the concertina width $w$ is given by $\ell$ -- in particular independent
on the thickness $t$ -- it is given by and $d^{2/3}\ell^{2/3}t^{-1/3}$
in our case, in qualitative accordance with our experimental observations
illustrated in
Figure~\ref{thickness}. 
\subsection{Experimental setup and samples}
\label{Sec:Experiments}
In the experiments, we investigated magnetic films of nano-crystalline
Permalloy (Ni$_{81}$Fe$_{19}$) and amorphous
Co$_{60}$Fe$_{20}$B$_{20}$ of various thicknesses and induced
magnetic anisotropy values. The films were deposited by magnetron
sputtering under ultra-high vacuum conditions. In order to control the
grain growth of the polycrystalline films, a Ta seed ($5$ nm) layer
was used for the Ni$_{81}$Fe$_{19}$ deposition. In all cases a
magnetic in-plane saturation field was applied during film deposition
to control strength and direction of the induced anisotropy. 
Using different magnetic field histories, films of different effective induced anisotropy were obtained. 

\begin{itemize}
\item A first set of samples was deposited in the presence of a
  homogeneous, static magnetic field. This results in a maximal and
  also well-aligned induced uniaxial anisotropy. A series of Permalloy and CoFeB samples was obtained by this method.
\item In a second set of samples the induced anisotropy was strongly
  reduced. In order to ensure this, the films were deposited in a
  magnetic field of alternating orthogonal alignment. The field
  direction was changed after approximately every $5$nm of film
  growth. The superposition of the so-obtained orthogonal anisotropy axes results in a strongly reduced induced anisotropy. 
\end{itemize}
The relevant material parameters -- for the comparison of the
experimental observations to the theoretical predictions -- are the following:
\begin{itemize}
\item Exchange length $d$: Permalloy $5$nm, CoFeB $3$nm.
\item For both materials the saturation polarization is $J_s \approx 1$T and the stray-field energy density is given by $K_d \approx 4 \times 10^5$ J/m$^3$. 
\item The uniaxial anisotropy coefficient is $K_u^\text{Permalloy} \approx 200$ J/m$^3$ for the high-anisotropy Permalloy and $K_u^\text{CoFeB}\approx 600$ J/m$^3$ for CoFeB, respectively. For the low-anisotropy Permalloy films we have $K_u^\text{Permalloy} \approx 50$ J/m$^3$.
\item Quality factor $Q = K_u/K_d$: High and low-anisotropy Permalloy
  $0.5 \times 10^{-3}$ and $0.125 \times 10^{-3}$, respectively, and CoFeB $1.5 \times 10^{-3}$. 
\item The average size of the individual grains of Permalloy is
  $\ell_\text{grain} \approx 12$ to $15$nm. It is assumed that up to a
  film thickness of about $30$nm, the grains display a column-like shape.
\item The film thicknesses range from $10$ to $150$nm, the film widths from $10$ to $100\upmu$m.
\end{itemize}
After film deposition, elongated stripes of various widths and a length of $2000\upmu$m were patterned by photolithography and subsequent ion beam etching. The stripes were aligned, both, parallel and orthogonal to the induced anisotropy direction.

The observation of domains and magnetization processes was carried out
in a digitally enhanced Kerr microscope, \cite{Hubert}. The
longitudinal Kerr effect was applied with its magneto-optical
sensitivity direction transversal to the stripe axis. The dominant
wavelength of the observed concertina patterns was computed by Fast
Fourier Transform. The result of the computation is in agreement with
the average wavelength determined by manually counting the folds in
the images, as soon as the concertina becomes discernible to the eye during field
reduction. The typical strength of the magnetic field, which is
applied for the saturation at the very beginning, is of the order of some mT.
\subsection{Nucleation}
We are interested in the magnetization pattern
in elongated thin-film elements of width $\ell$
(in $x_2$-direction) and thickness $t\ll\ell$ (in $x_3$-direction)
that forms under the variation of an external field aligned with the long axis (the $x_1$-axis),
that is, of the form $H_\ext=(-h_\ext,0,0)$. (To simplify the notations,
the minus is introduced which entails a positive critical field, see below.) 
%
%
As indicated above, we observe no influence of the sample's short edges on the
formation of the concertina {\it away} from the
short edges. Since it greatly simplifies the theoretical treatment, we therefore
henceforth will assume that the sample is infinite in $x_1$-direction
(and occasionally, for instance in the numerical treatment, impose
a large, but artificial period in that direction). 
One consequence of that assumption is that the uniform 
magnetization $m^*=(1,0,0)$
is a stationary point of the energy functional -- that is, satisfies
the corresponding Euler-Lagrange equations that express a torque
balance at every point of the sample -- for {\it all} values $h_\ext$
of the external field $H_\ext$ of the form above.
\begin{figure}[htb]
\psset{unit=0.2cm,arrowinset=0,linearc=0.06}
\begin{pspicture}(-8,-.5)(25,7)
\rput{0}(-1.3,.45) {\psscaleboxto(.6,1.8){\(\{\)}} \rput[c](-2.2,.45){$t$}
\rput{-45}(1.2,4.1) {\psscaleboxto(.6,7){\(\{\)}} \rput[c](0,4.4){$\ell$}
\psset{linecolor=black,linewidth=.8pt}
\psline(0,-0.5)(15,-0.5)
\psline(0,1.5)(15,1.5)
\psline(5,6.5)(20,6.5)
\psset{linestyle=dashed,linewidth=.8pt}
\psline(5,4.5)(20,4.5)
\pspolygon(0,-0.5)(0,1.5)(5,6.5)(5,4.5)
\put(15,0){\pspolygon(0,-0.5)(0,1.5)(5,6.5)(5,4.5)}
\psset{linecolor=blue,linewidth=1pt}
\rput{0}(3.25,1.){\psline{->}(-1,0)(1.,0)}
\rput{0}(5.25,3){\psline{->}(-1.,0)(1.,0)}
\rput{0}(7.25,5){\psline{->}(-1.,0)(1.,0)}
\put(5,0){
\rput{0}(3.25,1.){\psline{->}(-1,0)(1.,0)}
\rput{0}(5.25,3){\psline{->}(-1.,0)(1.,0)}
\rput{0}(7.25,5){\psline{->}(-1.,0)(1.,0)}
}
\put(10,0){
\rput{0}(3.25,1.){\psline{->}(-1,0)(1.,0)}
\rput{0}(5.25,3){\psline{->}(-1.,0)(1.,0)}
\rput{0}(7.25,5){\psline{->}(-1.,0)(1.,0)}
}
\psset{linewidth=1pt,linecolor=black,linestyle=solid}
\put(-7,0){
\psline{->}(-.5,0)(4,0)
\rput[c](4,-1){$x_1$}
\psline{->}(0,-.5)(0,4)
\rput[r](-1,4){$x_2$}
\psline{->}(-.3,-.3)(2.5,2.5)
\rput[l](3,3){$x_3$}
}
\psline[linecolor=gray,linewidth=2pt,arrowsize=1]{<-}(25,2.5)(20,2.5)
\rput[c](22.5,1){$H_\ext$}
\end{pspicture}
\caption{The idealized geometry of the sample. The homogeneous
  external field $H_\ext$ is parallel to the long axis of the sample.} 
\end{figure}
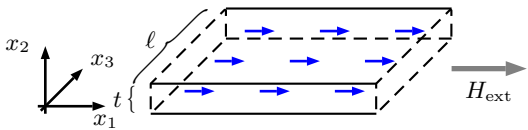  

The experiments suggest that as the strength of the field is reduced
starting from saturation, i.e., $h_\ext<0$,
and finally reversed, a bifurcation at some critical value
$h_\ext^*>0$ of the external 
field $H_\ext=(-h_\ext,0,0)$ is at the origin of the concertina pattern, see
Figure \ref{fig:coarsening} and Figure \ref{CoFeB}.
Due to the unit-length constraint \eqref{constraint},
infinitesimal variations of $m^*$ are of the form  
$\delta m=(0,\delta m_2,\delta m_3)$. 
Since the uniform magnetization only generates
Zeeman energy, the linearization of the energy in $m^*$ -- neglecting anisotropy --  is given
by the exchange energy and the stray-field energy of the
infinitesimal variation, that is, $d^2\int_\Omega |\nabla \delta m|^2\dd x+ \int_\text{all space}
|H_\text{stray}(\delta m)|^2\dd x$, augmented by the linearization of
the Zeeman energy. The latter is due
to the constraint \eqref{constraint} given by $-h_\ext\int_\Omega
(\delta m_2^2 +\delta m_3^2)\d x$ which is a consequence of the expansion
$m_1=(1-\delta m_2^2-\delta m_3^2)^{1/2}\approx 1-\frac{1}{2}(\delta
m_2^2+\delta m_3^2)$.
\subsection{Unstable modes}
\label{modes}
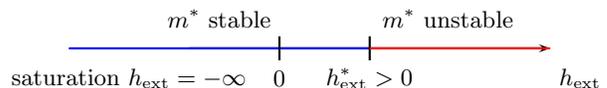
\begin{figure}[htb]
  \centering
\psset{unit=.8}
  \begin{pspicture}(-1.9,0)(6,1.9)
\psline{->}(0,1)(6,1)
\psline[linecolor=blue]{-}(-2,1.)(3,1.)
\rput[c](.5,1.5){$m^*$ stable}
\psline(3,1.2)(3,.8)
\psline[linecolor=red]{-}(3,1.)(6,1.)
\psline(1.5,.8)(1.5,1.2)
\rput[c](1.5,.5){$0$}
\rput[c](3,.5){$h_\ext^*>0$}
\rput[c](4.3,1.5){$m^*$ unstable}
\rput[c](-1,.5){saturation $h_\ext=-\infty$}
\rput[c](6.5,.5){$h_\ext$}
\end{pspicture}
  \caption{The scale of the external field $h_\ext$.}
\label{fieldordering}
\end{figure}

We start with the linear stability analysis of the uniform
magnetization by discussing potentially
unstable modes on the level of the linearization of the energy.
We argue at which value of the external field $h_\ext$ each
of the modes becomes unstable. At this so-called critical field  
the infinitesimal release of Zeeman energy becomes larger than the
infinitesimal contributions due to exchange and stray-field energy.
We neglect uniaxial anisotropy (i.e., we set $Q=0$) for the moment,
since on the level of this infinitesimal discussion, a longitudinal or
transversal anisotropy just leads to a shift of the critical field
$h_\ext^*\leadsto h_\ext^*+ Q$, see Section \ref{anisotropy}.
Since the shift entails
that the sign of the critical field can change, we note in that context that if we speak about {\it reducing} the 
strength of the external field we usually mean that the critical field
is approached from saturation ($h_\ext=-\infty$) if not stated differently. Similarly we
say that the external field is {\it increased} after the
critical field is passed. In this sense, the critical field is
interpreted as the {\it zero point} on the scale of the external field, cf.\
Figure \ref{fieldordering}.

\smallskip

\begin{figure}[htb]
  \centering
\input{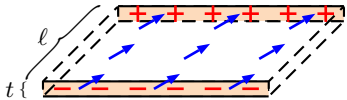}
\caption{Coherent rotation and generated surface charges.}
\label{coherent}
\end{figure}

I) The first mode we discuss is a coherent rotation, i.e.,
$\delta m=(0,\delta m_2,\delta m_3)$ is constant in space. Such a mode releases Zeeman energy
per length in $x_1$-direction of the infinitesimal amount $h_\ext\ell t \delA^2$,
where $\delA=(\delta m_2^2+\delta m_3^2)^{1/2}$ is the infinitesimal amplitude
of the coherent rotation.
A coherent rotation necessarily generates surface charges.
Since the top and bottom surfaces have larger area than the two lateral 
surfaces, an in-plane rotation ($\delta m_3\equiv0)$ is favored, cf.\ Figure \ref{coherent}.
This mode generates surface charges of infinitesimal density $\pm
\delA$. Over distances $\ell$ much larger than $t$,
these surface charges act like two oppositely charged wires at distance $\ell$ of line
density $t \delA$ -- also in the following if not mentioned otherwise always
infinitesimally and per length in $x_1$-direction.
Hence the mode generates an infinitesimal stray field of order $\sim t^2 (\ln{\ell}{t}^{-1}) \delA^2$.
Therefore, this mode becomes unstable when $h_\ext\sim
{t}\ell^{-1}(\ln\ell{t}^{-1})$.

\begin{figure}[htb]
  \centering
\input{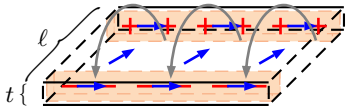}
\caption{Buckling mode and generated volume charges and
  stray field --  for reasons of a clear presentation only drawn in
  the region above the sample.}
\label{buckling}
\end{figure}

II) The second mode we consider is buckling, cf.\ Figure \ref{buckling}. The magnetization avoids the lateral surface charges
by just laterally buckling in the middle of the cross section, i.e., 
$\delta m=(0,\delA\sin(\pi \frac{x_2}{\ell}),0)$. However, since 
$\nabla\cdot\delta m=\pi\ell^{-1}\delA \cos(\pi \frac{x_2}{\ell})$,
the surface charges of the coherent rotation turn into
volume charges. At distances much larger than $t$ from the cross section, these
volume charges act like surface charges of amplitude $\sim \ell^{-1}t \delA$.
Since these surface charges change sign over a distance $\ell$, they generate
a stray field which extends a distance $\sim\ell$ away from the cross section,
and which is of the magnitude $\sim \ell^{-1}t\delA$.
Hence this mode generates a stray field energy $\sim t^2 \delA^2$,
which is only smaller by a logarithm 
than in case of the previous mode of coherent rotation. Moreover, since
$|\nabla\delta m|^2=\pi^2\ell^{-2}\delA^2\cos^2(\pi \frac{x_2}{\ell})$, the mode generates exchange
energy $\sim d^2 \ell^{-1} t \delA^2$. Since the release of Zeeman
energy scales as $\sim h_\ext\ell t \delA^2$ as in case of the first
mode above, this mode becomes unstable
at $h_\ext\sim {d^2}{\ell}^{-2}$ in the regime
$t\ll{d^2}\ell^{-1} $ and at $h_\ext\sim
{t\ell}d^{-2} $ in the regime
$t\gg{d^2}\ell^{-1}$. 
\smallskip

\begin{figure}[htb]
  \centering
  \input{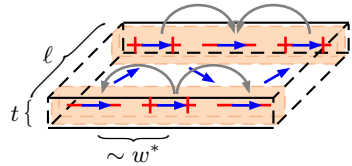}
\caption{Oscillatory buckling mode and generated surface charges
  and stray field --  for reasons of a clear presentation only drawn in
  the region above the sample.}
\label{oscbuckling} 
\end{figure}

 III) The third mode we discuss is oscillatory buckling, cf.\ Figure \ref{oscbuckling}. This mode
reduces the stray-field energy through a modulation of the lateral buckling in $x_1$-direction, i.e.,
$\delta m=(0,\delA\sin(\pi\frac{x_2}{\ell})\sin(2\pi\frac{x_1}{w}),0)$
with a wavelength $w$ that satisfies $t\ll w\ll \ell$. Since $w\gg t$, the
volume charges generated by this mode act like surface charges of amplitude
$\sim\ell^{-1} t\delA$ over distances much larger than $t$ from the cross section. 
However, these surface charges change sign over
a distance $w\ll\ell$, so that the generated stray field only extends over
a distance $\sim w$ away from the cross section. Hence this mode generates
a stray-field energy $\sim \ell^{-1}t^2w\delA^2$, which is substantially 
less than the stray-field energy of the two prior modes for $w\ll \ell$. However, since $w\ll\ell$, the exchange
energy is now dominated by the oscillation in $x_1$-direction, which
leads to an infinitesimal exchange energy $\sim d^2 \ell w^{-2} t \delA^2$.
Hence the wavelength $w$ which leads to the minimal infinitesimal
total stray-field energy and exchange energy of order $\sim d^{2/3}\ell^{-1/3}t^{5/3}\delA^2$ is given by $w^*\sim d^{2/3}\ell^{2/3}t^{-1/3}$.
This is consistent with our assumption $t\ll w\ll \ell$ provided
$d^2\ell^{-1}\ll t \ll (d\ell)^{1/2}$. The oscillatory
buckling mode becomes unstable at a field strength of order
$h_\ext\sim d^{2/3} \ell^{-4/3} t^{2/3}$. 

IV) The fourth mode we consider is curling. This mode avoids charges altogether
by an $x_3$-dependent magnetization, i.e.,
$\delta m=(0,\delA\sin(\pi\frac{x_2}{\ell})\cos(\pi\frac{x_3}{t}),
\delA\ell^{-1}t\cos(\pi\frac{x_2}{\ell})\sin(\pi\frac{x_3}{t}))$.
The exchange energy in this case is dominated by
the gradient in $x_3$-direction which scales as $\sim d^2 \ell t^{-1} \delA^2$.
Hence the curling mode becomes unstable at $h_\ext\sim d^2 t^{-2}$.
\smallskip

\begin{figure}[htb]
\psset{unit=1.8}
\begin{pspicture}(0,.8)(2.2,3)
\pspolygon[linestyle=none,fillstyle=solid,fillcolor=gray](1,1)(2,1)(2,2)
 \psset{plotpoints=10}
\psaxes[labels=none,ticks=none]{->}(0,1)(2.2,3)
\psline(1,0.9)(1,1.1)
\rput[c](1,0.7){1}
\rput[r](-0.2,1){1}
\infixtoRPN{1/(x*log(1/x^2+4))*log(5)}\psplot{0.135}{1}{\RPN}
\infixtoRPN{1/x}\psplot[linestyle=dashed]{0.34}{1}{\RPN}
\infixtoRPN{x^2}\psplot[linestyle=dotted]{1}{1.7}{\RPN}
\infixtoRPN{x}\psplot{1}{2}{\RPN}
\rput[c](.35,1.5){I}
\rput[c](.35,1.3){coherent}
\rput[c](.35,2.2){II}
\rput[c](.35,2.0){buckling}
\rput[c](1,2.5){III}
\rput[c](1,2.3){osc.\ buckling}
\rput[c](1.8,2.3){IV}
\rput[c](1.8,2.1){curling}
\rput[c](2.2,.7){$\frac{t}{d}$}
\rput[r](-.2,3){$\frac{\ell}{d}$}
\end{pspicture}
  \caption{Phase diagram with the four regimes for the nucleation.}
\label{phasediagram}
\end{figure}
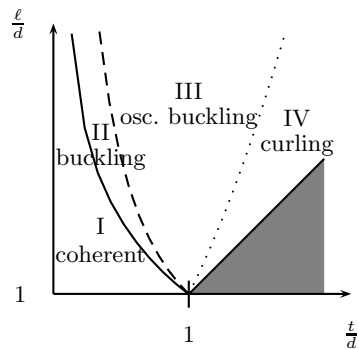

The discussion above shows that there are (at least) four different parameter
regimes for the nucleation -- expressed in terms of the two
non-dimensional parameters $t/d\ll \ell/t$. These regimes are
characterized by a certain scaling of the critical field $h_\ext^*$
in the sense that one of the modes becomes
unstable as the external field passes the corresponding
field, while the other three modes are still stable, cf.\ Figure
\ref{phasediagram}.      
In particular, the oscillatory buckling mode is the first mode to
become unstable at a field 
$h_\ext^*\sim d^{2/3} \ell^{-4/3} t^{2/3}$ in the regime ${d^2}{\ell^{-1}}\ll t \ll
(d\ell)^{1/2}$.
By a refinement of the above discussion, it can be rigorously shown that there
are exactly four regimes, cf.\ Theorem 1 in \cite[p.357]{CAO04}.

\subsection{Period of the unstable mode: Experiment vs. theory}
\label{period} 
\begin{figure}[htb]
  \centering
\includegraphics[width=0.4\textwidth]{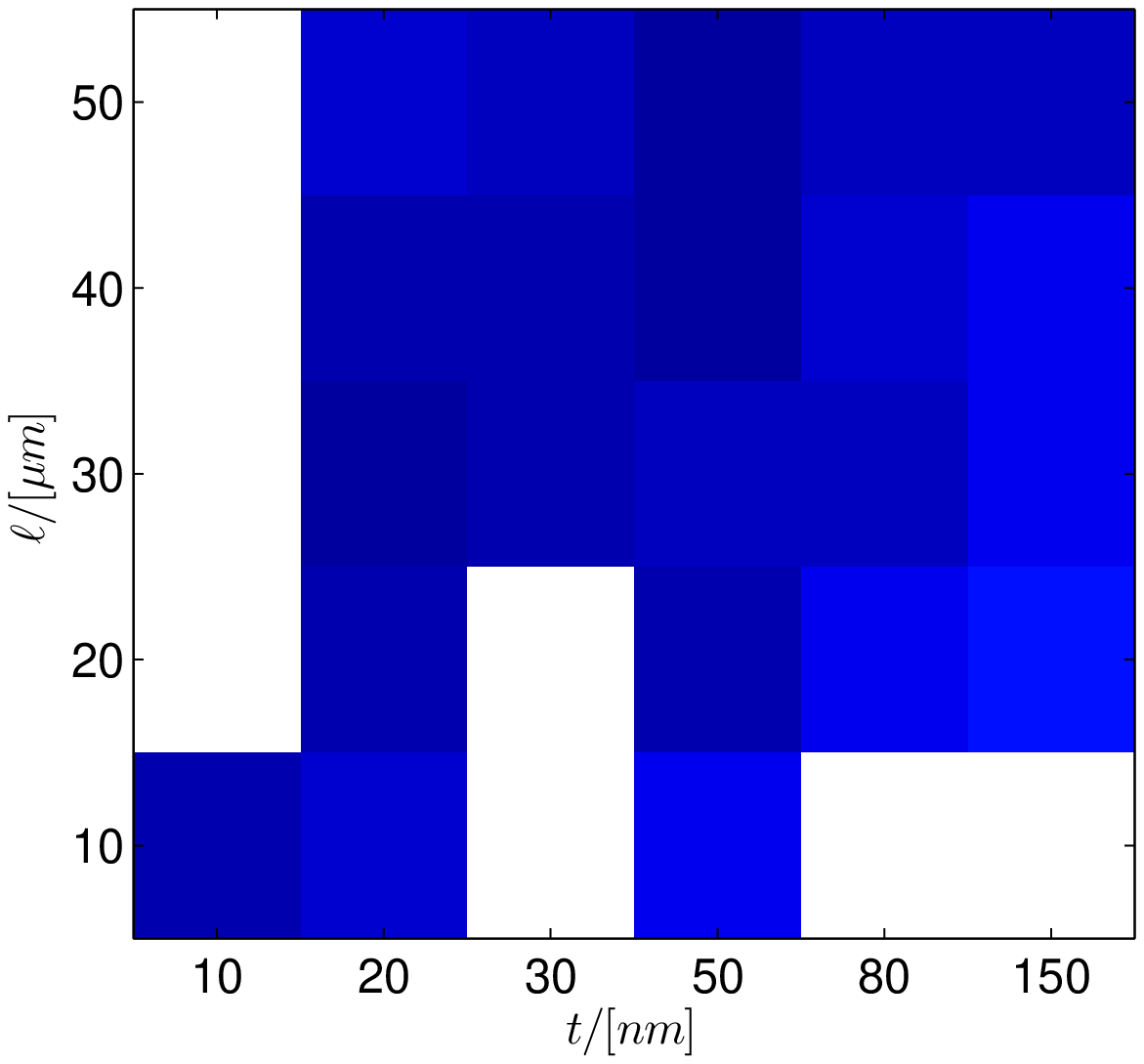}
\includegraphics[width=0.4\textwidth]{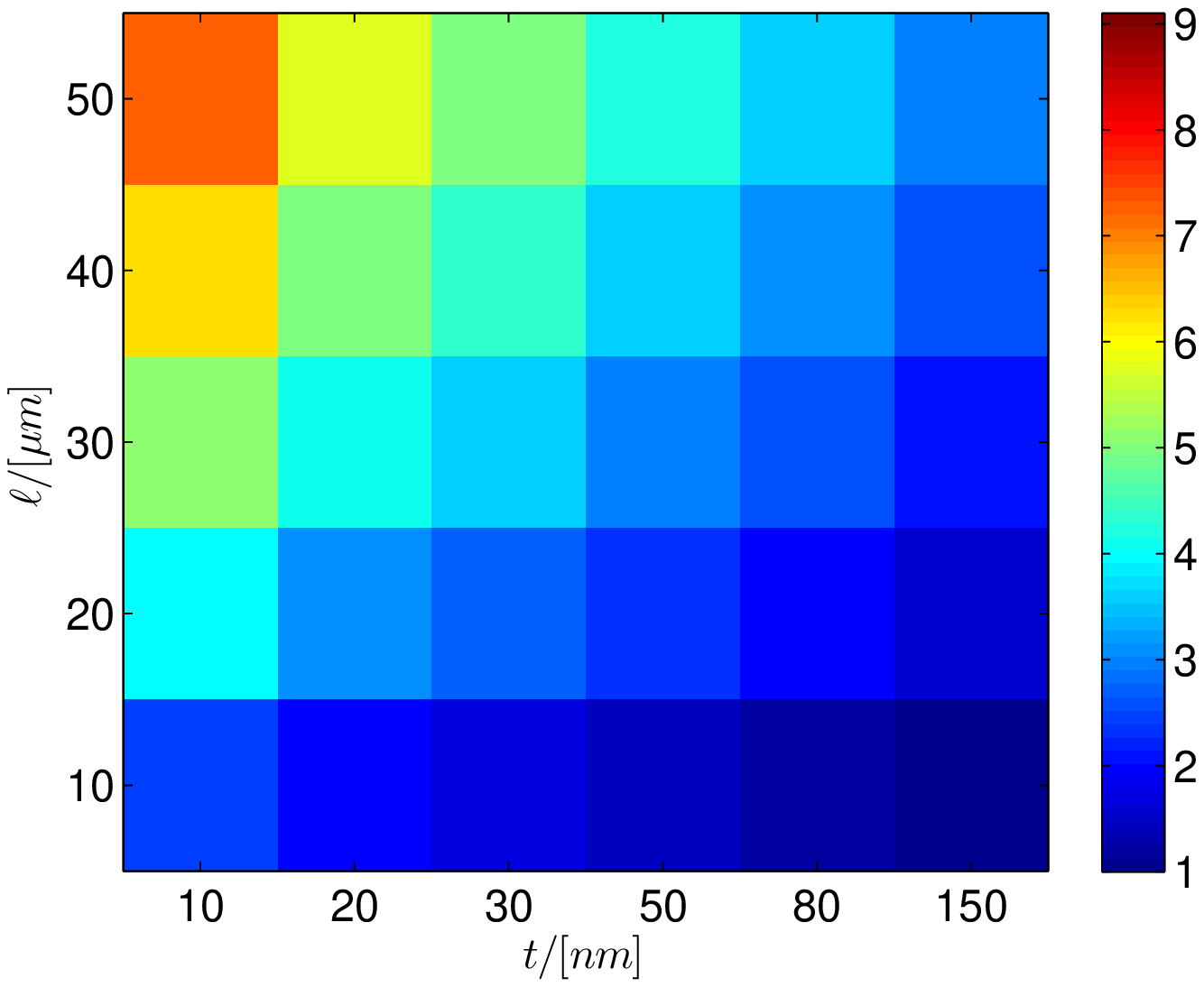}
  \caption{The theoretical period of the unstable mode is in good
    correspondence to the measurements: The upper image shows
the ratio of the experimentally observed period and
the period of the unstable mode. The white patches correspond to broken or defect-ridden samples. The lower displays the ratio of the
period $w^*$  and the smallest period, i.e., $w^*(\ell=50
\upmu\text{m},t=150\text{nm})$, at
all. Both images share the same color map.}
\label{expinstab}
\end{figure}

Clearly, the regime of interest for us is the Regime III,
i.e., the oscillatory buckling regime characterized by ${d^2}\ell^{-1}\ll t \ll (d\ell)^{1/2}$.
In this regime, an asymptotic analysis of the linearization of the
energy on the basis of the above discussion shows that the (first) unstable mode is indeed asymptotically of the form $\delta
m=(0,\delA\sin(\pi\frac{x_2}{\ell})\sin(2\pi\frac{x_1}{w}),0)$, cf.\
Theorem 1 in \cite[p. 389]{CAO05} and see
also below. Based on a refinement of the prior linear stability
analysis one can moreover determine the asymptotic behavior of $w^*$
{\it including} the numerical factor that is given by
\begin{equation}
\label{wavenumber}
  w^*\approx(32\pi)^{1/3}d^{2/3}\ell^{2/3}t^{-1/3}.
\end{equation}
So far we have learned that in Regime III at field strengths
$h_\ext^*\sim d^{2/3}\ell^{-4/3} t^{2/3}$ there is a bifurcation in direction of the unstable mode $\delta
m=(0,\delA\sin(\pi\frac{x_2}{\ell})\sin(2\pi\frac{x_1}{w^*}),0)$. 
We claim that the concertina pattern grows out of this
unstable mode. If so, the experimentally observed period
$w^*_\text{exp}$ should be close to the period $w^*$ of
the unstable mode. Defining and determining $w^*_\text{exp}$ is delicate:
As $h_\ext$ 
increases (after the critical field $h_\ext^*$ is passed), there is a continuous transition from the magnetization
ripple, see Subsection \ref{ripplesec}, to the concertina pattern, which is far from
exactly periodic, and which coarsens subsequently, see Section
\ref{coarsening section}.
As $w^*_\text{exp}$ we take the average period as soon as the concertina
pattern is discernible to the eye.
Figure \ref{expinstab} shows the result of this comparison for a
broad range of sample dimensions $\ell$ and $t$ and (therefore) a
fairly broad range of periods
$w^*$: The ratio of
the smallest width $\ell$ compared to the largest is $5$, the ration of the
smallest thickness $t$ compared to the largest is
$15$. The smallest period $w^*$ is
expected for a thick film of small width, the largest period for a
thin film of large width, differing by a factor close to six
(neglecting the prediction for the broken or defect samples).
The ratio $\frac{w^*_\text{exp}}{w^*}$ of the experimental period with
respect to the
prediction ranges around two. We basically see this as a confirmation of our
hypothesis, namely that the concertina grows out of the oscillatory buckling. Notice that the deviation has a clear trend: $w_\text{exp}^*$
is larger than $w^*$. We give an explanation for this systematic
deviation in Section \ref{coarsening section}. 
\subsection{A reduced energy functional}\label{Reduziertes Modell}
In the forthcoming section we start with the investigation of the type
of the bifurcation. For the 
moment we continue to neglect anisotropy, although it may affect the 
type of bifurcation as we shall discuss in Section \ref{anisotropy}. 
In order to understand the type of bifurcation
we now first pass to a reduced model adapted to our Regime III: The
form of the unstable mode suggests that the out-of-plane component 
and the dependence on the thickness variable are negligible, i.e., we assume \
$m_3\equiv0$ and $m=m(x_1,x_2)$ respectively. Since the
unstable mode varies faster in $x_1$-direction than in
$x_2$-direction, we neglect $|\partial_2 m|^2$ with respect to
$|\partial_1 m|^2$ in the exchange energy density.
Since the oscillation in the sign of the charge density is on smaller
length scales in $x_1$-direction than in $x_2$-direction, we neglect
$h_2^2$ with respect to $h_1^2+h_3^2$ in the stray-field energy density, where $H_\stray=(h_1,h_2,h_3)$. Finally, since we are interested
in small deviations from $m^*=(1,0,0)$, we expand $m_1=\sqrt{1-m_2^2}\approx1-\frac{m_2^2}{2}$,
so that we may neglect $|\nabla m_1|^2$ with respect to $|\nabla m_2|^2$ in the
exchange energy density. We also use this approximation in the
charge density and in the Zeeman energy. Hence (up to an additive constant)
we are left with the {\it reduced energy functional}
\begin{multline}
E(m_2)\approx d^2 t \int_{\Omega'} (\partial_1m_2)^2 \dd x_1\dd x_2\\
+\int_\text{all space}( h_1^2+h_3^2) \;\dd x_1\dd x_2\dd x_3
-h_\ext \;t \int_{\Omega'} m_2^2 \;\dd x_1\dd x_2,
\label{reduziertesmodell}
\end{multline}
where the stray field $H_\stray=(h_1,0,h_3)$ is determined via
\begin{align}
&\partial_3 h_1-\partial_1 h_3=0,\notag\\
\int_\text{all space} (h_1\partial_1\zeta+h_3\partial_3&\zeta) \dd x_1\dd x_2\dd x_3\notag\\
\quad=t\int_{\Omega'} (-\tfrac{m_2^2}{2}&\partial_1\zeta+m_2\partial_2\zeta)
\dd x_1\dd x_2\text{ for all } \zeta,\label{field}
\end{align}
which is a consequence of the alternative formulation \eqref{alternative}.
Here, $\Omega'$ denotes the in-plane cross section of our sample
$\Omega=\Omega'\times(0,t)$.
\smallskip

We note that the stray-field energy is only finite if $m_2$ vanishes
at the lateral long edges,
i.e., $m_2(x_1,x_2)=0$ for $x_2=0,\ell$ (as is true for the unstable mode).
Notice that \eqref{field} can be written as
\begin{align*}
\partial_1 h_1+\partial_3 h_3=0\quad\mbox{for}\;x_3\not=0,\\
[h_3]\;=\;t(-\partial_1\tfrac{m_2^2}{2}+\partial_2m_2)\quad\mbox{for}\;x_3=0,
\end{align*}
where $[h_3]$ denotes the jump $h_3$ experiences across $x_3=0$.
This formulation shows that $x_2$ is just a parameter in the equations for the stray field,
which behaves like a two-dimensional stray field (in the $x_1x_3$-plane) generated
by the ``line charge'' $t(-\partial_1\frac{m_2^2}{2}+\partial_2m_2)$.
\smallskip

We note that the only non-quadratic term in the energy comes from
the non-linear charge distribution
$t(-\partial_1\frac{m_2^2}{2}+\partial_2m_2)$. This allows us to derive the scaling of the amplitude of the magnetization: It should be such
that both terms in the charge distribution balance. Since in view of the
unstable mode the typical $x_1$-scale of the variations of $m_2$ is given by 
$w^*\sim d^{2/3}\ell^{2/3}t^{-1/3}$,
whereas the typical $x_2$-scale of variations of $m_2$ is given by the sample width $\ell$,
the contributions $\partial_1\frac{m_2^2}{2}$ and $\partial_2m_2$ balance provided the
amplitude of $m_2$ scales as $d^{2/3}\ell^{-1/3}t^{-1/3}$. This suggests the
following non-dimensionalization of length, and reduced units for
the stray field and the magnetization:
\begin{align}\label{rescaling1}
\begin{split}
x_1=d^{2/3}\ell^{2/3}{t}^{-1/3} \hat x_1,\; x_2&=\ell \hat
x_2, \; x_3= d^{2/3}\ell^{2/3}{t}^{-1/3}\hat x_3,\\
h_1=d^{2/3}\ell^{-4/3}t^{2/3}\hat h_1,&\quad
h_3=d^{2/3}\ell^{-4/3}t^{2/3}\hat h_3,\quad\\
m_2=d^{2/3}&\ell^{-1/3}t^{-1/3}\hat m_2.
\end{split}
\end{align}
After the reduction leading to \eqref{reduziertesmodell}, only the stray-field depends on the $x_3$-component for which the relevant length scale is
of course the wavelength
of the oscillation, cf.\ Section \ref{modes}, III and Figure \ref{oscbuckling} -- the relevant length scale in case of the
magnetization leading to the reduction \eqref{reduziertesmodell} is
of course the film thickness $t$.  
If we also rescale the external field -- in the same way as the stray
field -- and the energy itself
according to
\begin{align}
h_\ext&=d^{2/3}\ell^{-4/3}t^{2/3}\hat h_\ext,\label{fieldscaling}\\
E &=d^{8/3}\ell^{-1/3} t^{2/3} \hat E,\label{energyscaling}
\end{align}
we obtain the reduced {\it rescaled} energy functional
\begin{multline}
\hat E(\hat m_2)= \int_{\hat \Omega'} (\hat \partial_1\hat
m_2)^2 \dd\hat x_1\dd\hat x_2
\\+\int_\text{all space} (\hat h_1^2+\hat h_3^2) \dd\hat x_1\dd\hat x_2\dd\hat x_3
-\hat h_\ext  \int_{\hat \Omega'} \hat m_2^2 \dd\hat x_1\dd\hat x_2,
\label{rescaledenergy}
\end{multline}
where the reduced {\it rescaled} stray field is determined by 
\begin{align*}
\hat \partial_1 \hat h_1+\hat \partial_3 \hat
h_3=0\quad\mbox{for}\;\hat x_3\not=0,\\
[\hat h_3]\;=\;(-\hat\partial_1\tfrac{\hat m_2^2}{2}+\hat
\partial_2\hat m_2)\quad\mbox{for}\;\hat x_3=0,
\end{align*}
The reduced {\it rescaled} formulation shows that the reduced
energy functional contains just {\it one} non-dimensional parameter,
namely the reduced external field $\hat h_\ext$ -- instead of {\it four} parameters (exchange length,
sample dimensions and $h_\ext$) for the full model.
Moreover, the {\it vector} field $m=(m_1,m_2,m_3)$, function of {\it three}
variables $(x_1,x_2,x_3)$,
has been replaced by the {\it scalar} function $\hat m_2$, function
of {\it two} variables $(\hat x_1,\hat x_2)$. Finally, the computation of the stray field
is a {\it two}-dimensional computation (in $(\hat x_1,\hat x_3)$ only
with $\hat x_2$ as a parameter) instead of a {\it three}-dimensional
one. 
All this simplifies both the
theoretical treatment and the numerical simulation. For clarity,
we will mostly discuss our results in the rescaled variables
\eqref{rescaledenergy}
-- and only occasionally return to the original variables,
mostly for comparison with the experiment and when we take into account anisotropy.
\smallskip

In Theorem 3 in \cite[p.233]{CAOS06}, we rigorously show that the reduced energy functional
is the scaling limit of the renormalized full micromagnetic energy in Regime III. 
\subsection{Bifurcation}
\label{bifurcationsection}
We now return to the issue of the type of bifurcation on the
level of the reduced model. Let us note that the Hessian of the reduced model in $\hat m_2\equiv
0$ can be explicitly diagonalized and the first unstable mode
is given by $\hat m_2^{*}=\sin(\pi \hat x_2)\sin(2\pi\tfrac{
  x_1}{\hat w^*})$, where $\hat w^*=(32\pi)^{1/3}$ in agreement with \eqref{wavenumber}.
The reduced critical field is given by 
\begin{equation}
\hat h_\ext^*=3\left(\tfrac{\pi}{2}\right)^{4/3}.\label{criticalfield}
\end{equation}
In order to determine the type of bifurcation, we have to investigate
the energy functional $\hat E$ close to the one-dimensional subspace
$\{A \hat m_2^*\}$ generated by the
unstable mode $\hat m_2^*$.
Because of the invariance
of both the energy $\hat E$ and the unstable mode $\hat m_2^*$ under the transformation 
($\hat m_2\leadsto -\hat m_2$ and $\hat x_2\leadsto 1-\hat x_2$), 
all odd terms in the amplitude $A$ in the expansion of $\hat E(A \hat m_2^*)$ vanish. 
In particular the cubic term vanishes so that the bifurcation is degenerate. 
\smallskip

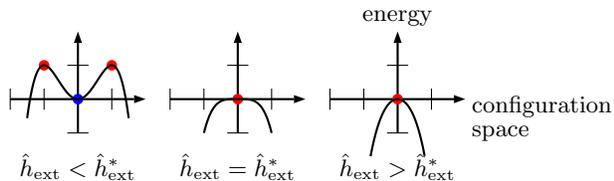
\begin{figure}[htb]
  \centering
  \psset{unit=0.45cm}
  \begin{pspicture}(-2,-3)(2,2.5)
 \psset{plotpoints=1000,labels=none,arrowinset=0}
 \psaxes*{->}(0,0)(-2,-1)(2,2)
 \infixtoRPN{4*(-1/4*x^4+1/2*x^2)}
\pscircle[fillstyle=solid,fillcolor=blue,linecolor=blue](0,0){2pt}
\pscircle[fillstyle=solid,fillcolor=red,linecolor=red](-1,1){2pt}
\pscircle[fillstyle=solid,fillcolor=red,linecolor=red](1,1){2pt}
 \psplot[linecolor=black]{-1.5}{1.5}{\RPN}
\rput[c](0,-2){$\hat h_\ext<\hat h^*_\ext$}  
 \end{pspicture}\quad
  \begin{pspicture}(-2,-3)(2,2.5)
 \psset{plotpoints=1000,labels=none,arrowinset=0}
 \psaxes*{->}(0,0)(-2,-1)(2,2)
 \infixtoRPN{-(x^4)}
\pscircle[fillstyle=solid,fillcolor=red,linecolor=red](0,0){2pt}
 \psplot[linecolor=black]{-1}{1}{\RPN}
\rput[c](0,-2){$\hat h_\ext=\hat h^*_\ext$}  
\end{pspicture}\quad
  \begin{pspicture}(-2,-3)(5,2.5)
 \psset{plotpoints=1000,labels=none,arrowinset=0}
 \psaxes*{->}(0,0)(-2,-1)(2,2)
\rput[l](2.2,-.2){configuration}
\rput[l](2.2,-1){space}
\rput[c](0,2.4){energy}
 \infixtoRPN{-2*(x^2)-x^4}
\pscircle[fillstyle=solid,fillcolor=red,linecolor=red](0,0){2pt}
 \psplot[linecolor=black]{-.8}{.8}{\RPN}
\rput[c](0,-2){$\hat h_\ext>\hat h^*_\ext$} 
 \end{pspicture}
  \caption{Energy landscape close to the bifurcation. The loss of
    stability at the critical field leads to a first-order phase
    transition -- on a large scale however the energy is coercive.}
\label{Energy landscape}
\end{figure}

This degeneracy of the bifurcation means that at the critical field 
strength $\hat h_\ext^*$, the first non-vanishing term in the expansion
of $\hat E(A \hat m_2^*)$ with respect to $A$ is at least quartic. Hence it is
not sufficient to consider $\hat E$ just along the linear space
$\{A  \hat m_2^*\}$ but it has to be analyzed along a curve
$\{A  \hat m_2^*+A^2\hat m_2^{**}\}$ in configuration space. Indeed, the
curvature direction $\hat m_2^{**}$ affects the
quartic term in the expansion and has to be determined such that 
$\hat E$ is minimal.
This minimization problem of the coefficient of the quartic term
is quadratic in $\hat m_2^{**}$ and thus can be solved explicitly. We obtain
\begin{equation*}
\hat  m_2^{**}=-\tfrac{1}{10}(\tfrac{2}{\pi})^{1/3}\sin(2\pi{ x_2})\sin(4\pi\tfrac{ x_1}{\hat  w^*}),
\end{equation*}
which leads to a negative coefficient of the quartic term in the
expansion of $\hat E$, namely
\begin{multline}\label{Ot.1}
 \hat E(A \hat m_2^*+A^2 \hat m_2^{**})\approx
(\hat h_\ext-
\hat h_\ext^*)\left(\tfrac{ \pi} {2}\right)^{1/3}A^2
-\tfrac{\pi}{640}A^4.
\end{multline}
The negative quartic coefficient implies that the bifurcation is
subcritical or of first order. Subcriticality entails that 
close to $\hat m_2\equiv 0$, there are only {\it unstable}
stationary points for $\hat h_\ext$ slightly below $\hat h_\ext^*$,
and {\it no} stationary points close to $\hat m_2\equiv 0$ for $\hat
h_\ext$ slightly above $\hat h_\ext^*$, cf.\ Figure \ref{Energy landscape}. 

\begin{figure}[htb]
\includegraphics[width=0.2\textwidth]{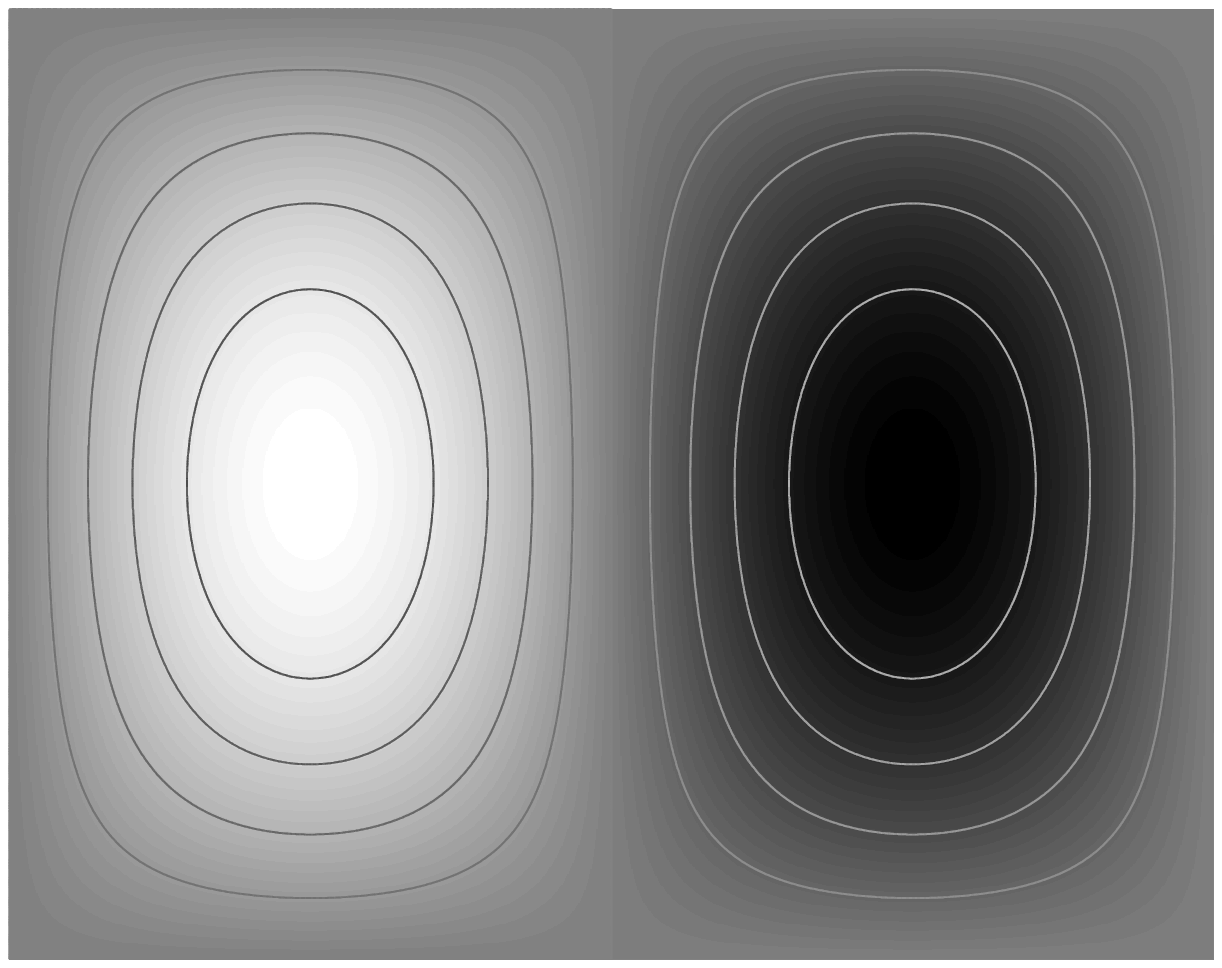}
\begin{pspicture}(0,0)
\psset{linecolor=red,linestyle=solid,linewidth=6.5}
\put(-2.73,2.1){
\psline[linewidth=1.pt](0,0)(.4,0)
\psline[linewidth=1.pt](0.2,-0.2)(0.2,0.2)}
\put(-1.35,2.1){
\psline[linewidth=1.pt](0,0)(.4,0)}
\put(-1.35,.6){
\psline[linewidth=1.pt](0,0)(.4,0)
\psline[linewidth=1.pt](0.2,-0.2)(0.2,0.2)}
\put(-2.73,.6){
\psline[linewidth=1.pt](0,0)(.4,0)}
\end{pspicture}
\includegraphics[width=0.2\textwidth]{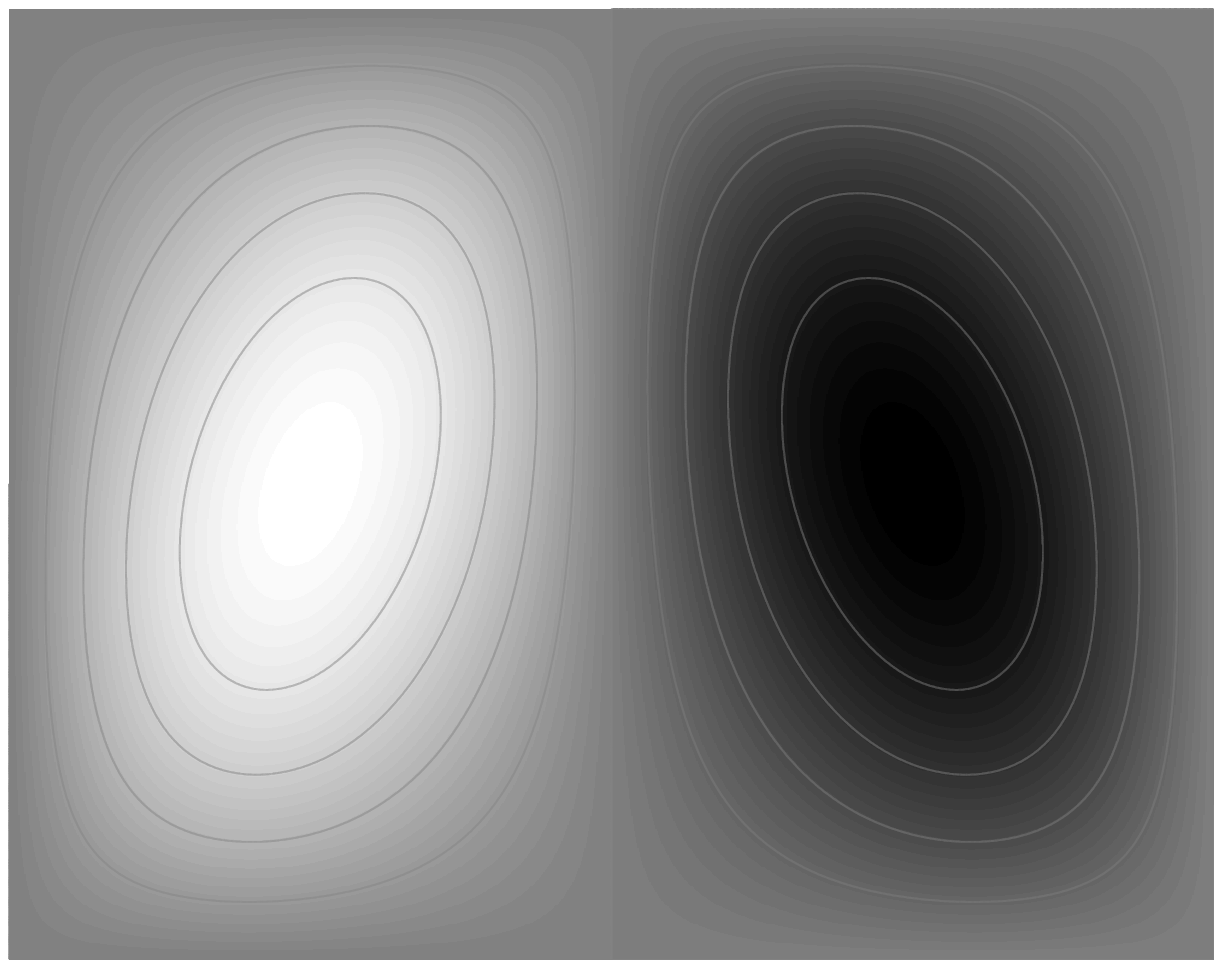}
\begin{pspicture}(0,0)
\psset{linecolor=red,linestyle=solid,linewidth=6.5}
\put(-2.3,2.1){
\psline[linewidth=1.pt](0,0)(.4,0)
\psline[linewidth=1.pt](0.2,-0.2)(0.2,0.2)}
\put(-1.7,2.1){
\psline[linewidth=1.pt](0,0)(.4,0)}
\put(-1,.6){
\psline[linewidth=1.pt](0,0)(.4,0)
\psline[linewidth=1.pt](0.2,-0.2)(0.2,0.2)}
\put(-3,.6){
\psline[linewidth=1.pt](0,0)(.4,0)}
\end{pspicture}
  \caption{Unstable mode $\{A\hat m_2^*\}$ and additional curvature
    correction $\{A\hat m_2^*+A^2\hat m_2^{**}\}$ with generated charges.}
\label{unstablemode}
\end{figure}

At first sight it is surprising that the stray-field energy
contribution to $\hat E$, which gives rise to the only quartic term in 
$\hat m_2$, and clearly is non-negative, may nevertheless allow for a negative
coefficient in front of the quartic term in the expansion (\ref{Ot.1}). 
This comes from the fact that the two terms in the charge density 
$-\hat \partial_1 \frac{\hat m_2^2}{2}+\hat \partial_2 \hat m_2$
interact, giving rise to a cubic term in $\hat m_2$ (quartic in $A$), which indeed allows for cancellations.
The way how this operates is better understood in physical space: 
The term $ \hat m_2^{**}$ in $A \hat m_2^*+A^2 \hat m_2^{**}$
(the curvature direction in configuration space) induces a 
tilt of the symmetric
charge distribution of $A \hat m_2^{*}$, see Figure \ref{unstablemode}. 
This tilt brings opposite charges closer together, 
thereby reducing the stray field energy -- while increasing the exchange energy
to a lesser amount.
\smallskip

Since the bifurcation is first order, it is not obvious whether 
minimizers of the reduced energy functional can be related to the
unstable mode. In particular, this finding sheds doubt on our
hypothesis that the concertina pattern inherits the period of the
unstable mode. It is even not obvious whether minimizers of the
reduced energy functional exist at all. However, one can show that the
reduced energy is coercive for all values of the external field $ \hat h_\ext$,
see Theorem 4 in \cite[p.236]{CAOS06}. This in particular
implies that there always exists a global minimizer of the reduced
energy -- which corresponds to a local minimizer of the original energy
\eqref{B1:Micromag}, see Theorem 5 in \cite[p.237]{CAOS06} -- in
particular for fields larger than the critical field. But it is not
immediately clear how these minimizers relate to the unstable mode.
\smallskip

\begin{figure}[htb]
\includegraphics[width=0.45\textwidth]{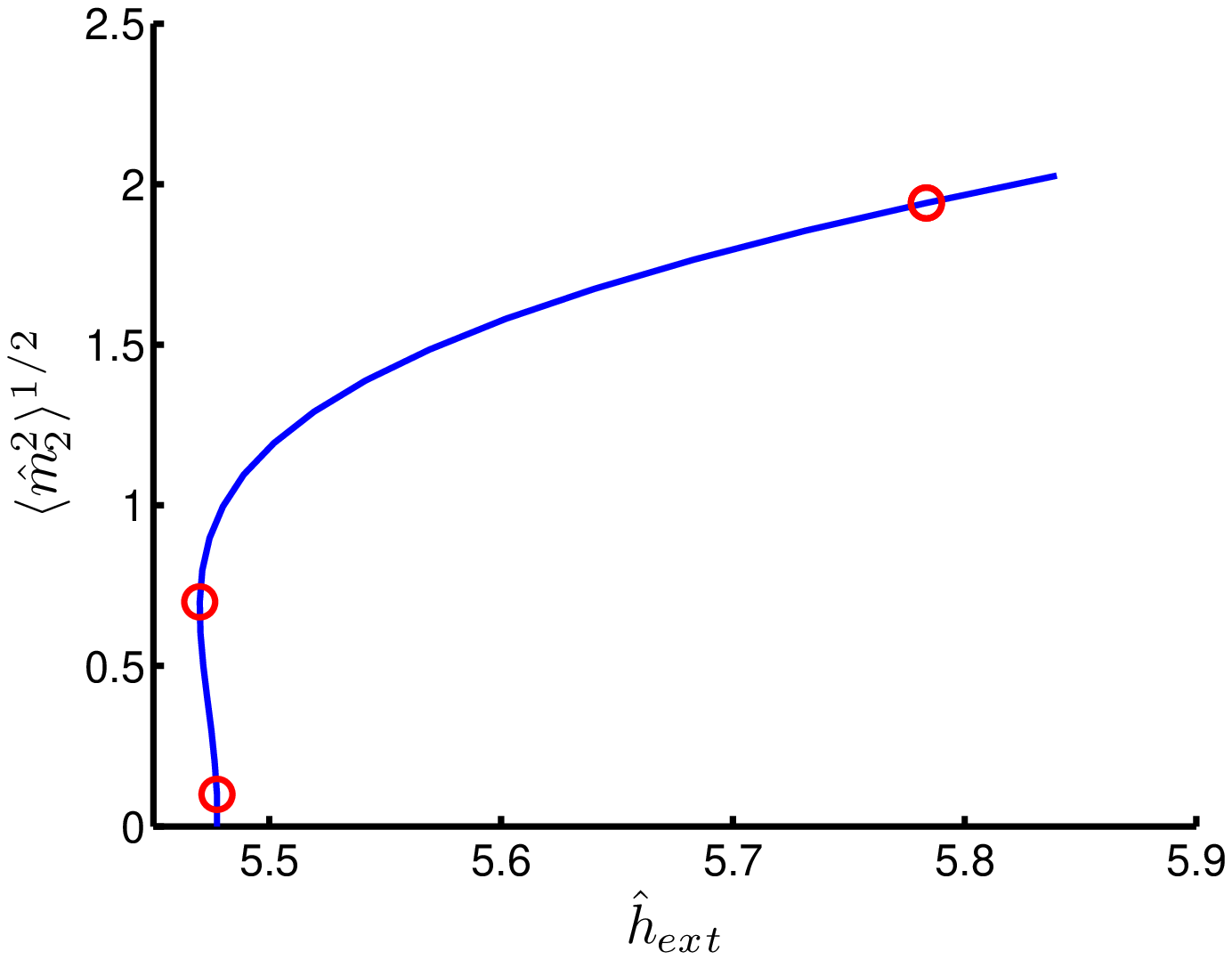}
\includegraphics[width=0.45\textwidth]{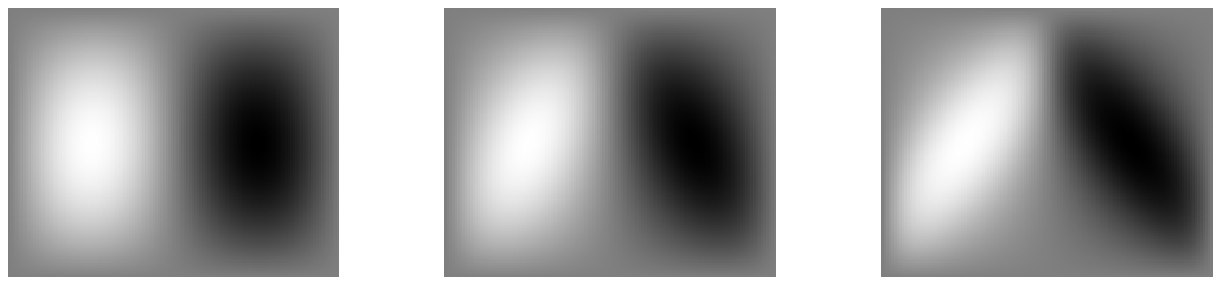}
\caption{Numerical simulations: The $\hat w^*$-periodic branch close to the bifurcation and
  the pattern at the indicated fields. The gray scales encode
  the $m_2$ component but are {\it not} comparable. The whole spectrum is
  exhausted so that the structure of the pattern can best be
  resolved. By $\langle \hat m_2^2 \rangle^{1/2}$ we denote the
  spatial root mean square
 of $\hat m_2$, i.e., the amplitude of the average magnetization.}
\label{wsternbranch}
\end{figure}

It is natural to resort to numerical simulations; details on the
discretization and the algorithms are provided in Section \ref{Numerics}.  
To confirm the conjecture that the unstable mode in Regime III is
indeed related to the concertina pattern, we use a numerical
path-following algorithm in order to compute the bifurcation branch.
Figure \ref{wsternbranch} shows the outcome of the numerical simulations.
As expected due to the coercivity of the energy functional, we find a
turning point as we follow the bifurcation branch. The turning
point is located at a field which is just slightly -- about
one percent -- smaller than the
critical field. After the turning point the configurations become stable,
at least under perturbations of the same period.  
\smallskip

\begin{figure}[htb]
\includegraphics[width=0.45\textwidth]{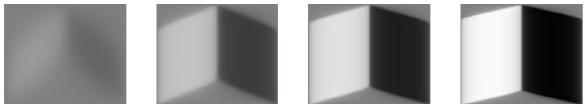}
\caption{Numerical simulations: The $\hat w^*$-periodic concertina pattern exhibits a clear scale
  separation as $\hat h_\text{ext}$ increases; $\hat h_\text{ext}=6.84,\,23.2,\,40.1,\,57.3$ from left to right. The gray scales linearly encode the
  $\hat m_2$-component and are comparable.}
\label{scalesep}
\end{figure}

As the field increases beyond the turning point, the unstable mode
develops into a domain pattern of concertina type with its typical scale
separation between the wall width and the domain size, cf.\ Figure
\ref{scalesep}. We thus find a continuous transition from the unstable mode to
the concertina pattern --  confirming our hypothesis.
\smallskip

The numerical simulations lead
to the conjecture that the
magnetization in a perfectly homogeneous, isotropic sample exhibits 
a first-order phase transition from the uniformly
magnetized state to the concertina state of period $w^*$ at the
critical field. Clearly this does not explain the deviation of the
average wavelength in the experimental measurements from the
theoretical prediction. Before we address this deviation we now
introduce a sharp-interface model, so-called {\it domain theory}, that is used to
investigate the further transformation of the concertina for large external fields
$\hat h_\ext\gg 1$, in particular the coarsening, see Subsections \ref{optimal period},
\ref{coarsesec}, and \ref{PredInstab}. Our explanation of the
coarsening will also provide an understanding of the initial deviation
of the period, cf.\ Subsection \ref{extendbif}.

\section{Domain theory}
In the numerical simulations, we observe for large external fields a clear scale separation
between domains, where the
magnetization is almost constant, and walls, in which the
magnetization quickly turns, cf.\ Figure
\ref{scalesep}. 
This suggests the application of a sharp-interface model, namely domain
theory. In the following we first
discuss admissible Ansatz functions and then derive their energy within
domain theory. This leads to a model which only depends on a small number of
parameters in configuration space that is used in Section \ref{coarsening section} and \ref{anisotropy} in order to get a
better understanding of the coarsening of the concertina.    
 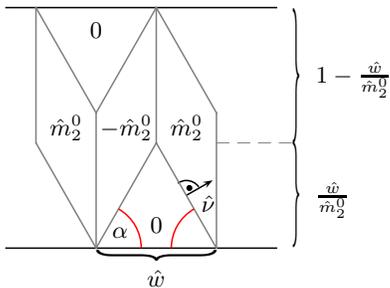
\begin{figure}[htb]
  \centering
\psset{unit=0.4} 
  \begin{pspicture}(0,-1.5)(12,8)
\psline[linecolor=gray,linestyle=dashed,linewidth=0.02](7,3.5)(9.5,3.5)
\psset{linewidth=0.05,linecolor=black,arrows=c-c}
    \put(6,1.75){\psline{->,unit=0.5}(0,0)(1.75,1)} 
\psarc[linecolor=black]{-}(6,1.75){.5}{35}{125}
\pscircle[fillstyle=solid,fillcolor=black](6.1,2){.1}
    \put(6.5,1.2){$\hat \nu$}
\rput[c](4,4){$- \hat m_2^0$}
    \rput[c](6,4){$\hat  m_2^0$}
    \rput[c](2,4){$\hat  m_2^0$}
    \rput[c](5,0.75){$0$}
    \rput[c](3,7.25){$0$}
    \put(9.5,4.6){{\psscaleboxto(.3,4.5){\(\}\)}}}
    \rput[l](10.3,1.6){$\tfrac{\hat  w}{\hat  m_2^0}$}
    \put(9.5,.9){{\psscaleboxto(.3,3.5){\(\}\)}}}
    \rput[l](10.3,5.6){$1-\tfrac{\hat  w}{\hat m_2^0}$}
    \rput[c](3.8,.5){$\alpha$}

\put(3,-0.5) {\rotateright{\psscaleboxto(.5,4){\(\}\)}}} 
\rput[c](5,-1){$\hat w$}

\psset{linewidth=0.05,linecolor=black,arrows=c-c}
        \pnode(0,0){a}
    \pnode(9,0){b}
    \pnode(0,8){c}
    \pnode(9,8){d}
    \pnode(1,8){e}
    \pnode(1,3.5){f}
    \pnode(5,8){g}
    \pnode(5,3.5){h}
    \pnode(3,0){i}
    \pnode(3,4.5){j}
    \pnode(7,0){k}
    \pnode(7,4.5){l}
\SpecialCoor
\psset{linewidth=0.05,linecolor=black,arrows=c-c}
    \ncline{a}{b}
    \ncline{c}{d}
\psset{linewidth=0.05,linecolor=gray,arrows=c-c}    
    \ncline{e}{f}
    \ncline{g}{h}
    \ncline{i}{j}
    \ncline{k}{l}

    \ncline{f}{i} 
    \ncline{e}{j}
    \ncline{j}{g}
    \ncline{i}{h}
    \ncline{l}{g}
    \ncline{k}{h}
\psarc[linecolor=red]{-}(3,0){1.5}{0}{60}
\psarc[linecolor=red]{-}(7,0){1.5}{120}{180}
\end{pspicture}
 \caption{Domain theory: The mesoscopic charge-free Ansatz function.}
 \label{ansatz}
 \end{figure}
\smallskip

On a mesoscopic scale, the computed magnetization is close to a
piecewise constant magnetization of amplitude $\hat m_2^0$, i.e., $\hat m_2=\pm \hat m_2^0$ in
the quadrangular domains and $\hat m_2=0$ in the triangular domains as
indicated in Figure \ref{ansatz}.
We observe that the angles in the pattern are
related to the amplitude of the magnetization
$\hat m_2^0$, cf.\ Figure \ref{scalesep}; approximately we have that $\sin
\alpha= 2\hat  m_2^0$. This is related to the fact that the (reduced) stray-field
energy is strongly penalized for large fields, as we shall explain
now. In fact, the piecewise constant
magnetization is a distributional solution of
\begin{equation}
\label{Burgers}
  -\hat \partial_1(\tfrac{\hat m_2^2}{2}) + \hat \partial_2 \hat m_2=0, 
\end{equation}
which means that the normal component of the vector field
$(-\tfrac{\hat m_2^2}{2},\hat  m_2)$ is continuous across the
interfaces -- this is a version of the Rankine-Hugoniot condition in the theory of conservation laws. This condition obviously holds in case of the vertical walls. In
case of the diagonal walls the condition
\begin{equation}
0=[\hat \nu \cdot (-\tfrac{\hat m_2^2}{2},\hat m_2)]=\hat \nu \cdot  (-\tfrac{1}{2}(\hat m_2^0)^2,\hat m_2^0), \label{jumpcond}
\end{equation}
where $\hat \nu$ denotes the normal of the diagonal wall as
depicted in Figure \ref{ansatz}, 
is equivalent to $\sin \alpha=2\hat m_2^0$.
Therefore the piecewise constant magnetization satisfying
\eqref{jumpcond} mesoscopically carries no stray-field energy.    
Of course, on a microscopic scale equation \eqref{Burgers} does not hold: The
continuous transition in the wall generates a right hand side,
i.e., dipolar charges. Note that walls have to form since \eqref{Burgers} does not
allow for non-trivial smooth solutions with boundary data $\hat m_2=0$.     
\smallskip

Within domain theory we therefore consider piecewise constant
magnetizations of concertina type of period $\hat w$
and of amplitude $\hat m_2=\pm\hat  m_2^0$ in the
quadrangular and $\hat m_2=0$ in the triangular domains, s.t.\ \eqref{jumpcond} holds. Since the angles
are fixed by \eqref{jumpcond}, admissible
configurations are characterized by two parameters, namely the
amplitude of the
magnetization $\hat m_2^0$ and the width of the folds $\hat w$. 
\begin{figure}[htb]
\includegraphics[width=0.45\textwidth]{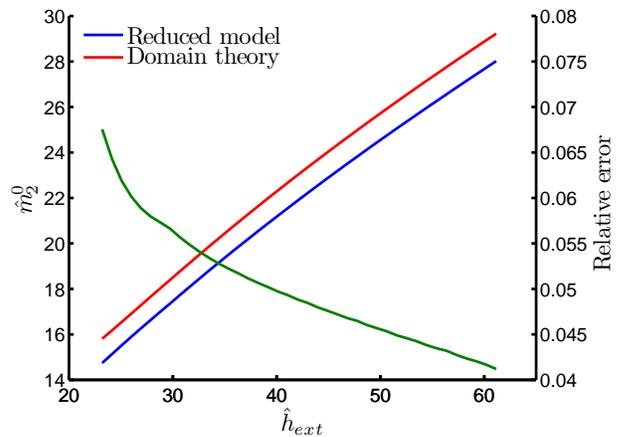}
\caption{Domain theory and numerical simulations: The domain theoretic prediction for the optimal amplitude (red) and
the computed amplitude based on the reduced model (blue). For the reduced model we display the amplitude, i.e., the maximal value which is attained in the quadrangular domain.}
\label{optimalampli}
\end{figure}
\smallskip

The energy which discriminates between these solutions is given by the
total wall energy, which is an
appropriate line-energy density integrated over the interfaces, augmented
by Zeeman energy. 
The specific line energy is a function of the jump
$[\hat m_2]=2\hat m_2^0$ of the magnetization across the wall -- an
infinitesimal version of the wall angle. 
Due to the shear invariance of the reduced energy, namely
\begin{equation}
  \hat x_1 \,=\,s\,\hat x_2 +\tilde x_1 , \quad \hat x_2 \,=\,\tilde
  x_2 ,\quad \hat m_2\,=\,\tilde m_2-s,
\label{shearinvariance}
\end{equation}
by which a diagonal wall can be transformed into a vertical wall -- for
the choice of $s=\pm \frac{ \hat m_2^0}{2}$ --
the specific line energy can
be obtained by restricting the reduced energy 
functional to one-dimensional transitions with boundary data $\pm
\hat m_2^0$ in case of the vertical walls and $\pm \frac{\hat m_2^0}{2}$ in
case of the diagonal walls. The optimal transition layers are low-angle N\'eel walls whose
line-energy density scales as 
\begin{equation*}
 \hat e_\wall\big(\tfrac{[\hat m_2]}{2}\big)\,=\,\hat e_\wall(\hat
 m_2^0)\,\approx\,\tfrac{\pi}{8} (\hat m_2^0)^4 \ln^{-1} \frac{
  \hat w_\text{tail}}{\hat w_\text{core}},
\end{equation*}
where $\hat w_\text{tail}$ and $\hat w_\text{core}$ are the two characteristic length
  scales of the N\'eel wall,
  \cite[Section 6]{DKMO05}. The tails of the N\'eel wall
  decay only logarithmicly and spread as much as possible. In case of the
concertina pattern, they are only limited by the neighboring walls --
thus $\hat w_\text{tail}\approx \frac{\hat w}{4}$. A more careful inspection shows that the
core width decreases with increasing jump size, more precisely
$\hat w_\text{core}\sim(\hat m_2^0)^{-2}$, see
\cite[
Subsection 3.5.5]{S06}.
Hence we obtain
\begin{equation}
\label{lineenergy}
\hat e_\wall(\hat m_2^0)\,\approx\,\tfrac{\pi}{8} (\hat m_2^0)^4 \ln^{-1} 
  (\hat w (\hat m_2^0)^2).
\end{equation}
With the rescaling \eqref{rescaling1} and \eqref{fieldscaling} this turns into
\begin{equation}
\label{lineenergy2}
 e_\wall(m_2^0)\,\approx\,\tfrac{\pi}{8} t^2(m_2^0)^4 \ln^{-1} 
  (d^{-2} t w (m_2^0)^2).
\end{equation}
Within the class of admissible magnetizations, the domain theoretic energy becomes a function of only three
parameters, namely $\hat m_2^0$, $\hat w$ and $\hat h_\ext$. To see
this, notice that one period of the pattern in Figure \ref{ansatz}
contains 
\begin{itemize}
\item  two vertical walls of height $1-\frac{\hat w}{\hat m_2^0}$ and of jump
  size $2\,\hat m_2^0$, leading to an energy contribution of
  $2\,(1-\frac{\hat w}{\hat m_2^0})\,\hat e_\wall(\hat m_2^0)$,
\item four diagonal walls of projected height $\frac{\hat w}{\hat m_2^0}$ and of
  jump size $\hat m_2^0$, leading to an energy contribution of
  $4\,\frac{\hat w}{\hat m_2^0}\,\hat e_\wall(\frac{\hat m_2^0}{2})$,
\item two quadrangular domains of total area $\hat w -\tfrac{\hat w^2}{\hat m_2^0}$,
  leading to a Zeeman energy of $-\hat h_\ext(\hat m_2^0)^2\,(\hat  w -\tfrac{\hat w^2}{\hat m_2^0})$.
\end{itemize}
Hence, the total domain energy per period in rescaled variables is given by
 \begin{align}
\hat  E_\domain(\hat m_2^0,&\hat h_\ext, \hat w)\,\notag\\=&\;2\,(1-\tfrac{\hat
  w}{\hat  m_2^0})\, \hat e_\wall(\hat m_2^0)+4\,\tfrac{\hat  w}{\hat
  m_2^0} \,\hat e_\wall\big(\tfrac{ \hat m_2^0}{2}\big)\notag \\
&-\hat h_\ext( \hat m_2^0)^2(\hat  w  -\tfrac{\hat  w^2}{\hat  m_2^0}).
\label{Edomain}
 \end{align}
Within the original scaling the domain theoretic energy takes the form of
  \begin{align}
  E_\domain(m_2^0,&h_\ext,w)\notag\\=&\;2\big(\ell-\tfrac{w}{m_2^0}\big)\, e_\wall(m_2^0)+4\,\tfrac{w}{m_2^0}
  \,e_\wall\big(\tfrac{ m_2^0}{2}\big)\notag\\
& -h_\ext(m_2^0)^2\,t\left( w\ell  -\tfrac{w^2}{m_2^0}\right).
 \label{Edomainun}
  \end{align} 
First of all we apply \eqref{Edomain} to derive the optimal
amplitude of the $\hat w^*$-periodic concertina pattern as a function of the external field $\hat h_\ext$ by optimizing the energy
in $\hat m_2^0$. Of course, domain theory is only applicable and thus a
good approximation for the reduced model for $\hat
h_\ext\gg 1$ in which case there is
a clear scale separation between walls and domains. Figure
\ref{optimalampli} shows that in this case domain theory is in
good agreement with our numerical simulations.
\smallskip

\begin{figure}[htb]
  \centering
\psset{unit=0.4} 
  \begin{pspicture}(0,-1.5)(13,8.5)
        \rput[l](10.3,1.6){$\tfrac{ 2|A_2-A1|}{\hat m_2^1+\hat m_2^2}$}
    \put(9.5,.9){{\psscaleboxto(.3,3.5){\(\}\)}}}

\put(3,-0.5) {\rotateright{\psscaleboxto(.3,2.2){\(\}\)}}} 
\rput[c](4.5,-1.3){$\frac{m_2^1|A_2-A_1|}{\hat m_2^1+\hat m_2^2}$}
 \rput[c](4.2,4){$-\hat  m_2^1$}
    \rput[c](6.8,4){$ \hat m_2^2$}
    \rput[c](5,0.75){$0$}
\pscircle[fillstyle=solid,fillcolor=black](5.2,3.5){.1}
    \rput[c](5.2,2.5){$A_0$}
\pscircle[fillstyle=solid,fillcolor=black](3,0){.1}
    \rput[r](2.8,.5){$A_1$}
\pscircle[fillstyle=solid,fillcolor=black](8.2,0){.1}
  \rput[l](8.4,.5){$A_2$}
\pscircle[fillstyle=solid,fillcolor=black](5,8){.1}
  \rput[c](5,8.5){$A_3$}

\psset{linewidth=0.05,linecolor=black,arrows=c-c}
    \pnode(0,0){a}

    \pnode(9,0){b}
    \pnode(0,8){c}
    \pnode(9,8){d}
    \pnode(1,8){e}
    \pnode(1,3.5){f}

    \pnode(5,8){g}

    \pnode(5.2,3.5){h}
    \pnode(3,0){i}
    
\pnode(3.24,5.2){j}

    \pnode(8.2,0){k}
    \pnode(7.7,4.85){l}
\SpecialCoor
\psset{linewidth=0.05,linecolor=black,arrows=c-c}
    \ncline{a}{b}
    \ncline{c}{d}
\psset{linewidth=0.05,linecolor=gray,arrows=c-c}    
    \ncline{g}{h}
    \ncline{i}{j}
    \ncline{k}{l}

    \ncline{j}{g}
    \ncline{i}{h}
    \ncline{l}{g}
    \ncline{k}{h}
\put(8.2,0){\psline(0,0)(1,2)}
\put(7.7,4.85){\psline(0,0)(1,2)}
\put(3,0){\psline(0,0)(-1,1.8)}
\put(3.24,5.2){\psline(0,0)(-1,1.8)}
\psline[linecolor=gray,linestyle=dashed,linewidth=0.02](5.2,3.5)(9,3.5)
\psline[linecolor=gray,linestyle=dashed,linewidth=0.02](5.2,3.5)(5.2,0)
\psset{linewidth=0.05,linecolor=black,arrows=c-c}
\end{pspicture}
 \caption{Domain theory: Generalized tilted Ansatz function.}
 \label{genansatz}
 \end{figure}
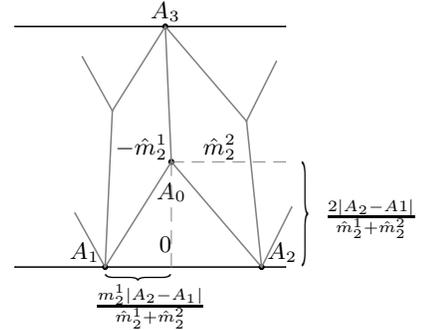

Before we go on with the analysis of domain theory let us emphasize
that the experimentally observed concertina is of course not of uniform
period and equal amplitude as our domain theoretic Ansatz above. As shown in Figure \ref{genansatz},
there are also oblique piecewise constant weak solutions of
\eqref{Burgers}. Nevertheless this class of
Ansatz functions is very rigid: An elementary calculation
shows that the location of the interior triplet $A_0$ is uniquely
determined by the jump condition \eqref{jumpcond}, if the distance between
the boundary triplet $A_2$ and $A_1$, and $\hat m_2^1$ and $\hat m_2^2$ on both sides are
given. Hence the continuation of the pattern is uniquely determined
if either the amplitude in the
next quadrangular domain or the location of the next triplet, i.e., the width of the next quadrangular domain, is
prescribed. 
\section{Coarsening of the concertina pattern}
\label{coarsening section}
\subsection{Domain theory: The optimal period of the concertina
  pattern for large external fields}
\label{optimal period}
\begin{figure}[htb]
\includegraphics[width=0.45\textwidth]{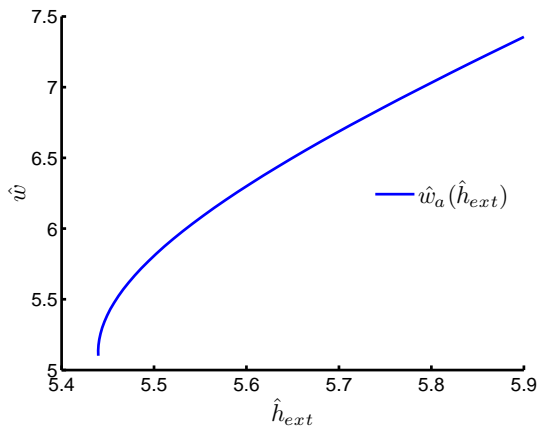}
\caption{Numerical simulations: The optimal period of the concertina pattern as a function of
the external field computed on the basis of the reduced model.}
\label{optimalperiod}
\end{figure}

Experiments show an increase in the average concertina period $w$
as the external field $h_\ext$ is increased after the
pattern has formed, see Figure \ref{fig:coarsening}.
The general tendency that the optimal period $w$ is an increasing function of $h_\ext$
can be understood on the basis of domain theory in the reduced
variables $\hat m_2^0$, $\hat h_\ext$ and $\hat w$.
By optimizing the energy per {\it unit length} with respect to the period
$\hat w$ {\it and}
the amplitude $\hat m_2^0$ of the transversal component, we obtain the
following scaling of the optimal period of the pattern as a function
of the external field
\begin{equation}
\hat w_a(\hat h_\ext)\sim \hat h_\ext\ln \hat h_\ext \quad
 \hat h_\ext\gg 1.\label{scalingperiod}
\end{equation}
In particular we find that the optimal period increases with increasing
field $\hat h_\ext$ -- the $a$ in $w_a$ stands for {\bf a}bsolute minimizer.
Domain theory also yields the (same) scaling behavior for the optimal transversal component of the magnetization
\begin{equation}
\hat {m_2}_{a}(\hat h_\ext)\sim \hat h_\ext\ln \hat h_\ext \quad
 \hat h_\ext\gg 1.\label{scalingmag}
\end{equation}
We note that both scalings have also been confirmed by a rigorous
asymptotic analysis of the reduced energy functional \eqref{rescaledenergy} which does not rely on a simple concertina
Ansatz, cf.\ Theorem 1 in \cite[p.147]{OS10}. Moreover, numerical simulations
of the reduced energy show that 
the optimal period increases with $\hat h_\ext$ also for external fields close to the critical field, see Figure
\ref{optimalperiod}. The optimal
period shown in this diagram was computed by minimizing the energy per
unit length both with respect to the magnetization and the period, for varying external field.  
\subsection{Coarsening: A modulation instability}
\label{coarsesec}
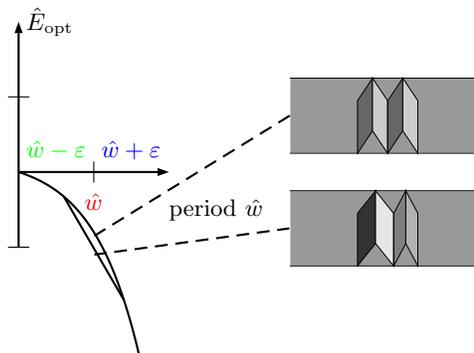
\begin{figure}[htb]
  \begin{pspicture}(0,-2)(3.5,2.5)
 \psset{plotpoints=1000,labels=none,arrowinset=0}
 \psaxes*{->}(0,0)(0,-1)(2,2)
 \rput[l](0.1,2){$\hat E_\opt$}
 \rput[l](2,-0.5){period $\hat w$}
 \rput[c](1,-0.4){\small\color{red}$ \hat w$}
 \rput[c](0.5,0.3){\small\color{green}$ \hat w-\varepsilon$}
 \rput[c](1.5,0.3){\small\color{blue}$ \hat w+\varepsilon$}
 \infixtoRPN{0.2*5^x*(-1)+0.2}
 \psplot[linecolor=black]{0}{1.6}{\RPN}
 \infixtoRPN{(-0.2*(5^(0.6))+0.2)*(x-1.4)/(-0.8)+(-0.2*(5^(1.4))+0.2 )*(x-0.6)/(0.8)}
 \psplot[linecolor=black]{0.6}{1.4}{\RPN}
 \end{pspicture}
\psset{unit=0.1cm}
\begin{pspicture}(0,-7.5)(20,20)
\put(0,15){
\pspolygon[linewidth=0,linecolor=white,fillstyle=solid,fillcolor=mgray](0,0)(25,0)(25,10)(0,10)
\psline[linewidth=.1](0,0)(25,0)
\psline[linewidth=.1](0,10)(25,10)
\put(10,0){\pspolygon[linewidth=.1,fillstyle=solid,fillcolor=ddgray](-1,0)(1,3)(1,10)(-1,7)
  \put(2,0){\pspolygon[linewidth=.1,fillstyle=solid,fillcolor=llgray](1,0)(-1,3)(-1,10)(1,7)}}
\psset{linecolor=blue,linewidth=1pt,arrowsize=1.1}
\psset{linecolor=black}
\put(14,0){\pspolygon[linewidth=.1,fillstyle=solid,fillcolor=ddgray](-1,0)(1,3)(1,10)(-1,7)
\put(2,0){\pspolygon[linewidth=.1,fillstyle=solid,fillcolor=llgray](1,0)(-1,3)(-1,10)(1,7)}}
}
\psline[linestyle=dashed](0,20)(-26,4)
\psline[linestyle=dashed](0,5)(-26,1.5)
\pspolygon[linewidth=0,linecolor=white,fillstyle=solid,fillcolor=mgray](0,0)(25,0)(25,10)(0,10)
\psline[linewidth=.1](0,0)(25,0)
\psline[linewidth=.1](0,10)(25,10)

\put(10,0){
  \pspolygon[linewidth=.1,fillstyle=solid,fillcolor=dddgray](-1,0)(1.4,3)(1.4,10)(-1,7)
  \put(2,0){\pspolygon[linewidth=.1,fillstyle=solid,fillcolor=lllgray](1.8,0)(-.6,3)(-.6,10)(1.8,7)}
}
\put(14,0){
\pspolygon[linewidth=.1,fillstyle=solid,fillcolor=mdgray](-.2,0)(1.4,3)(1.4,10)(-.2,7)
\put(2,0){\pspolygon[linewidth=.1,fillstyle=solid,fillcolor=mlgray](1,0)(-.6,3)(-.6,10)(1,7)}

}
  \end{pspicture}
\caption{Concavity of the minimal energy per period implies an
  instability under modulation of the wavelength.}
\label{concavity}
\end{figure}

Although the above analysis predicts that the optimal period $\hat w_a$ increases
as the field $\hat h_\ext$ increases, it does not explain why and in which way a
concertina pattern of period $\hat w$ becomes {\it unstable} as $\hat h_\ext$ increases. We will see that both the increasing period for large fields and the
deviation of the initial period close to critical field from the one
of the unstable mode are due to an instability under long wavelength
modulations of the pattern. The mechanism behind the instability is the following: Given
$\hat h_\ext$ and a period $\hat w$, an optimization in the
transversal component $\hat m_2$ yields that the optimal energy {\it per period} $\hat E_\opt(\hat h_\ext,\hat w)$ is a concave function in $\hat w$ if $\hat h_\ext$ is sufficiently large.
The concavity suggests -- as depicted in Figure \ref{concavity} -- that
the concertina
pattern of a uniform period $\hat w$ becomes unstable towards 
perturbations which increase the period to $\hat w+\epsilon$ and the
corresponding amplitude of the transversal component to $\hat m_2^0(\hat w+\epsilon)$
in some folds, and decrease the period to $\hat  w-\epsilon$ and the
amplitude to $\hat m_2^0(\hat w-\epsilon)$
it in other folds. This modulation eventually leads to the collapse of
the smaller folds, i.e., the coarsening.
However, in view of the non-local character of the
stray-field energy, it is not clear whether this simplified picture, i.e.,
that the energy of the modulation amounts to the modulation of the energy,
applies. As we shall see in Subsection \ref{PredInstab}, a modulation
of the period on a very long length scale overcomes this objection. Thus the concavity of the minimal energy implies an
instability under long wavelength modulations of the pattern.
\smallskip

In order to derive the concavity of the minimal energy we apply domain
theory for large external fields in Subsection \ref{PredInstab} and an
extended bifurcation analysis close to the critical field in
Subsection \ref{extendbif}. We will see that both asymptotics match
the results of the numerical simulation of our reduced model. 
\smallskip

Let us mention that the modulation instability of the concertina
pattern is closely related to the so-called Eckhaus instability which
was discovered in the context of non-linear instabilities in
convective systems leading to a change in wavelength of the observed
periodic pattern, cf.\ \cite{MR1173866}. 
 
\subsection{Bloch-wave theory: Instability with increasing field}
\label{PredInstab}
As indicated above, not only the optimal period but also the coarsening can be explained on the basis of domain theory
for large external fields $\hat h_\ext \gg 1$. This relies on the optimal energy {\it per period} $\min_{\hat
  m_2}\hat E_{\domain}(\hat m_2,\hat
h_\ext,\hat w)$. For periods $\hat w$ much smaller than the optimal
period at some value of the external field $\hat h_\ext$, i.e., $ \hat w\ll  \hat
h_\ext\ln  \hat h_\ext$, we find that 
\begin{multline}
\label{ws}
\min_{\hat
  m_2}\hat E_{\domain}(\hat m_2,\hat
h_\ext,\hat w)\sim-\hat  h_\ext^2 \hat w^2\ln(\hat  h_\ext \hat
w^2).
\end{multline}
In particular, the optimal energy per period in \eqref{ws}
is {\it concave} in the period $\hat w$.
Although domain theory therefore suggests an instability under
wavelength modulation for periods which are much smaller
than the optimal period, it is too rigid to allow for
such a type of perturbation, even in the
class of generalized Ansatz functions, cf.\ Figure \ref{genansatz}.       
\smallskip

\begin{figure}[htb]
\includegraphics[width=0.45\textwidth]{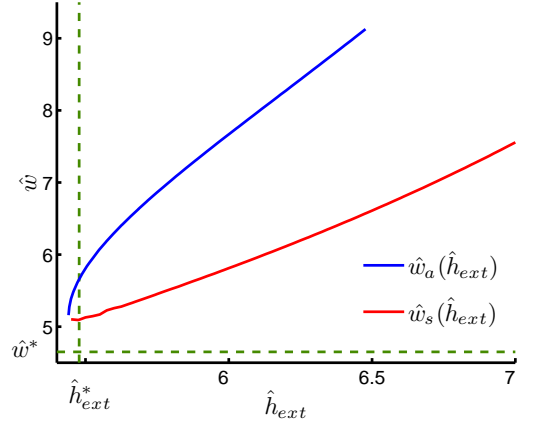}
\caption{Numerical simulations: Comparison of the optimal and marginally stable period of the
  concertina pattern as a function of the external field -- both
  computed on the basis of the reduced model. In the region below the
  red curve the minimal energy per period is concave and thus a
  concertina of that period is unstable and coarsens.}
\label{figstability}
\end{figure}

It is rather on the level of the reduced model that it can be seen
that the concavity translates into an instability (despite the 
potentially long-range interactions via the stray field).
Indeed, a so-called Bloch-wave analysis of the reduced model shows that the
concavity is in a one-to-one correspondence with an instablitity under long wavelength 
modulations of the pattern. In the Bloch-wave analysis one considers $N\hat w$-periodic eigenfunctions of the Hessian of the form
\begin{equation*}
\delta  \hat m_2\;=\;e^{-i \hat x_1 \hat k_1}\, \delta \hat m_2^{\hat k_1}\notag
\end{equation*}
with wavenumber $\hat k_1=\frac{2\pi}{N\hat w}$ and $N$ some large integer and where $\delta \hat m_2^ {\hat k_1}$ is $\hat w$-periodic with
  respect to $\hat  x_1$, i.e., one condisders sinusiodal modulations of some suitable $\hat w$-periodic function.
An asymptotic expansion of 
\begin{equation}
\operatorname{Hess} \hat E(\hat m_2)(e^{-i \hat x_1 {\hat k_1}}\delta  \hat m_2^{\hat k_1})=\lambda^{\hat k_1}
e^{-i\hat x_1{\hat k_1}}\, \delta \hat m_2^{\hat k_1}\label{Bloch} 
\end{equation}
for small wavenumbers $\hat k_1\ll 1$, i.e., $N\gg1$, shows that the
first eigenvalue can be related to the second derivative of
the optimal energy per period $\hat E_\opt=\min_{\hat m_2}\hat
E$. More precisely, one can show that the
eigenvalue possesses the expansion 
\begin{equation*}
\lambda^{\hat k_1}\approx c_0\;{\hat k_1}^{2}\frac{\dd^2}{\dd \hat w^2}\hat E_\opt(\hat h_\ext,
\hat w) \quad\text{for}\quad {\hat k_1} \ll 1,
\end{equation*}
where $c_0$ denotes a constant that depends on $\hat m_2$,
see \cite[Theorem 5.1]{S10}.
This shows that the concavity of $\hat E_\opt(\hat h_\ext, \hat w)$
with respect to the period $\hat w$ implies that the concertina pattern of a given period
$\hat w$ is unstable. Domain theory predicts that the
marginally stable period $\hat w_s$, i.e., $\hat w_s$ such that \ $\frac{\dd^2}{\dd\hat
  w^2}\hat E_\opt (\hat h_\ext,\hat w_s)=0$, scales as $\hat w_s \sim \hat h_\ext \ln
\hat h_\ext$, cf. \eqref{ws} -- we note that the $s$ in $w_s$ stands for marginally
{\bf s}table. Figure \ref{figstability}
displays the optimal and the
marginally stable period computed on the basis of the reduced energy
functional. 
\smallskip

\begin{figure}[htb]
\includegraphics[width=0.45\textwidth]{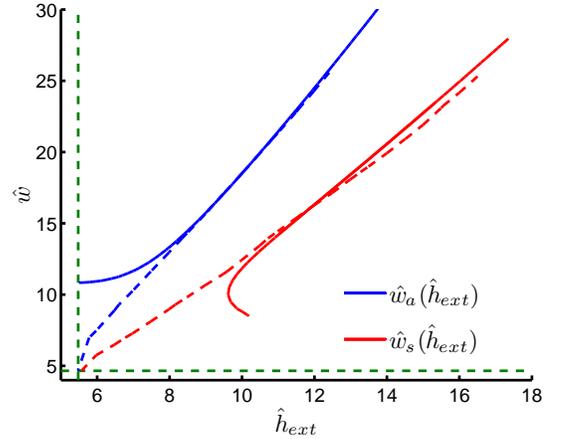}
\caption{Numerical simulations and domain theory: The optimal and marginally stable period computed on the basis
  of the reduced model (dashed) match the predictions on the basis of
  domain theory in the regime $\hat h_\ext \gg 1$.}
\label{comp_red_dom_w}
\end{figure}

Figure \ref{comp_red_dom_w} shows that the computation of the optimal
and the marginally stable period on the basis of domain theory matches the
numerical simulations on the basis of the reduced model.

\subsection{Bifurcation analysis: Instability for small fields}
\label{extendbif}
The numerical computations, cf.\  Figure \ref{figstability}, show that the optimal energy per
period is concave not only for large external
fields as predicted by domain theory. In fact, we extract from our
numerical data that $\frac{\dd^2}{\dd
  \hat w^2}\hat E_\opt(\hat h_\ext,\hat  w^*)$ is negative also for small external fields
up to the turning point. This is consistent with the numerical computation of
the eigenvalue $\lambda^N$ based on the asymptotic expansion of equation
\eqref{Bloch}.
Hence, the Bloch-wave analysis implies that the $\hat w^*$-periodic 
concertina pattern is unstable under long wavelength
modulations close to the critical field. 
\smallskip

\begin{figure}[htb]
\includegraphics[width=0.45\textwidth]{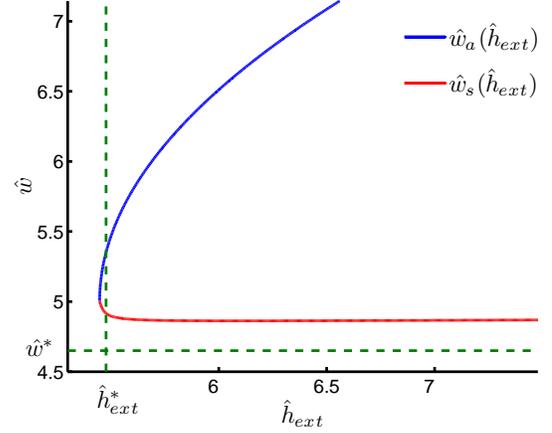}
\caption{Bifurcation analysis: The optimal and marginally stable period as a function of the
  external field obtained on the basis on the extended bifurcation analysis.}
\label{figstability2}
\end{figure}

This qualitatively explains the trend in the deviation
of the initial concertina period $w^*_\text{exp}$ from the period of the
unstable mode, see Subsection \ref{period}.
Close to the critical field, the concavity can be confirmed with
the help of an asymptotic
bifurcation analysis. To see this, we extend our Ansatz from Section \ref{bifurcationsection} and take into account small deviations of the wavenumber
$\hat k=\hat k^*+\hat{\delta k}$. As we have seen in \eqref{Ot.1} in Section
\ref{bifurcationsection}, the quartic coefficient in the energy
expansion, namely $\frac{\pi}{640}$, is small  compared to the second order
coefficient and the scale of the reduced external field. Due to that degeneracy it
is necessary to additionally take into account a contribution of cubic order in the perturbation
of $\hat m_2=0$, i.e., we use the extended Ansatz
\begin{equation*}
\hat m_2\approx A\hat m_2^*+A^2\hat m_2^{**}+A^3\hat m_2^{***}.
\end{equation*}
Optimizing the coefficients in the expansion of the energy with respect to $A$ subsequently in $\hat m_2^{**}$ and $\hat m_2^{***}$ leads to an expansion of
the energy density of the form
\begin{align*}
\tfrac{\hat k}{2\pi} \hat E(A  &\hat m_2^*+A^2 \hat m_2^{**}+A^3 \hat
m_2^{***})\\\approx&\tfrac{1}{4} (\hat h_\ext^*(\hat k)-\hat h_\ext) A^2 - c_4(\hat k)A^4+c_6(\hat k)A^6 ,
\end{align*}
where $c_4(\hat k^*)=\frac{\pi}{640}\frac{\hat k^*}{2\pi}$ in
accordance with \eqref{Ot.1}.
Hence under the assumption that $c_4(\hat k^*)\approx 0.00105$ is small, the
energy density to leading order can be approximated by
\begin{align}
\tfrac{\hat k}{2\pi} \hat E&(A  \hat m_2^*+A^2 \hat m_2^{**}+A^3 \hat
m_2^{***})\notag\\\approx&\tfrac{1}{4}
\big(\tfrac{\dd^2}{\dd \hat k^2}\hat h_\ext^*(\hat k)_{|\hat k=\hat k^*}\tfrac{\hat{\delta
    k}^2}{2}+\delta \hat h_\ext\big)A^2 \notag\\&-(c_4(\hat k^*)+
\tfrac{\dd}{\dd\hat k}c_4(\hat k)_{|\hat k=\hat k^*}\hat {\delta 
k})A^4+c_6(\hat k^*)A^6.\label{eckmod}
\end{align}
The numerical values of the coefficients are given by
\begin{align*}
  &\tfrac{\dd^2}{\dd \hat k^2}\hat h_\ext^*(\hat k)_{|\hat k=\hat k^*}=3,\\
&\tfrac{\dd}{\dd\hat k}c_4(\hat k)_{|\hat k=\hat k^*}\approx-0.0217,\\
&c_6(\hat k^*)\approx 0.000207.
\end{align*}
Notice that $c_6(\hat k^*)$ is positive,
confirming the numerically observed turning point of the
$\hat w^*$-periodic branch. Obviously, the asymptotic expansion displays an
asymmetric behavior in $\hat{\delta k}$; the energy decreases for $\hat{\delta  k}<0$. 
Based on the expansion \eqref{eckmod}, one can characterize the optimal wavenumber
and the optimal period. We note that the concavity of the
minimal energy {\it per period} as a function of the {\it period} is equivalent to the concavity of
the energy {\it density} as a function of the {\it wavenumber} $\hat k$:
\begin{equation*}
  \tfrac{\dd^2}{\dd\hat w^2}\hat E(\hat w)\;=\;\tfrac{\hat
    k^3}{(2\pi)^2}\tfrac{\dd^2}{\dd\hat k^2}\Big(\hat k \hat E\big(\tfrac{2\pi}{\hat k}\big)\Big).
\end{equation*} 
Figure \ref{figstability2} shows the optimal period and the
marginally stable period calculated on the basis of \eqref{eckmod}.
We read off that the $\hat w^*$-periodic concertina
pattern is indeed unstable at the critical field.
\smallskip

\begin{figure}[htb]
\includegraphics[width=0.45\textwidth]{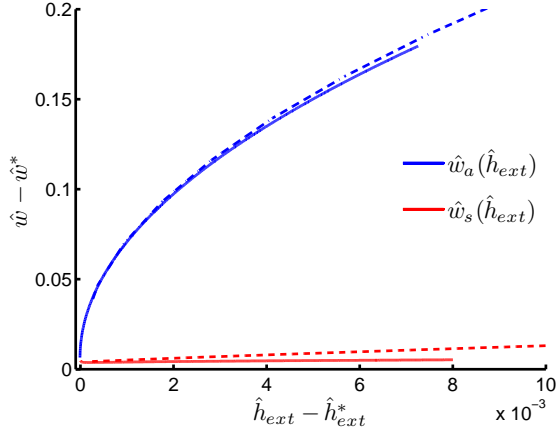}
\caption{Numerical simulations and bifurcation analysis: The prediction on the basis of the reduced model (dashed)
  matches the prediction on the basis of the extended bifurcation
  analysis for a near-degenerate value of $\hat Q=0.0295$ close to
  $\hat Q^*\approx 0.03$, cf.\ Section
\ref{anisotropy}.}
\label{agreestability}
\end{figure}

A comparison between Figure \ref{figstability} and Figure
\ref{figstability2} shows that the predictions on the basis of the
asymptotic expansion differ from the optimal and the marginally
stable period computed on the basis of the reduced model, compare for
example the scale of the external field. This deviation is related to
our assumption that the quartic coefficient is small so that the
energy can be approximated by \eqref{eckmod}. On the other hand, Figure \ref{agreestability} shows that the
asymptotics match the reduced model if we add a quartic contribution $+\frac{\hat Q}{4}\int
m_2^4$ to the reduced energy where the value of the parameter $\hat Q$
is such that the contribution cancels $c_4(\hat k^*)$ in
\eqref{eckmod} ( which happens $\hat Q \approx 0.03$). We will see later that such an
additional quartic contribution has a physical meaning if we take into
account a uniaxial anisotropy, see Section \ref{anisotropy}. It turns out that
$\hat Q$ corresponds to an appropriately rescaled quality factor $Q$. 

\subsection{Numerical bifurcation analysis: Type of secondary
  instability and downhill path in energy landscape}
\begin{figure}[htb]
\includegraphics[width=0.45\textwidth]{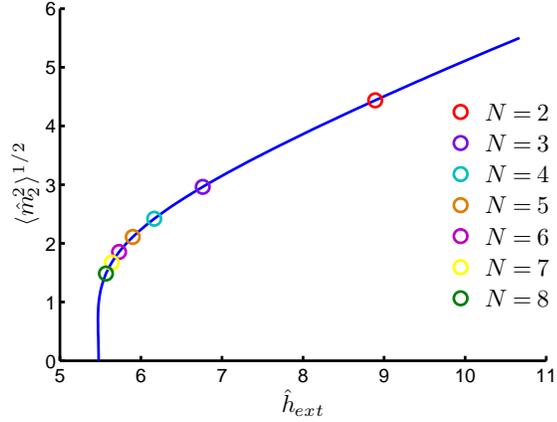}
\caption{Numerical simulations: The appearance of the secondary instability under
  $N\hat w^*$-periodic perturbations as a function of $N$. The
  critical field for $N=8$ is given by $5.602$.}
\label{dep_on_N2}
\end{figure}

With the help of a bifurcation-detection algorithm we are able to
compute at which field the $\hat w^*$-periodic concertina becomes unstable under $N\hat
w^*$-periodic perturbations while we follow the primary branch. Figure
\ref{dep_on_N2} shows the secondary critical fields; as expected (cf.\ Subsection \ref{extendbif} and Figure
\ref{figstability}) the secondary instability approaches the turning
point as the integer $N$ increases. We note that it is reached for
finite $N$. 
\smallskip

In the following we want to study in which {\it way} the concertina
pattern becomes unstable. We first present the outcome of the
computation of the secondary bifurcation branches. 
We point out that due to the symmetries
of the pattern, the bifurcations are not simple in the sense that more
than one branch bifurcates. 
\begin{figure}[htb]
\includegraphics[width=0.45\textwidth]{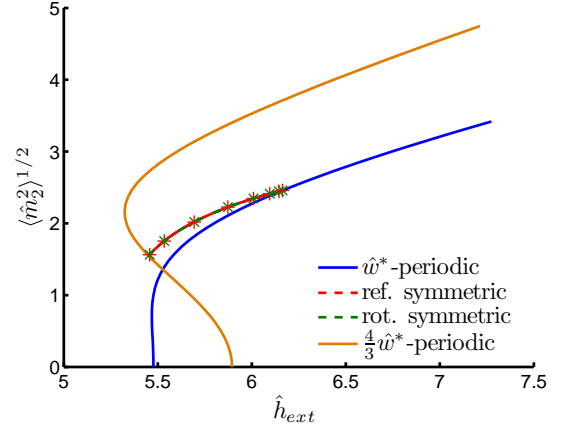}
\caption{Numerical simulations: Bifurcation diagram for $2\hat w^*$ perturbations: The bifurcation branches that connect the $\hat w^*$-periodic
  (blue) and the $2\hat w^*$-periodic branch (orange). The
  magnetization patterns at the indicated fields are shown in Figure
  \ref{reflectional}.}
\label{diagramN4}
\end{figure}
\smallskip

\begin{figure}[htb]
\includegraphics[width=0.45\textwidth]{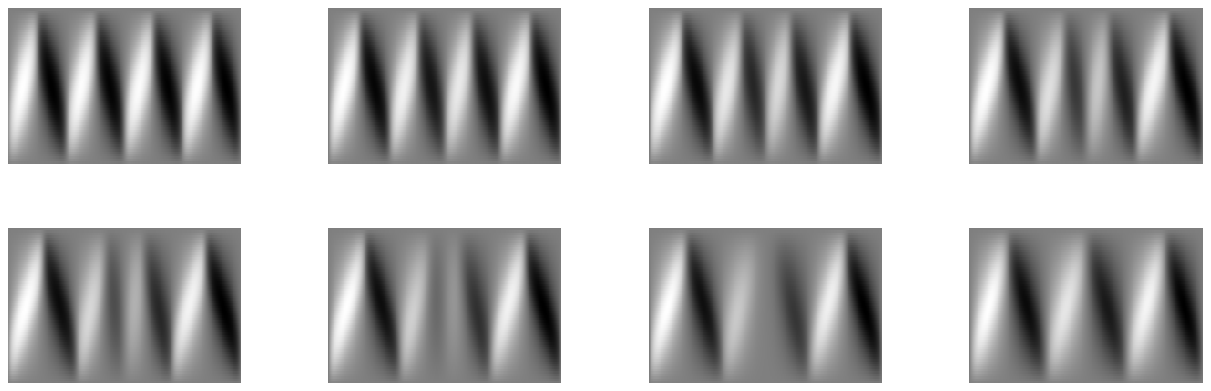}\\
\psline[linecolor=red](3.325,2.77)(3.325,3.55)
\includegraphics[width=0.45\textwidth]{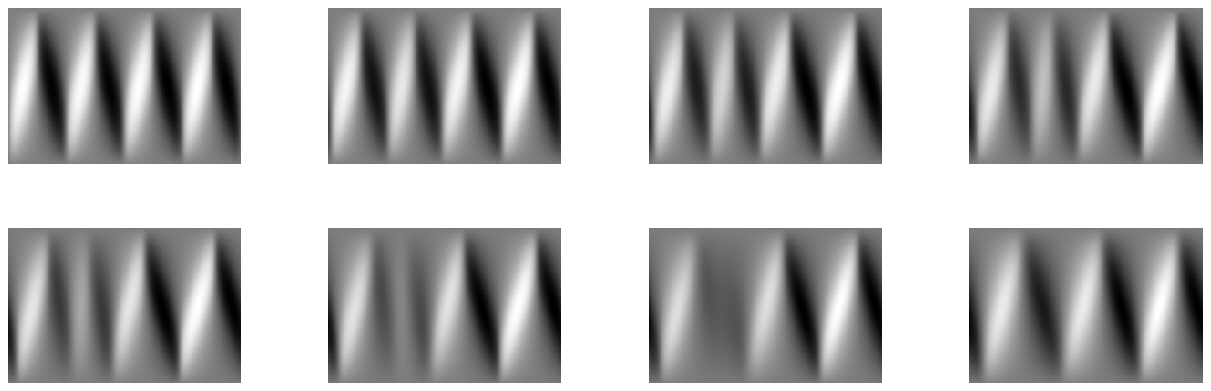}\\
\psarc[linecolor=dgreen]{->}(-.94,1.05){.2}{10}{340}
\pscircle[fillstyle=solid,linecolor=dgreen,fillcolor=dgreen](-.94,1.05){.01}
\caption{Numerical simulations: Reflectional symmetric with respect to center wall (top) and
  rotational symmetric with respect to the midpoint of white face (bottom) magnetization
  pattern on the unstable bifurcation branch connecting
  the $\hat w^*$-periodic and the $\tfrac{4}{3}\hat  w^*$-periodic branch. The
  central fold collapses (top); white face disappears and two adjacent
  black faces merge (bottom).}
\label{reflectional}
\end{figure}

The symmetries of the
pattern can be identified as linear representations of the dihedral
group $D_{2N}$, where $N$ indicates the number of folds. The secondary
bifurcation branches are computed with the help of a numerical
branch switching algorithm which is adapted to the problem of multiple
bifurcations. Generically,
there are two distinct types of branches: Branches along which rotational
symmetry is broken and reflectional symmetry is conserved and vice versa, see Figure \ref{reflectional}. In case of the
first type of branches, a fold collapses as two neighboring faces
disappear; in case of the
second type of branches, the number of folds decreases as one face
disappears and the two adjacent faces merge. 
During
the coarsening process, the width of the remaining folds is
adjusted. Let us point out that the first instability of
the $\hat w$-periodic concertina under $N\hat w$-periodic perturbations in the
end leads to the collapse of exactly one fold -- reducing the total
number of folds from $N$ to $N-1$, see Figure \ref{dep_on_N2}.
\begin{figure*}[htb]
  \begin{center}
\vspace{1cm}
 \includegraphics[width=.9\textwidth,height=.18\textwidth]{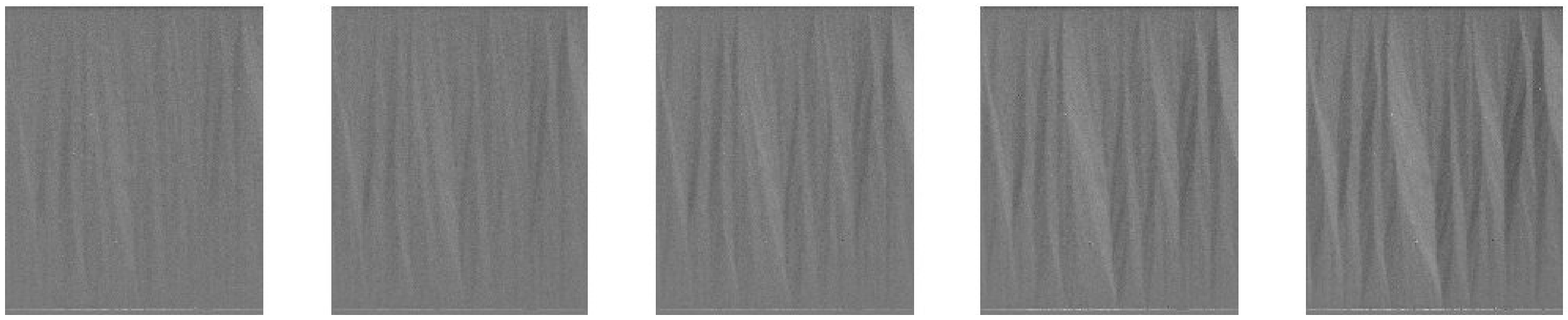}
\vspace{.5cm}
\includegraphics[width=.9\textwidth,height=.18\textwidth]{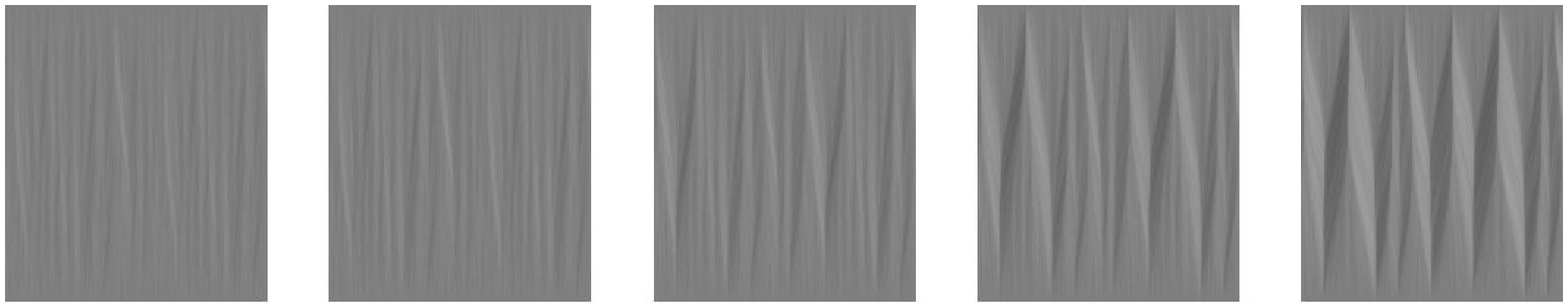}
  \end{center}
\caption{Experiment and numerical simulations: The
  coarsening of the concertina pattern in a Permalloy sample (top row)
  of $30$nm thickness and $70\upmu$m width 
  compared to the numerical simulations (bottom row). A ripple-like
  structure grows into the concertina pattern. Within the numerical
  simulations we iteratively increment the external field and minimize
the energy. The computational domain is of period $6\hat w^*$. The
numerical images are scaled according to \eqref{rescaling1}. The
numerical image
hence display 1.8 times the unit cell; the numerical images therefore appear to be more uniform
than the experimental concertina.}
\label{coarseningripple}
\end{figure*}

\subsection{Wavelength modulation in the experiments}
In the experiments, the $x_1$-wavelength of the
modulation is restricted by the finite extension of the
sample. Moreover, inhomogeneities and defects of
the material, in particular those at the edges of the cross section, strongly affect the
formation. This is reflected by the
fact that walls occur at the same pinning sites when
the experiment is rerun. The existence of pinning sites hence leads to
an effective modulation wavelength that is just a small multiple $N$ of the wavelength of the pattern. In
particular we expect that pinning sites have a stabilizing effect and therefore prevent coarsening. 
Therefore, the seemingly artificial numerical simulation for small and
moderate $N$, cf.\ Figure \ref{dep_on_N2}, may be more relevant for
the experiment than the Bloch-wave analysis, i.e., $N\nearrow
\infty$, cf. Section \ref{PredInstab}.
\subsection{Domain Theory: Instability for decreasing field}
\begin{figure}[htb]
 \includegraphics[width=0.45\textwidth]{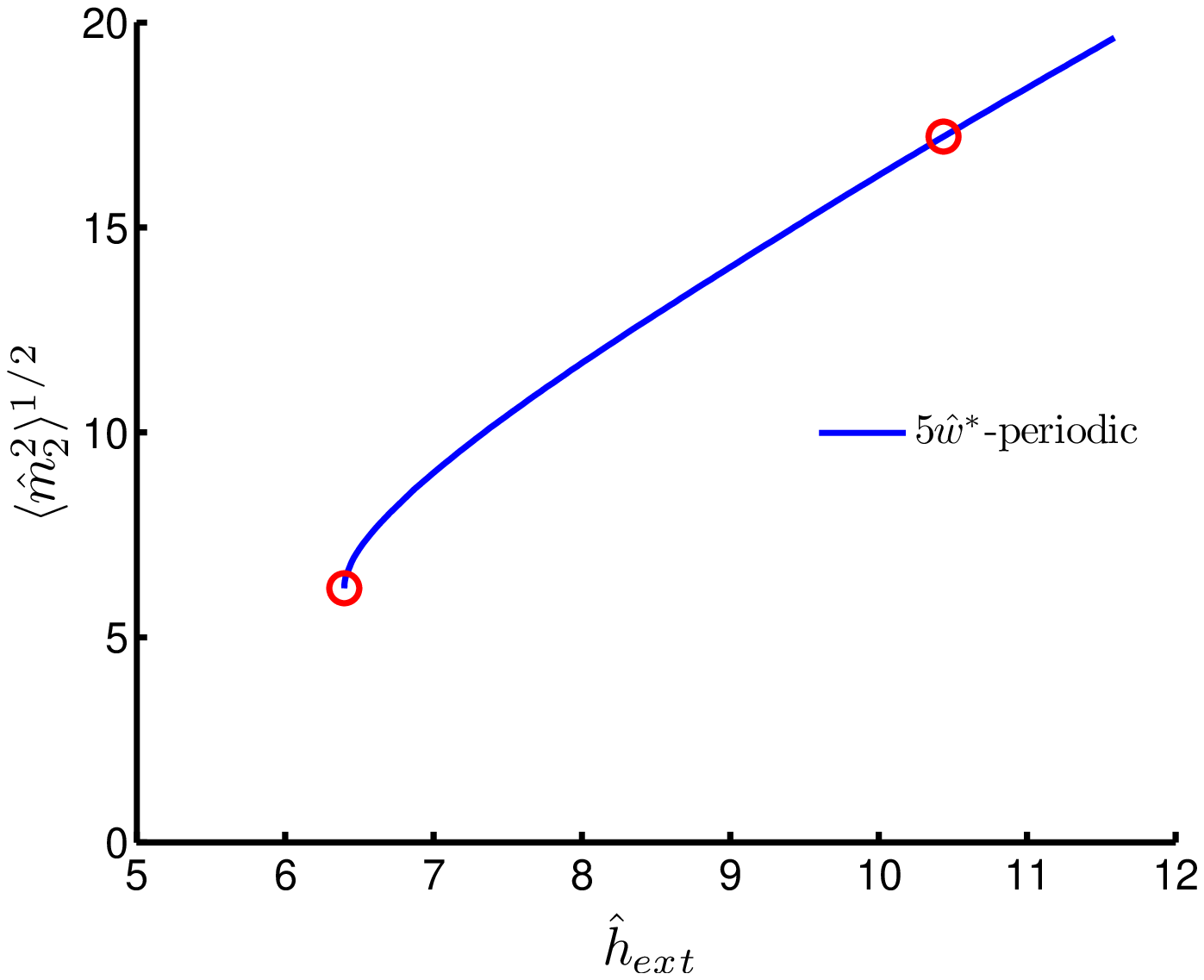}
\begin{pspicture}(0,0)
\psset{linecolor=blue,linestyle=solid,linewidth=6.5}
\rput{27}(-4.2,3.85){
\psline[linewidth=1.pt](0,0)(.3,.2)
\psline[linewidth=1.pt](0,0)(.3,-.2)
}
\end{pspicture}
 \includegraphics[width=0.45\textwidth, height=.095\textwidth]{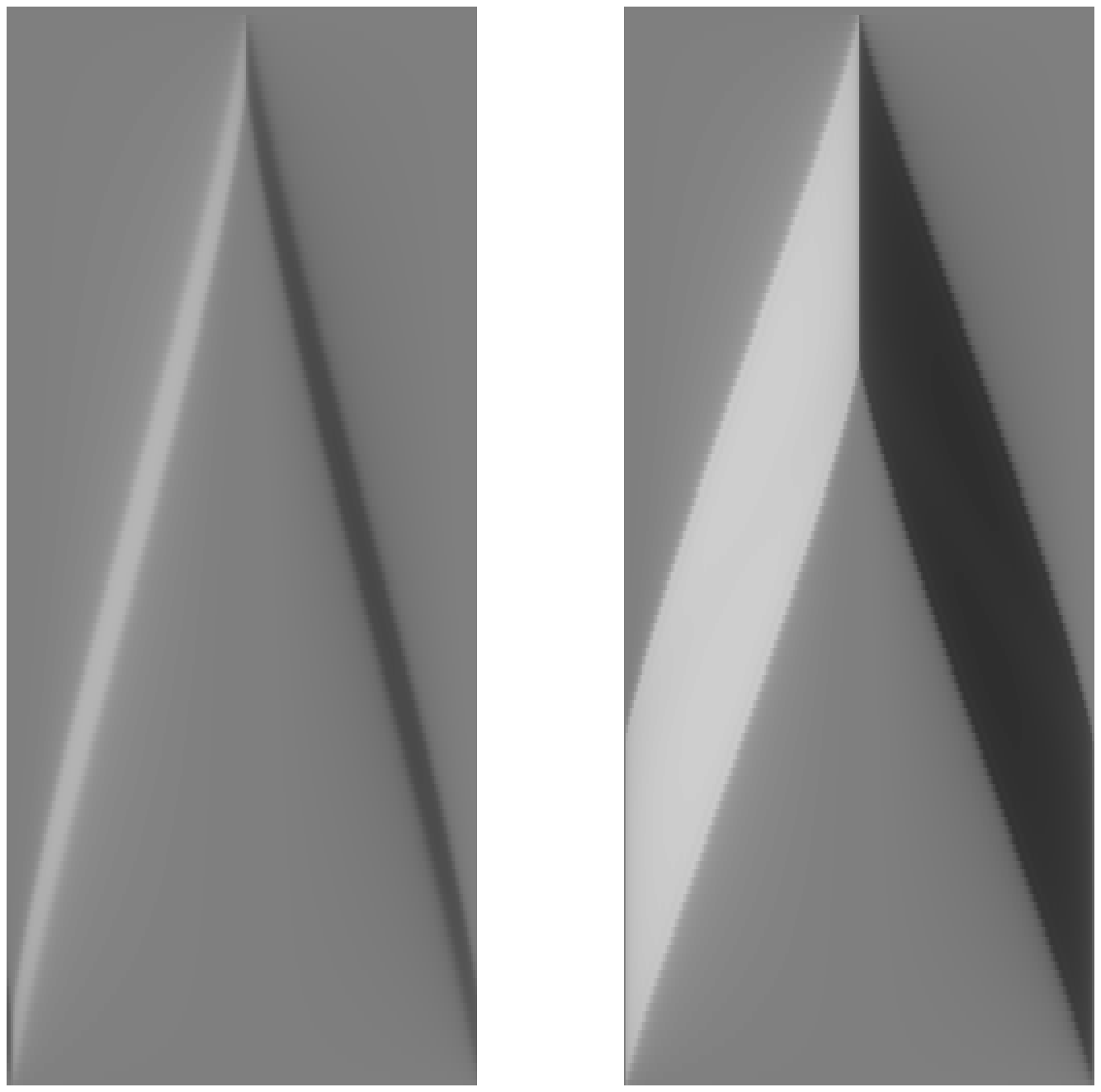}
 \caption{Numerical simulations: The coarsened concertina pattern degenerates as the external
 field is reduced. The numerical simulations confirm the prediction
 based on domain theory: The pattern degenerates at the turning point
 of the branch.}
 \label{refining}
 \end{figure}

The experiments also show that the concertina period $\hat w$ decreases
with {\it decreasing } external field $\hat h_\ext$. 
This has a simple
explanation on the level of domain theory, too. Suppose that the
concertina period had increased at several coarsening events during
the increase of the field. As
the decreasing external field $\hat h_\ext$ 
drops below its optimal scaling given the period $\hat w$, that is, for
$ \hat w\gg  \hat  h_\ext\ln  \hat h_\ext$, the optimal concertina
pattern does not suffer a long wavelength instability, but instead degenerates in the sense that the closure domains invade the
whole cross section.
Simulations of the reduced model confirm this scenario predicted by
domain theory, see Figure \ref{refining}, which shows a
pattern of period $5\hat w^*$ close to the turning point:
The numerical backward cycle, in which we start at the multiply coarsened state
and then after minimization repeatedly decrease the external field by a fixed increment,
shows that the coarsened pattern stays stable up to the turning point
that coincides with the moment at which the pattern degenerates as
mentioned above. 
Depending on
the initial level of coarsening, the period is then either refined or we
reach the uniformly magnetized state after the minimization. 

\subsection{Conclusion: Hysteresis and scattering of data}
\begin{figure}[htb]
\includegraphics[width=0.45\textwidth]{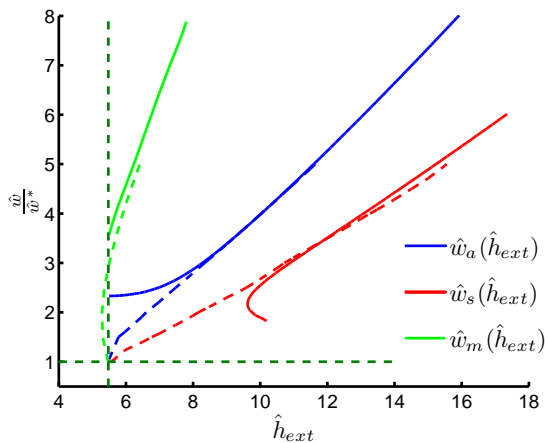}
\caption{Numerical simulations: The marginally stable (red), optimal
  (blue) and maximal period (green) of the concertina
  pattern as a function of the external field $\hat h_\ext$. The
  dashed and solid curves depict the result on the basis of the reduced
  model and on the basis of domain theory, respectively.}
 \label{hysloop1}
 \end{figure}

Summing up, domain theory in conjunction with a Bloch-wave argument
indicates that the concertina pattern of period $w$ is present or stable
at a given field $ h_\ext$
if and only if $w\sim\ell^2t^{-1}  h_\ext\ln (d^{-2/3}\ell^{4/3}t^{-2/3} h_\ext)$, which is
confirmed by the numerical simulations. In particular we expect that
the height of the triangular domains ($\sim \frac{w}{m_2^0}$) is close
to constant as the external field increases, cf.\ \eqref{scalingperiod}
and \eqref{scalingmag}.  
If the period deviates
by a (large) factor from that expression, it becomes unstable. 
On the other hand, this analysis also suggest that there is a {\it range}
of $w\sim \ell^2t^{-1} h_\ext\ln  d^{-2/3}\ell^{4/3}t^{-2/3}h_\ext$
for which the concertina pattern
is stable, see Figure \ref{hysloop1}. This may explain some of the scatter in the experimental data
and the pattern's hysteresis. 
\smallskip

Figure \ref{hysloop1} displays the marginally
stable period (below the red curve the minimal energy per period is concave
and thus the concertina of smaller period unstable as the field increases) and the optimal period depending on the
external field. The upper green curve indicates the turning points of
the $\hat w_m$-periodic branches, i.e.,
the smallest external field for which a
concertina of a certain maximal period $\hat w_m$ exists -- clearly
the $m$ in $\hat w_m$ stands for {\bf m}aximal. Observe that the maximal period $\hat w_m$ on the basis of
domain theory and on the basis of the reduced model coincide for large
external field, too.  
The region bounded by $\hat w_s$ and $\hat w_m$ corresponds to the
range of stable periods.

 \begin{figure}[htb]
 \includegraphics[width=0.45\textwidth]{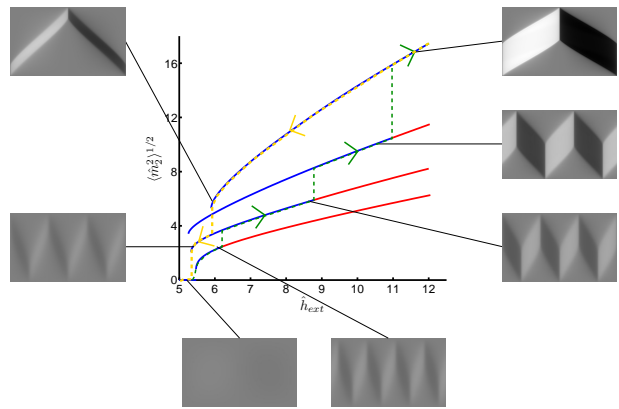}
  \caption{Numerical simulations: The hysteresis loop. As we increase the external field we
    follow the red path: The concertina pattern coarsens if the period
  is smaller than the stable period. As we decrease the field we
  follow the yellow path: Starting
from a coarsened concertina the pattern degenerates as we reach the
turning point of the branch. The pattern refines towards the optimal
period until it finally disappears.}
 \label{hysloop2}
 \end{figure}
 \begin{figure*}[htb]
 \includegraphics[width=0.2\textwidth]{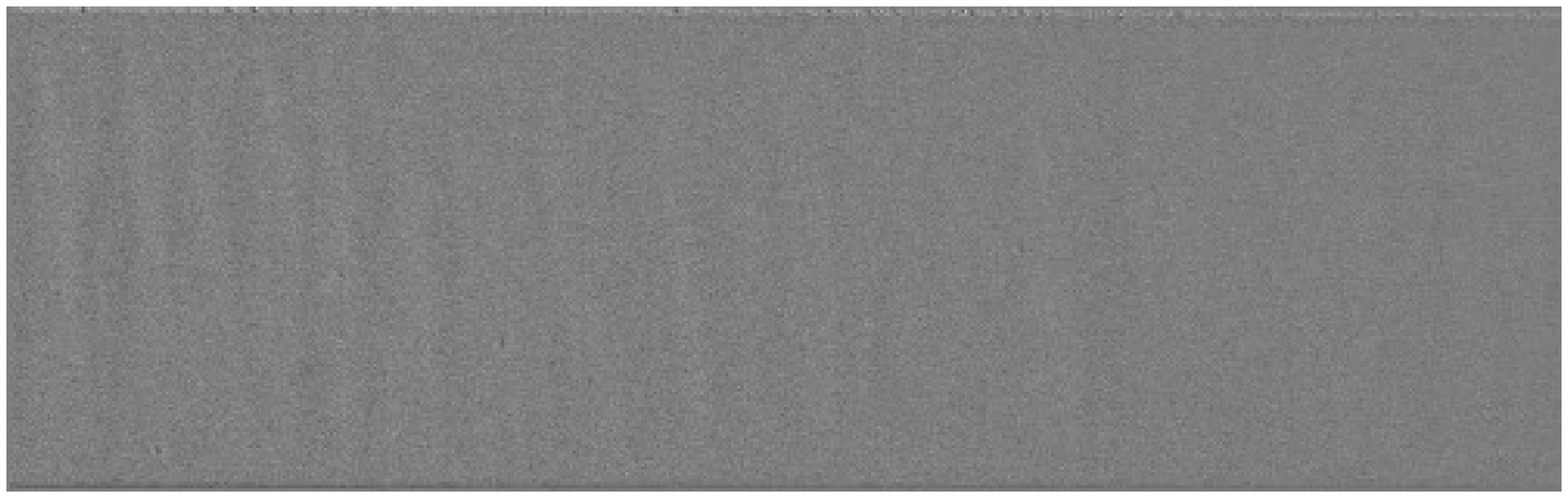}
 \includegraphics[width=0.2\textwidth]{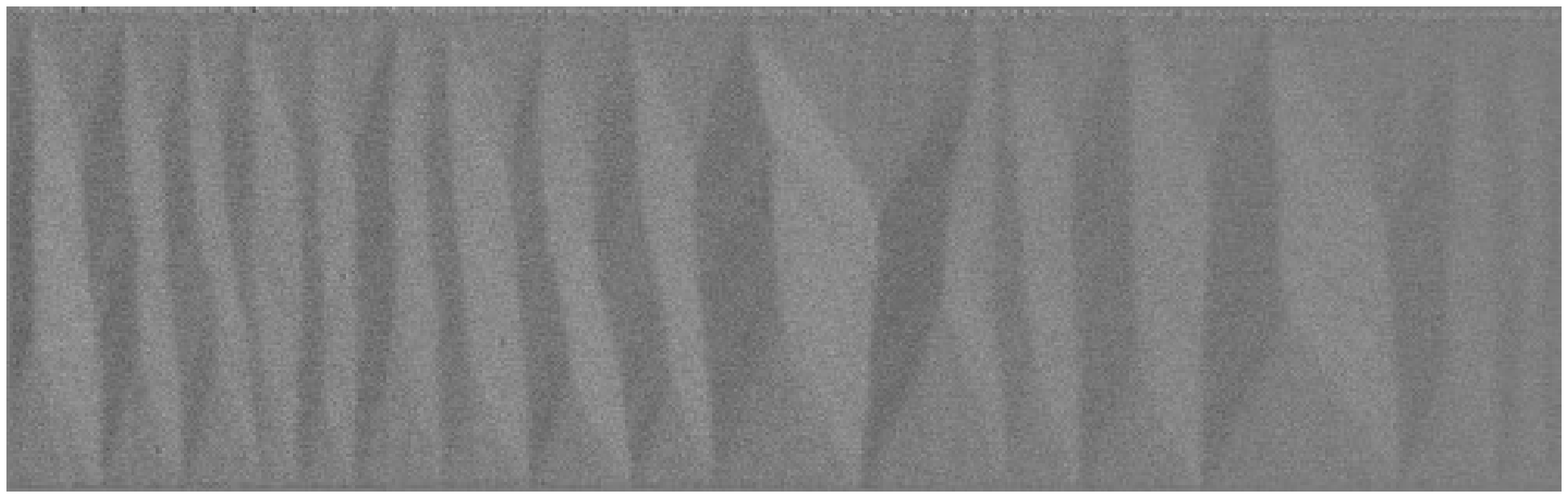}
 \includegraphics[width=0.2\textwidth]{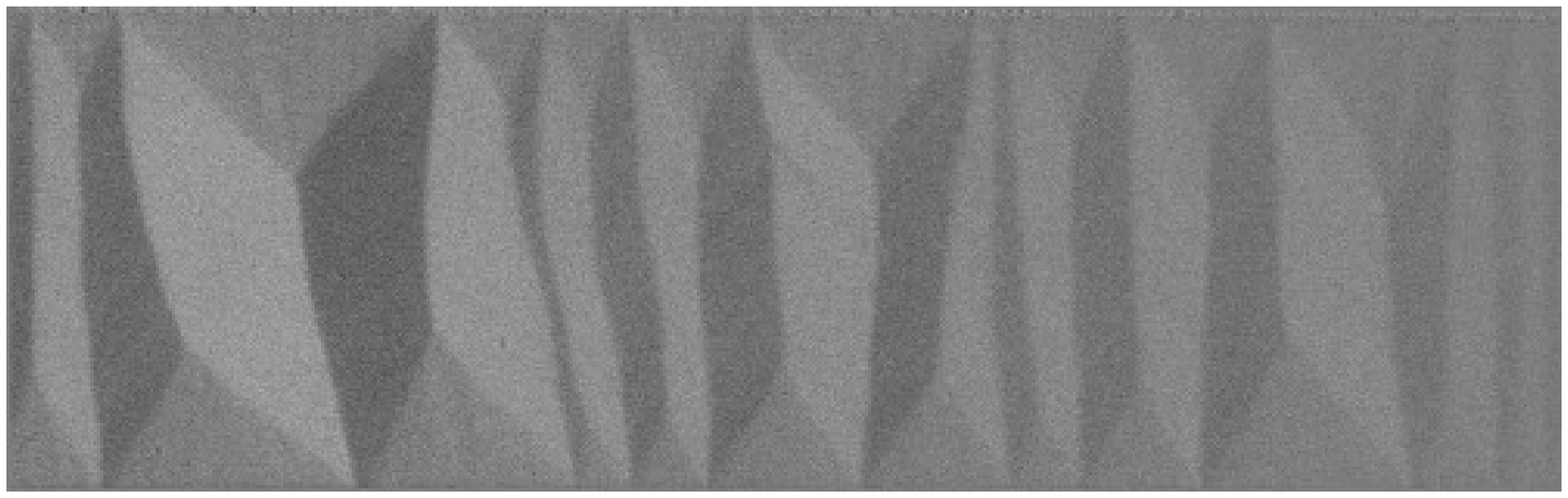}
 \includegraphics[width=0.2\textwidth]{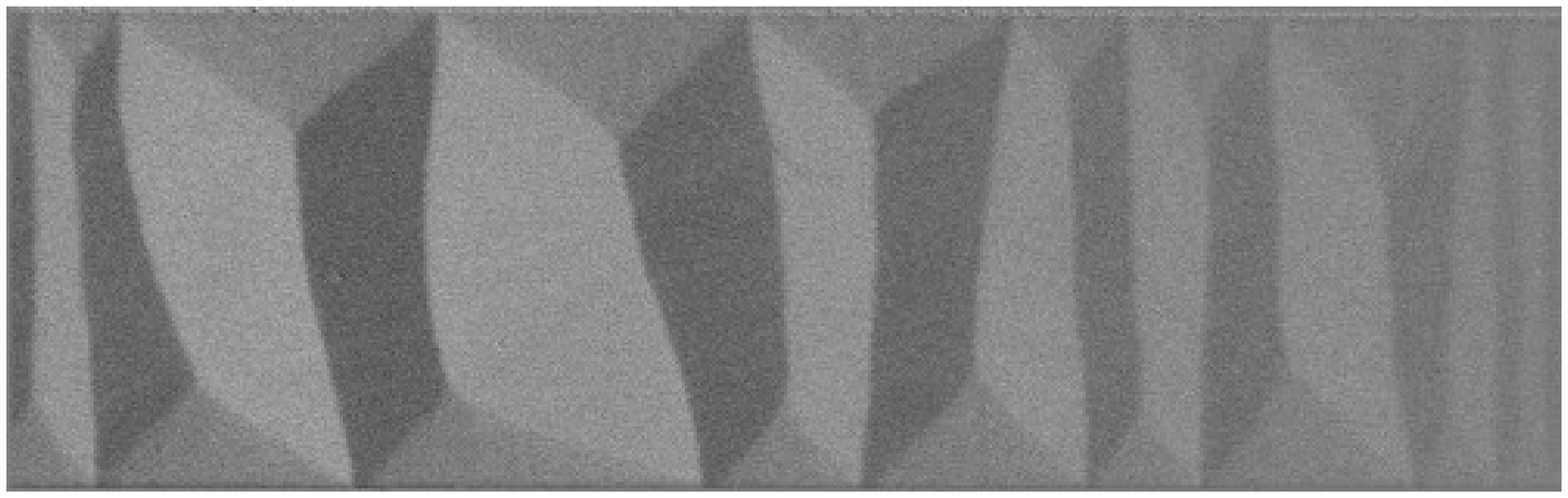}\\
 \includegraphics[width=0.2\textwidth]{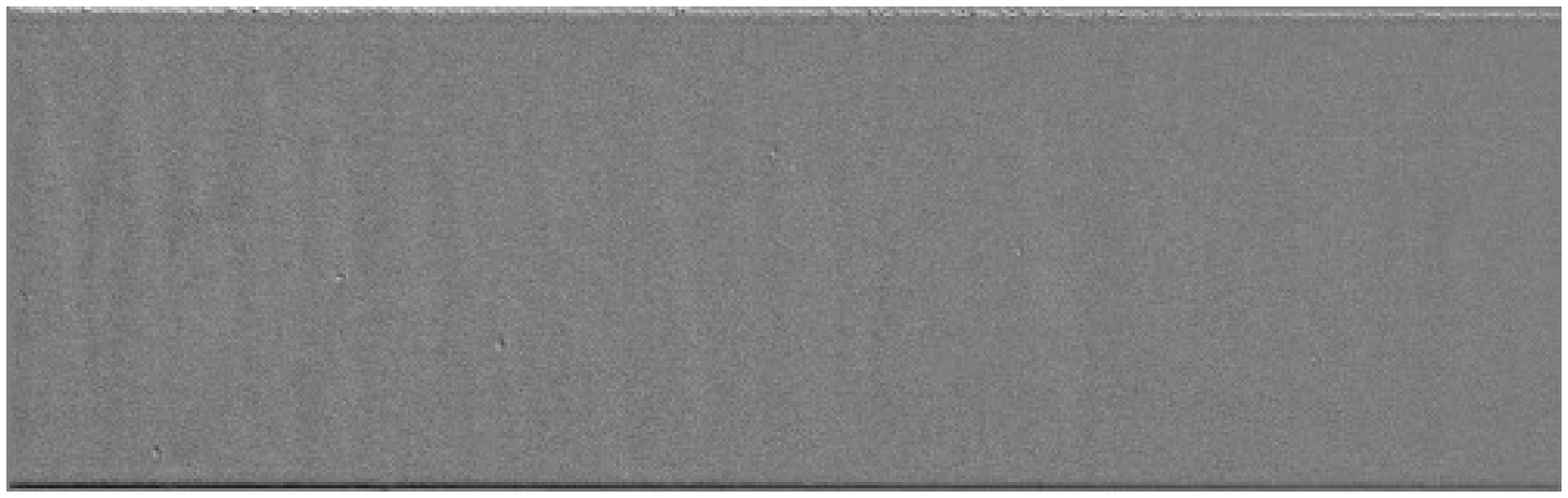}
 \includegraphics[width=0.2\textwidth]{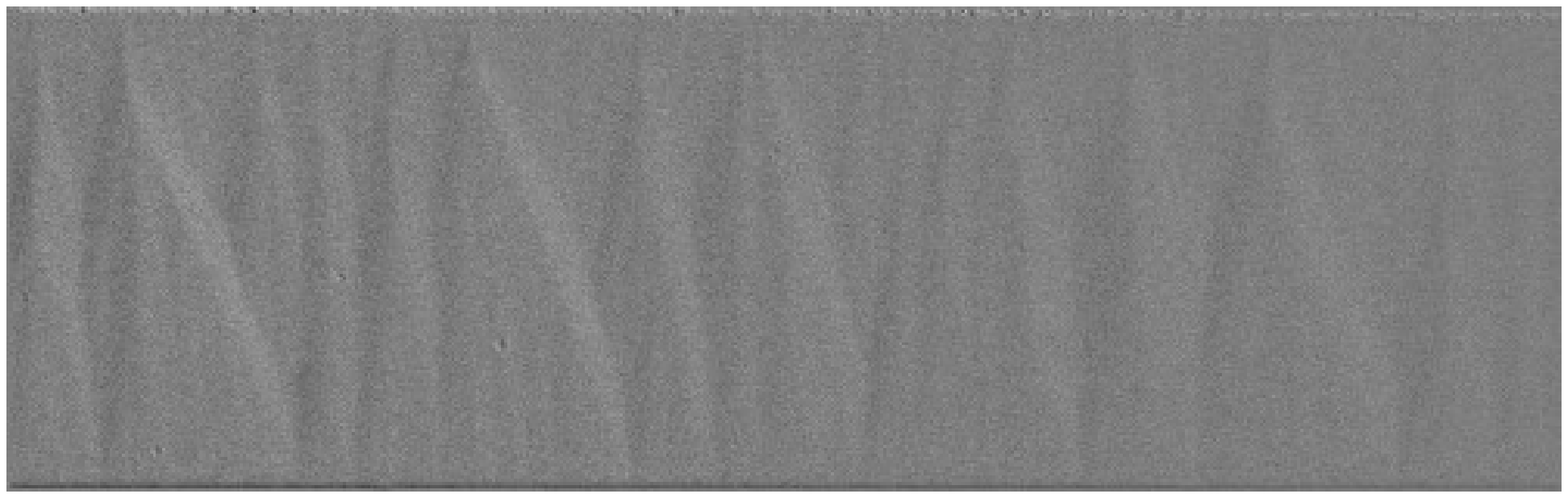}
 \includegraphics[width=0.2\textwidth]{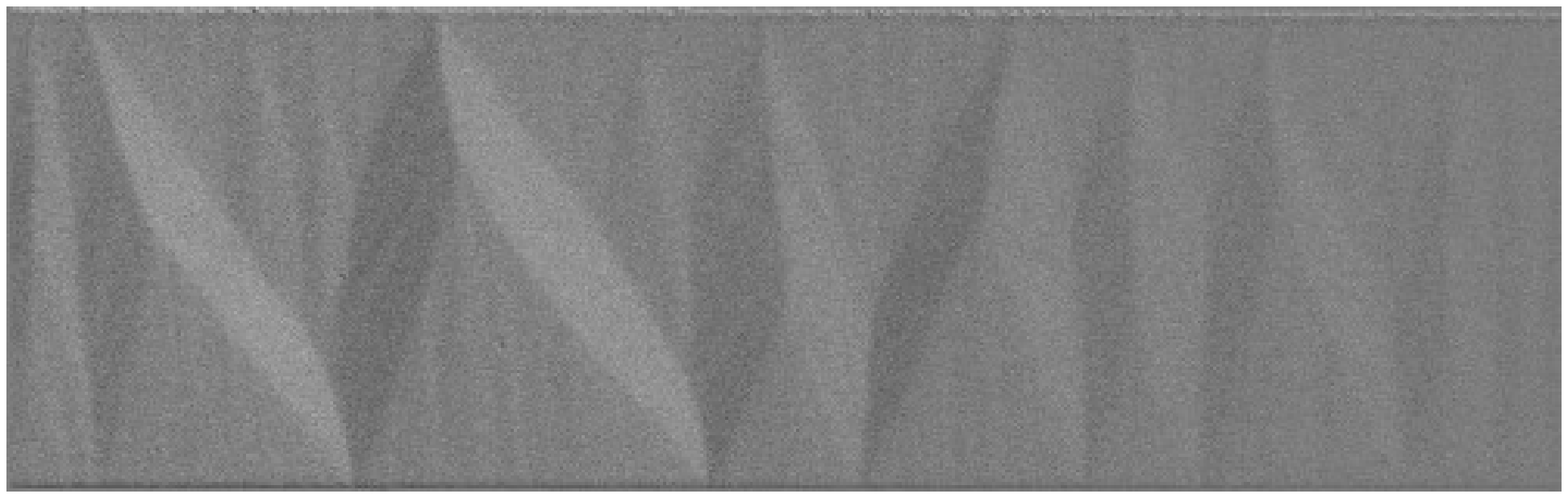}
 \includegraphics[width=0.2\textwidth]{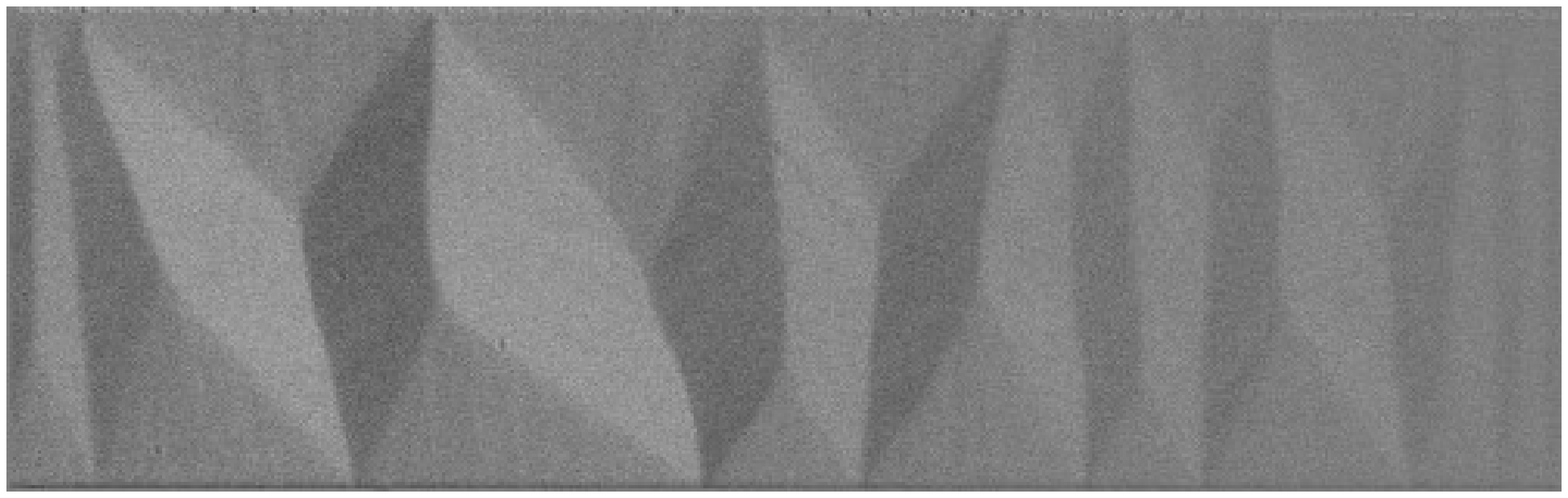}
  \caption{Experiment: The hysteresis cycles of a Permalloy sample of
    $30$nm thickness and $50\upmu$m width. The upper row shows the
    pattern as the external field increases (from left to right), the
    lower row shows the pattern as the external field decreases (from
    right to left).}
 \end{figure*}

\section{Discretization and numerical simulations}
\label{Numerics}
The numerical simulations are based on a finite difference
discretization of the reduced rescaled energy functional
\eqref{rescaledenergy}. The transversal component $\hat m_2$ is
approximated on a uniform Cartesian grid. The discretization of the
exchange, anisotropy and Zeeman energy is straight forward. In case of
the non-linear charge density $\hat\sigma=-\hat \partial_1\frac{\hat m_2^2}{2}+\hat \partial_2
\hat m_2$ our choice of a finite difference
stencil is motivated by the inheritance of the shear invariance
\eqref{shearinvariance}.  
The stray-field energy can efficiently
be computed using Fast Fourier Transform with respect to $\hat
x_1$. For an introduction of the discretization scheme see
\cite[Subsection 3.2]{S06}. Note that the computation of the energy
and related quantities, such as gradient or Hessian, can be
parallelized since the non-locality is only with respect to one
dimension. For the parallelization we decompose the computational
domain into horizontal slices with respect to $\hat x_2$. 

\begin{figure}
\psset{unit=.8}
\begin{center}
\begin{pspicture}(0,0)(6,3.2)
\pnode(.7,-0.2){A}\pnode(2,0.75){C}\pnode(5.5,1.8){E}
\nccurve[angleA=40,angleB=200]{A}{C}
\nccurve[angleA=20,angleB=230]{C}{E}
\psline{->}(1.5,0.5)(5.25,2.75)
\pscircle[fillstyle=solid,fillcolor=black](4,2){.1}
\psline[linestyle=dashed](3.6,2.7)(4.8,.6)
\pscircle[fillstyle=solid,fillcolor=black](1.5,.5){.1}
\pscircle[fillstyle=solid,fillcolor=black](4.47,1.2){.1}
\rput[r](1.4,0.6){$(\hat m_2^n,\hat h_\ext^n)$}
\rput[c](5.3,3){$t^{n}$}
\rput[c](3.5,2.1){$p^{n+1}$}
\rput[c](4.2,.8){$(\hat m_2^{n+1},\hat h_\ext^{n+1})$}
\end{pspicture}
\end{center}
\caption{Tangent predictor-corrector continuation method.}
\label{predictor}  
\end{figure}
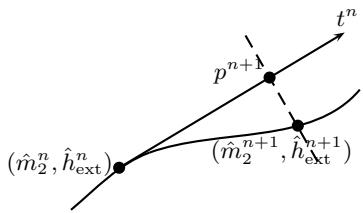

We apply numerical simulations to compute (local) minimizers and
stationary points. The naive approach using steepest descent algorithms for the computation
of minimizers is slow and even fails close to bifurcation
points. The iterative path-following techniques that we apply in order to compute
an approximation to a branch of stationary points are adapted to such situations,
cf.\ \cite{Georg}. The local tangent $t^n$ in a stationary
point $(\hat m^n,\hat h_\ext^n)$ of the branch
is used to obtain a predictor for the next point on the
branch $(m_2^{n+1},h_\ext^{n+1})$, cf.\ Figure \ref{predictor}. Within the corrector step the predictor is orthogonally (to
the tangent) projected onto the branch. This step amounts to the
solution of a non-linear equation, more precisely an augmented Euler-Lagrange
equation:
\begin{equation*}
\left(  \begin{array}{c}
    
\nabla_{\hat m_2}\hat E(\hat m^{n+1}_2,\hat h^{n+1}_\ext)\\
( ( \hat m^{n+1}_2,\hat h^{n+1}_\ext)-p^{n+t})\cdot t^n
  \end{array}\right)=0.
\end{equation*}

A bifurcation point can be detected with the help
of an appropriate indicator function, cf.\ \cite{Georg}. However, both the bifurcation detection and the branch-switching technique which are described in that
reference are applicable for simple bifurcations points only. As described
in detail in \cite{S10}, both methods can be modified in order to cope with multiple
bifurcation points. This extension relies on the fact that multiple
bifurcations which occur due to symmetries of the primary solution generically can
be reduced to simple bifurcation points, cf. \cite{SymmPers}. 
\section{Polycrystalline anisotropy}
\label{ripplesec}
The experiments usually do not show a clear-cut critical field with
 a first-order transition (i.e., subcritical bifurcation). 
This can be due
 to lack of experimental resolution (the amplitude of the transversal
 component $\langle m_2^2\rangle^{1/2}=d^{2/3} \ell^{-1/3} t^{-1/3} \langle \hat
 m_2^2\rangle^{1/2}$ at the turning point ranges between $0.015$ to $0.063$ for
 typical sample dimensions, namely widths $\ell$ between $10$ to $50\upmu$m and
 thicknesses $t$ between $30$ and $100$nm) or due to the 
 presence of the so-called ripple that smoothes out the transition, as
 we shall explain in this section.
The ripple is the in-plane small-scale oscillation of the
magnetization -- 
perpendicular to its average direction -- in extended films. In this section, we show how the linear ripple
theory developed in \cite{Hoffmann1,Harte} can be incorporated into our theory
for the concertina -- and explains the smoothing-out of
the first-order transition encountered in Section \ref{bifurcationsection}.

The ripple is triggered by an effective field 
of random direction on a small scale.
Several origins for this effective field have been
proposed in the literature, see for instance \cite[Section C]{Harte}; 
in polycrystalline thin films, the random orientation of the 
grains (via crystalline anisotropy) and local stresses 
(via magnetostriction) are seen as the main causes.
In our discussion, we focus on the former.
\smallskip

Hoffmann \cite{Hoffmann1} and Harte \cite{Harte}, 
based on the torque equilibrium, linearized around
a spatially constant magnetization (solely determined by
the external field and anisotropy). Hereby they identified
the linear response to (for instance) such a small-scale,
small-amplitude random effective field. The main finding
is that the stray field -- which penalizes transversal
more than longitudinal perturbations of the magnetization
because the former lead to a stronger charge oscillation --
results in a strong anisotropy of the response.  
Hoffmann \cite{Hoffmann1} characterized this response in terms of
the Green's function, whereas Harte \cite{Harte} characterized
it in terms of the multiplier in Fourier space, i.e., $k$-space. 
Since Hoffmann chose to expand the Green's
function (in a self-consistent way on the level of the length scale)
in terms of a Bessel function, see \cite[(5)]{Hoffmann1}, it deviates
order one from the exact expression in \cite{Krey,Brown}.
\smallskip

Clearly, the anisotropic rescaling \eqref{rescaling1} leading to
our reduced model and the anisotropic response have the same
origin. 
We will see that both the ripple and the
transition between ripple and concertina can be explained within the
framework of an extension of our reduced model. We note that our analysis of the
ripple is mainly a reformulation of the classical results. However,
the new insight is that the finite width $\ell$ of the sample leads
to a (continuous) transition from the ripple to the concertina.     
\smallskip

We now explain how to extend our reduced model.
We start from the 3-d model \eqref{B1:Micromag} with a uniaxial anisotropy
of strength $Q$ and position-dependent easy axis $e(x)$, i.e.,
with the term $-Q\int (m\cdot e)^2 \dd x$. In the approximation
of our reduced model, i.e., $m_3\equiv 0$, $m=m(x_1,x_2)$, and the
linearization $m_1\approx 1-\frac{m_2^2}{2}$ due to
$m_2^2\ll 1$, this term is, up to additive constants,
to leading order approximated by $-2Qt\int m_2\overline{e_1 e_2}\dd x_1\dd x_2$,
where $\overline{e_1 e_2}(x_1,x_2)$ denotes the vertical average of the
product of the first two components of the easy axis $e=(e_1,e_2,e_3)$. A random
anisotropy therefore acts to leading order as a random transversal external
field 
\begin{equation}
-2t\int_{\Omega'} h_\text{ripple} m_2 \dd x_1\dd x_2, \label{stochastic field}
\end{equation}
where $h_\text{ripple}=Q\overline{ e_1e_2}$. 
As mentioned, the position
dependence of $e$ arises from the random orientation of the grains
of size $\ell_{grain}$. Provided $t\ll\ell_{grain}\ll w^*$ (where
we take $w^*$ as a typical length scale of the magnetization pattern),
the stationary statistics of $\overline{e_1 e_2}$ are characterized by
\begin{multline}
\langle \overline{e_1 e_2}(0,0)\overline{e_1 e_2}(x_1,x_2)\rangle\\
=\ell_{grain}^2\delta(x_1)\delta(x_2)\langle \overline{e_1e_2}(0,0)^2\rangle,
\label{statistics}
\end{multline}
where $\langle \cdot\rangle$ denotes the ensemble average and $\delta$
the Dirac function. 
\smallskip

For subcritical fields $h_\ext<h_\ext^*$, we neglect the non-linear
term in the stray-field energy in \eqref{reduziertesmodell}. The resulting energy
functional is quadratic and linear in $m_2$, hence it is conveniently
expressed in terms of ${\mathcal F} m_2( k_1,k_2)$, which denotes the
Fourier transform of $ m_2$ in $x_1$ and the Fourier sine series in $x_2$:
\begin{multline*}
E( m_2) \\\approx \int_{-\infty}^\infty \sum_{k_2\in \frac{\pi\mathbb Z}{\ell} }
(d^2 k_1^2+\tfrac{1}{2}tk_2^2
k_1^{-1}-h_\ext) |\mathcal F m_2 |^2\\-2\mathcal{F} h_\text{ripple}
\mathcal F^{-1} m_2 \dd  k_1.
\end{multline*}
The explicit minimization yields
\begin{equation}\label{Ot.5}
{\mathcal F} m_2( k_1, k_2)
=\tfrac{1}{(d^2 k_1^2+\tfrac{1}{2}t k_2^2
k_1^{-1}- h_\ext)}{\mathcal F} h_\text{ripple}( k_1, k_2).
\end{equation}
We interpret this $m_2$ as the ripple.
Since (\ref{statistics}) on the level of ${\mathcal F} \overline{e_1 e_2}$ reads
$\langle |{\mathcal F} \overline{e_1 e_2}( k_1, k_2)|^2\rangle=\ell^2_\text{grain}$, (\ref{Ot.5}) is best
expressed in terms of the energy spectrum:
\begin{equation}\label{Ot.6}
\langle|{\mathcal F}m_2(k_1,k_2)|^2\rangle=Q^2\tfrac{\ell^2_\text{grain}}{(d^2 k_1^2+\tfrac{1}{2}t k_2^2
k_1^{-1}- h_\ext)^2}.
\end{equation}
This formula clearly displays the afore mentioned
anisotropic response of $ m_2$ to the isotropic field $h_\text{ripple}$.
\smallskip

From formula (\ref{Ot.6}) one can infer the predominant
wavenumber of the ripple, that is,
\begin{equation}\label{Ot.7}
\langle | k_1|\rangle = \frac{\sum_{k_2}\int_{-\infty}^\infty | k_1|
  \langle |\mathcal
F m_2|^2\rangle \dd  k_1}{\sum_{k_2}\int_{-\infty}^\infty \langle|\mathcal
F m_2|^2\rangle \dd  k_1}.
\end{equation}
For moderate stabilizing fields $t^2d^{-2}\gg
-h_\ext \gg d^{-2/3}\ell^{4/3}t^{-2/3} $,
we obtain from \eqref{Ot.7} that the average wavenumber scales as $\langle
|k_1|\rangle\sim (-h_\ext)^{1/2}d^{-1} \ll t d^{-2}$. This is the scaling
of the predominant wavenumber of the ripple in an extended film
\cite[p.34, (7)]{Hoffmann1}. 
Notice that the lower bound characterizing Regime III is equivalent to
$t^2d^{-2} \gg d^{-2/3}\ell^{4/3}t^{-2/3}$.
For large stabilizing fields $-h_\ext\gg t^2d^{-2}$ one can show that the average amplitude of the
ripple, given by $\int \sum_{k_2}\langle |\mathcal F m_2|^2\rangle \dd k_1$, tends to
zero.
Moreover, from (\ref{Ot.7}), because of the discreteness of $ k_2$, we
can infer
\begin{equation*}
\lim_{h_\ext\uparrow h_\ext^*}\langle  |k_1|\rangle
=\frac{2\pi}{w^*},
\end{equation*}
which is the wavenumber of the unstable mode \eqref{wavenumber}.
We thus learn that, as the strength $ h_\ext$ of the external 
field increases from negative values towards the critical value, 
the average wavelength of the ripple continuously increases 
\begin{itemize}
\item from the values characteristic for a film which is infinite 
in both $x_1$ and $x_2$-directions
\item to the wavelength of the
unstable mode that is at the origin of the concertina pattern
(which depends on the sample width).
\end{itemize}
Due to this transition it is thus not surprising that ripple and small-amplitude
concertina are difficult to distinguish.
\smallskip

We now address the numerical simulation of our augmented model
(\ref{Ot.3}). Let us therefore first rewrite the additional term
\eqref{stochastic field} in the rescaled variables
\eqref{rescaling1}. The rescaled reduced model \eqref{rescaledenergy} is
augmented by 
\begin{equation}\label{Ot.3}
-2\int \hat h_\text{ripple} \hat m_2 \dd\hat x_1\dd\hat x_2,
\end{equation}
where $\hat h_\text{ripple}$ is a stationary Gaussian field of vanishing mean and of
variance
\begin{equation}\label{Ot.4}
\langle \hat h_\text{ripple}(0,0)\hat h_\text{ripple}(\hat x_1,\hat x_2)\rangle
=(\sigma^*)^2\delta(\hat x_1)\delta(\hat x_2),
\end{equation}
with $\sigma^*=d^{-10/6}\ell^{5/6}t^{-1/6}Q\ell_{grain}\langle
\overline{e_1e_2}(0,0)^2\rangle^{1/2}$. In case of a uniform
distribution of the anisotropy axis in the plane, we have for example that $\langle
\overline{e_1e_2}(0,0)^2\rangle=\frac{1}{8}.$
\smallskip

On the level of the discretization, the field $\hat h_\text{ripple}$ is modeled as
a Gaussian random variable of mean zero, which is identically 
and independently distributed from grid point to grid point and
has variance $(\sigma^*)^2{\Delta\hat x_1}^{-1}{\Delta\hat x_2}^{-1}$, where $\Delta\hat x_i$ denotes
the grid size in direction $\hat x_i$. For the numerical simulations
we thus have to determine the value
of $\sigma^*$ for a typical sample. Let us consider a film of $30$nm
thickness and $70 \upmu$m width with typical grain size $\ell_\text{grain}=15nm$. For a local strength of anisotropy $Q=5\times
10^{-3}$ we obtain that $(\sigma^*)^2=125.87$.   
For the value of $(\sigma^*)^2=110.83$, our numerical simulation
indeed shows a continuous transition from the ripple to the
concertina pattern instead of a first-order phase transition due
to a subcritical bifurcation, see Figure \ref{coarseningripple}.
\smallskip

We also believe that our reduced model is the appropriate framework
to analyze the non-linear corrections to the linear ripple
theory. Indeed, we have seen in Section \ref{bifurcationsection} that it captures
the transition from the unstable mode to low-angle symmetric
N\'eel walls. We thus believe it also captures the transition from
the ripple to the blocked state that is related to hysteresis in extended
thin films \cite{Feldtkeller}.

Closing this section, we contrast the ripple, that can be seen
as a consequence of {\it quenched disorder}, to the effects of {\it thermal
fluctuations}. Thermal
fluctuations can be modeled by a random external field term 
in the Landau-Lifschitz-Gilbert
equation that is white noise in space {\it and time}.
The reason for modeling thermal fluctuations by a 
space-time {\it white} noise torque in the Landau-Lifschitz-Gilbert
equation is that the stationary measure of this Langevin equation 
is given by the Gibbs distribution
\begin{equation}\label{Gibbs}
\frac{1}{Z}\exp(-E(m))\dd m,
\end{equation}
where $E(m)$ is the 3-d micromagnetic energy functional and $\dd m$ is best
thought of as the high, but finite-dimensional measure after 
spatial discretization of $m$.
Following \cite[Subsection 2.4]{Berkov2005442}, we consider a situation where
the constant magnetization, say $m=(1,0,0)$, is a strict global minimizer
of $E$ (because of a sufficiently strong external field in direction
$(1,0,0)$). This justifies to replace $E(m)$ and $\dd m$ in (\ref{Gibbs}) by its
Hessian ${\rm Hess}(\delta m,\delta m)$ in $(1,0,0)$ and
$\dd\delta m_2 \dd\delta m_3$, respectively. This has the advantage that we obtain
a Gaussian measure that can be explicitly analyzed. The outcome is the
following: In case of a bulk material,
the expected value of $\delta m_2^2+\delta m_3^2$ diverges
as the mesh size $\Delta  x$ goes to zero; the expected value of the
wavenumber $|k|$ behaves as $\Delta  x^{-1}$. The same holds for thin films,
although the divergence is just logarithmic. 
\smallskip

This simple analysis highlights the need of
a renormalization in case of thermal fluctuations.
As we have seen, quenched disorder coming from polycrystallinity
can be modeled by a random field term in the micromagnetic energy that
is white noise {\it only} in space. As opposed to thermal fluctuations, 
there is no divergence in the amplitude
of the excitations in case of such a field term that is white noise
in space only -- the critical dimension for this random effect
is four. Moreover, in thin films, the dominant wavelength of the
in-plane fluctuations excited by such a field is determined by both
exchange and stray-field energy and turns out to be much larger than
the atomistic length scale $d$ and the typical grain size $\ell_\text{grain}$.
         
\section{Uniaxial anisotropy}
\label{anisotropy}
We now address the effect of uniaxial anisotropy -- constant throughout the
sample -- on the formation of the concertina pattern. 
We  focus on the two cases in which the easy axis coincides
with the $x_2$-axis (transversal anisotropy
$e=(0,1,0)$ in \eqref{B1:Micromag}) or in which the easy axis coincides with the
$x_1$-axis (longitudinal anisotropy $e=(1,0,0)$ in
\eqref{B1:Micromag}). Clearly, such type of anisotropy has no effect on the stationary point of the energy, i.e., the uniform magnetization. 
On the level of the reduced model 
both cases can be represented (up to an additive constant) by the additional quadratic term 
\begin{equation}
  \label{addaniso}
-Q \;t\int m_2^2\, \dd x_1 \dd x_2 
\end{equation}
with a {\it signed} quality factor $Q$. Transversal anisotropy corresponds
to $Q>0$, longitudinal anisotropy corresponds to $Q<0$. 
\smallskip

As will become clear below, when considering the 
effects of anisotropy, it is appropriate to expand the
Zeeman term to quartic order, i.e.,
\begin{equation*}
-h_\ext\;t\int (m_2^2+\tfrac{m_2^4}{4}\,) \dd x_1\dd x_2 .
\end{equation*}
The following Gedankenexperiment is helpful in understanding the
sequel: In extended thin films, i.e., $\ell=\infty$,
there is no incentive for a spatially varying magnetization
so that we may consider a constant magnetization $m_2$
in which case the relevant energy
per volume is given by 
$-Qm_2^2-h_\ext(m_2^2+\frac{m_2^4}{4})$. In this case the
critical field is given by $h_\ext^*=-Q$. For longitudinal
anisotropy, the bifurcation is subcritical, whereas for transversal anisotropy,
the bifurcation is supercritical and yields 
\begin{equation}
m_2=\pm (2(1+Q^{-1}h_\ext))^{1/2}.\label{predaniso2}
\end{equation}
Hence for finite $\ell$,
there are two competing mechanisms which lead
to a bifurcation and a selection of an amplitude for $m_2$:
uniaxial anisotropy and shape anisotropy in form of the stray-field energy.
\smallskip

\begin{figure}[htb]
  \centering
  \begin{pspicture}(0,0)(6,2.5)
\psline{->}(0,1)(6,1)
\psline[linecolor=red]{->}(1,1.)(6,1.)
\psline(1,1.2)(1,.8)
\rput[c](1,.5){$d^{2/3}\ell^{-4/3}t^{2/3}$}
\rput[c](1,0){Linear effect}
\psline(3,1.2)(3,.8)
\psline[linecolor=blue]{->}(3,1.07)(6,1.07)
\rput[c](3,1.5){$\ell^{-1}t$}
\rput[c](3,2){Strongly non-linear effect}
\psline(5,1.2)(5,.8)
\psline[linecolor=dgreen]{->}(5,1.14)(6,1.14)
\rput[c](5,.5){$d^{-2/3}\ell^{-2/3} t^{4/3}$}
\rput[c](5,0){Weakly non-linear effect}
\rput[c](6.5,1){$|Q|$}
\end{pspicture}
  \caption{The order of the different effects of anisotropy}
\label{ordering}
\end{figure}
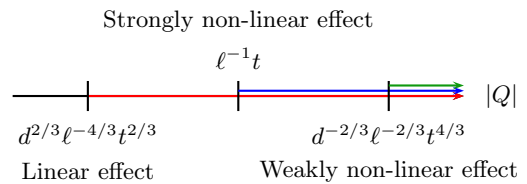

As we will see in the sequel, there are essentially three different
effects of anisotropy: 
a {\it linear} one, a {\it weakly non-linear} one, and a {\it strongly non-linear} one,
which we list and characterize below.
Notice that the order at which these effects arise with
increasing anisotropy does not agree with their ordering
with increasing non-linearity, cf.\ Figure \ref{ordering}: The linear effect becomes pronounced
for $|Q|\gg d^{2/3}\ell^{-4/3}t^{2/3}$,
the strongly non-linear one for $|Q|\gg \ell^{-1}t$,
and the weakly non-linear one only for $|Q|\gg d^{-2/3}\ell^{-2/3} t^{4/3}$.
Note that we have that $d^{2/3}\ell^{-4/3}t^{2/3}\ll \ell^{-1}t\ll
d^{-2/3}\ell^{-2/3} t^{4/3}$ provided $d^2\ell^{-1}\ll t$, 
which is the lower
bound on the film thickness which characterizes Regime III.
\smallskip

We mainly focus on the case of transversal anisotropy $Q>0$. In case of longitudinal anisotropy $Q<0$ we give an explanation for the experimental fact that the concertina cannot be observed at all.
\smallskip

\paragraph{Linear effect for weak anisotropy} $|Q|\gg d^{2/3}\ell^{-4/3}t^{2/3}$.
An obvious effect of anisotropy
is a shift of the critical field $h_\ext^*$ by the amount $-Q$;
we call it the ``linear effect'' of anisotropy
since it arises on the level of the linearization 
at $m_2\equiv 0$. In view of the scaling of the critical field
$h_\ext^*$ at $Q=0$, i.e., \eqref{fieldscaling}, we infer that the
value of
the critical field is dominated by the uniaxial anisotropy, i.e.,
\begin{equation}
\label{smallaniso}
h_\ext^*\approx -Q \quad \text{for}\quad |Q|\gg d^{2/3}\ell^{-4/3}t^{2/3}.
\end{equation}
We note that a transversal anisotropy decreases the distance between
the two critical fields $\pm h_\ext^*$ corresponding to the stationary states $\pm m^*$;
in particular, for $Q\sim d^{2/3}\ell^{-4/3}t^{2/3}$,
the critical field changes sign and thus the order between
the two critical fields switches. (Likewise for longitudinal anisotropy
the distance decreases.)
Although a clear-cut critical field
cannot be observed in the
experiments due to the
polycrystalline structure which
triggers the ripple, and since the
value of the effective external
field at the investigated sample
section is not available, the linear effect could be qualitatively confirmed: For
Permalloy samples of high (transversal) anisotropy the oscillatory instability
occurs before the external field is reversed. In accordance
with \eqref{smallaniso}, we observe for relatively wide
films that the relative strength of anisotropy increases and the
critical field decreases (theoretically approaching $-Q$). On the other hand for low-anisotropic Permalloy the
first oscillation is observed close to zero external field.     
\smallskip

\begin{figure}[htb]
\includegraphics[width=0.45\textwidth]{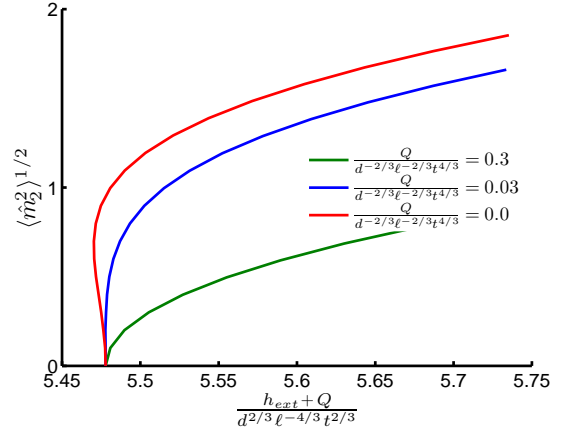}
\caption{Numerical simulations: Transition from sub- to supercritical bifurcation as strength
  of transversal anisotropy increases. For $Q=0.03\approx Q^*$ the
  bifurcation degenerates.}
\label{differentQs}
\end{figure}

\paragraph{Weakly non-linear effect} for strong anisotropy 
$|Q|\gg t(w^*)^{-1}\sim d^{-2/3}\ell^{-2/3}t^{4/3}$.
For sufficiently strong anisotropy $Q$, the quartic
term coming from the stray-field energy no longer
dominates the quartic term coming from the Zeeman energy
near the bifurcation. We call this effect the
``weakly non-linear effect'' of anisotropy, since it
can be analyzed on the level of an expansion of the reduced energy near
$m_2\equiv 0$ and $h_\ext=h_\ext^*$, cf.\ \eqref{Ot.1}, where we take into account the quartic Zeeman term
$-\frac{h_\ext}{4}tA^4 \int
(m_2^*)^4\dd x_1 \dd x_2$. The shift of the critical field suggests the following rescaling for the reduced external field
\begin{equation*}
  \hat h_\ext=d^{-2/3}\ell^{4/3} t^{-2/3}(h_\ext+Q).
\end{equation*}
In addition we set 
\begin{equation*}
  \hat Q=-\tfrac{1}{4}d^{2/3}\ell^{2/3} t^{-4/3} h_\ext
\end{equation*}
so that we obtain with the same rescaling of energy, length and
magnetization as in \eqref{rescaling1} and \eqref{energyscaling} the reduced energy functional augmented by
 \begin{align*}
+\hat Q  \int_{\hat \Omega'} \hat m_2^4 d\hat x_1d\hat x_2.
 \end{align*}
Therefore the energy close to the bifurcation takes the form of
 \begin{align*}
 \hat  E&(A  \hat m_2^*+A^2 \hat m_2^{**})\\&\approx-(\tfrac{\pi}{2})^{1/3}(\hat h_\ext - \hat h_\ext^*)\,A^2+( \tfrac{9}{64} \hat Q -\tfrac{\pi}{640})A^4.
 \end{align*}
For $|Q|\gg d^{-2/3}\ell^{-2/3} t^{4/3}\gg
d^{2/3}\ell^{-4/3}t^{2/3}$ the critical field asymptotically behaves as $h_\ext^*\approx
- Q$, cf.\ \eqref{smallaniso}, so that the reduced quality factor behaves as $\hat Q\approx
\frac{1}{4}d^{2/3}\ell^{2/3} t^{-4/3} \,Q$ close to the critical field.  
From the latter we read off that in the regime  $Q\gg 
d^{-2/3}\ell^{-2/3} t^{4/3}$ the quartic coefficient becomes
positive and therefore the bifurcation becomes
supercritical, cf.\ Figure \ref{differentQs}. 
Essentially it is a perturbation of the constant-magnetization
bifurcation in infinitely extended films mentioned above, cf.\ 
\eqref{predaniso2}.
In particular, the selected amplitude in this case scales as
$m_2\sim A\sim(1+h_\ext Q^{-1})^{1/2}$. 
On the level of the extended bifurcation analysis one finds that the
period of the unstable mode $ w^*$
lies in the stable region in the neighborhood of the critical
field. In agreement with this, for increasing external fields the numerical simulations show
that no modulation instability occurs and that there is no
coarsening. We note that domain theory is consistent with the numerical
simulations, too.

\begin{figure}[htb]
\includegraphics[width=0.45\textwidth]{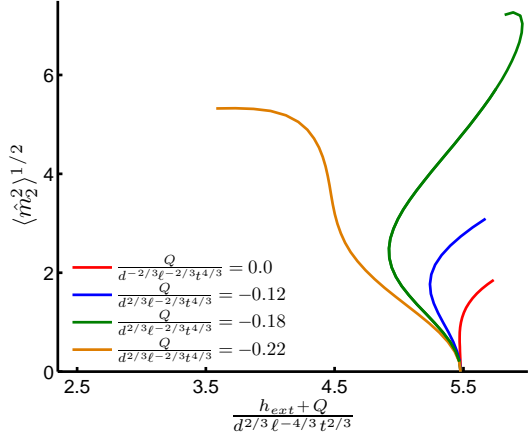}
\caption{Numerical simulations: Loss of the turning point as strength of longitudinal anisotropy increases}
\label{decreasingQs}
\end{figure}

\smallskip
On the other hand, for large longitudinal anisotropy, i.e.,
$-Q\gg d^{-2/3} \ell^{-2/3} t^{4/3}$, we expect that
there is no turning point on the bifurcating branch so
that it remains unstable -- to the effect that
no concertina pattern forms in the first place. The numerical
simulations in Figure \ref{decreasingQs} show a second turning point which
coincides with the break-up of the concertina pattern. For even larger
longitudinal anisotropy the first turning point is destroyed, cf.\
Figure \ref{decreasingQs}.
\smallskip

This observation can also be confirmed on the level of domain theory
where we take into account anisotropy and the quartic term in the Zeeman energy, cf.\ \eqref{Edomainun}: 
\begin{align}
 e_\domain(m_2^0,w)\,=\,&2\left(\ell-\tfrac{w}{ m_2^0}\right) e(m_2^0)+4\,\frac{ w}{ m_2^0}
 \,e\left(\tfrac{ m_2^0}{2}\right)\notag\\
&-(h_\ext+Q)( m_2^0)^2t\left( w \ell   -\tfrac{ w^2}{
    m_2^0}\right)\notag\\
&-h_\ext\tfrac{1}{4}(m_2^0)^4t\left( w \ell   -\tfrac{ w^2}{
    m_2^0}\right).\label{domainaniso}
 \end{align}
The quartic wall energy cannot compensate the destabilizing quartic Zeeman
contribution provided $h_\ext \,t\,w\gg \,t^2$ (up to a
logarithm). Therefore due to $h_\ext^* \sim -Q$ and $w\sim d^{2/3}\ell^{2/3}t^{-1/3}$ close to the bifurcation there are no (local) minimizers of the energy. 
\smallskip

Typical values for our Permalloy samples of strong uniaxial anisotropy range from
$\hat Q=\frac{|Q|}{4 d^{-2/3}\ell^{-2/3}t^{4/3}}\approx 2.1 \times 10^{-4}$ to
$0.023$ depending on the sample's width and thickness ($Q=5\times10^{-4}$, $t=10$nm to $150$nm, $\ell=10 \upmu$m to $50\upmu$m).
Typical values for CoFeB range from $\hat Q=7.8 \times \times 10^{-4}$ to $0.011$ ($Q\approx1.5\times 10^{-3}$, $t=30$nm-$100$nm, $\ell=10 \upmu$m-$50\upmu$m).
%
The uniaxial anisotropy is thus too
small to cause the weakly non-linear effect. However, although local minimizers
of the energy might exist in case of longitudinal anisotropy,
still the energy is not coercive as soon as the external field is
reversed.   
\smallskip

\paragraph{Strongly non-linear effects} for moderate anisotropy
$|Q|\gg \ell^{-1}t$. In that case one can distinguish two different
scenarios in the formation of the concertina:
\begin{itemize}
\item Scenario I:
If the amplitude (and shape) of the concertina pattern would not be affected
by anisotropy (besides the
critical field at which it bifurcates), like in an infinitely extended film, its optimal
amplitude would scale as  
\begin{align}{m_2}_a\sim \ell t^{-1} (h_\ext-h_\ext^*)&\overset{\eqref{smallaniso}}{\approx}\ell t^{-1} (h_\ext+Q)
\notag\\
&=\ell t^{-1} Q (1+Q^{-1} h_\ext),\label{scalingwall}
\end{align}
up to a logarithm for $h_\ext-h_\ext^*\gg d^{2/3} \ell^{-4/3} t^{2/3}$, as we have seen in \eqref{scalingmag} in Section \ref{optimal period}. 
\item Scenario II: If the amplitude of the
concertina pattern would be dominated by transversal anisotropy,
it would behave as 
\begin{align}
{m_2}_a\overset{\eqref{predaniso2}}{\sim}
(1+Q^{-1}h_\ext)^{1/2} \label{scalingbulk}
\end{align} for
$0<(1+Q^{-1}h_\ext)\ll 1$. 
\end{itemize}
Hence we expect
that for $Q\gg \ell^{-1}t$, the concertina pattern
is limited by stray field effects as long as
$0<1+Q^{-1}h_\ext\ll (Q^{-1}\ell^{-1}t)^2$ and by anisotropy effects once 
$(Q^{-1}\ell^{-1}t)^2\ll 1+Q^{-1}h_\ext\ll 1$.
Loosely speaking, the effect
of anisotropy kicks in for a large amplitude and is
most prominent close to field strength where the concertina
pattern vanishes. We call this the ``strongly non-linear effect''
of anisotropy. (Also this provides a reason to expand
the Zeeman term to higher order.)
\smallskip

\begin{figure*}[htb]
\psset{unit=2.7}
\begin{pspicture}(-0.5,-.5)(5,3)
\psset{linewidth=1pt}
 \psset{plotpoints=10}
\psaxes[labels=none,ticks=none]{->}(-.2,0)(4,2.5)
\psaxes[labels=none,ticks=none]{->}(3.5,0)(4,2.5)
\infixtoRPN{((x-1.4)/2)^(1/2)+1.05}\psplot[linecolor=black]{1.55}{3.3}{\RPN}
\infixtoRPN{(1+1.5)^(1/4)*(x)/1.5}\psplot[linecolor=black]{0.3}{1.45}{\RPN}
\infixtoRPN{((x-1.3)/2)^(1/4)+.6}\psplot[linecolor=blue]{1.55}{3.3}{\RPN}
\rput[c](4,-.3){$h_\ext$}
\rput[c](.75,2.5){Scenario I}
\rput[c](2.5,2.5){Scenario II}
\rput[c](.5,1){$\frac{\ell}{t}Q(1+Q^{-1}h_\ext)$}
\rput[c](1.5,.5){$\frac{\ell^2}{t}Q(1+Q^{-1}h_\ext)$}
\rput[c](1.5,.3){optimal $\sim$ stable}
\rput[c](2.5,2){$ (1+Q^{-1}h_\ext)^{1/2}$}
\rput[c](2.7,1.3){$ (t\ell)^{1/2}Q^{-1/2}(1+Q^{-1}h_\ext)^{1/4}$}
\psline(0,-.1)(0,.1)
\rput[c](0,-.3){$-Q$}
\psline(0.3,-.1)(0.3,.1)
\rput[c](0.8,-.3){$-Q+\mathcal O\big((\frac{d^2 t^2}{\ell^4})^{1/3}\big)$}
\psline(1.5,-.1)(1.5,.1)
\rput[c](2,-.3){$-Q+\mathcal O\big((\frac{t}{\ell})^{2}\frac{1}{Q}\big)$}
\psline(3.5,-.1)(3.5,.1)
\rput[c](3.5,-.3){$0$}
\psline(-.3,.3)(-.1,.3)
\rput[r](-.4,.3){$\Big(\frac{d^2}{\ell t}\Big)^{1/3}$}
\psline(-.3,1.2)(-.1,1.2)
\rput[r](-.4,1.2){$\frac{t}{\ell}\frac{1}{Q}$}
\psline(-.3,2.05)(-.1,2.05)
\rput[r](-.4,2.05){$1$}
\rput[c](3.5,2.7){$w$}
\psline(3.4,1.2)(3.6,1.2)
\rput[l](3.7,1.2){$\frac{t}{Q}$}
\psline(3.4,1.6)(3.6,1.6)
\rput[l](3.7,1.6){$\frac{(t\ell)^{1/2}}{Q^{1/2}}$}
\psline(3.4,2.05)(3.6,2.05)
\rput[l](3.7,2.05){$\ell$}
\psline(3.4,.3)(3.6,.3)
\rput[l](3.7,.3){$\Big(\frac{d^2\ell^2}{t}\Big)^{1/3}$}
\rput[c](-.2,2.7){$\langle m_2^2\rangle^{1/2}$}

\infixtoRPN{(1+1.5)^(1/4)*(x-.1)/1.5}\psplot[linecolor=blue]{0.3}{1.45}{\RPN}
\infixtoRPN{(1+1.5)^(1/4)*(x-.1)/1.5-.1}\psplot[linecolor=red, dash=.05 .05]{0.3}{1.45}{\RPN}
\infixtoRPN{1.1}\psplot[linecolor=red]{1.55}{3.3}{\RPN}
\psline[linecolor=red](4.3,2)(4.5,2)
\rput[l](4.3,1.9){marg.\ stable period $w_s$}
\psline[linecolor=blue](4.3,1.6)(4.5,1.6) 
\rput[l](4.3,1.5){opt.\ period $w_a$}
\psline[linecolor=black](4.3,1.2)(4.5,1.2)
\rput[l](4.3,1.1){opt.\ trans.\ comp.\ $m_2$}
\end{pspicture}
\caption{Table of scaling behavior of the optimal and marginally stable
  period and the amplitude of the transversal component in the regime $t\ell^{-1}\ll Q \ll d^{-2/3}\ell^{-2/3}t^{4/3}$.}
\label{scenario}
\end{figure*}
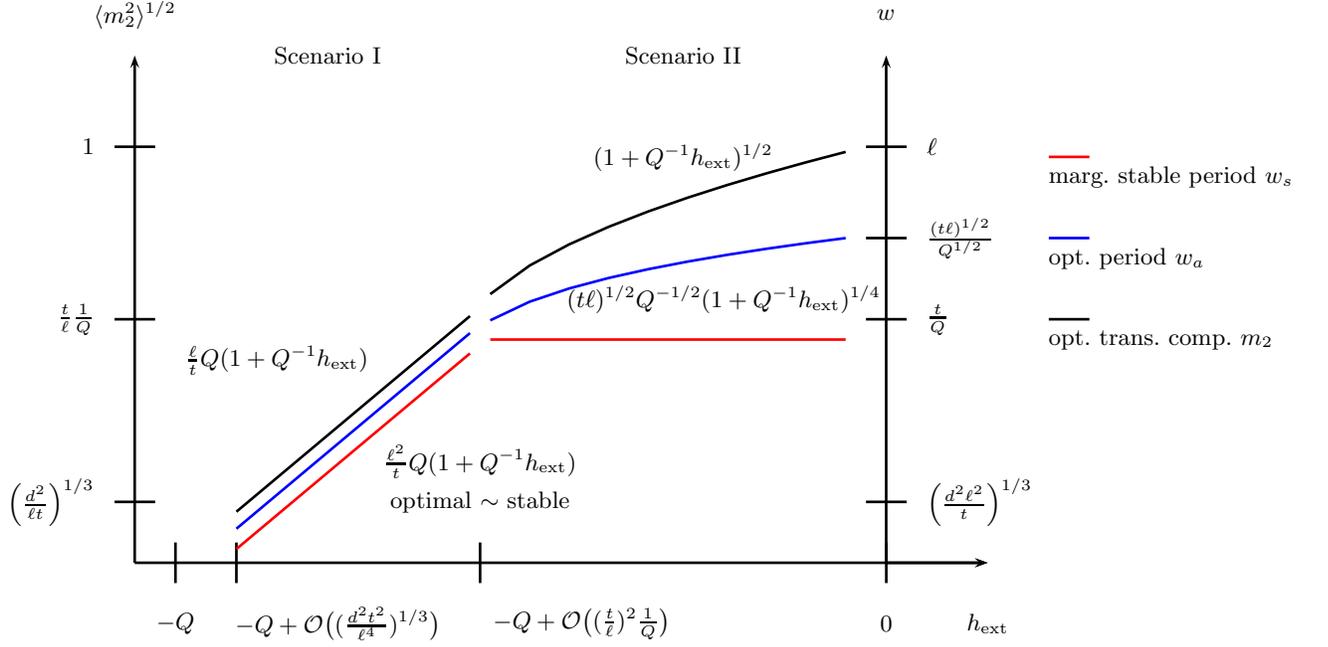

We note that we have to take into account the lower order wall energy
in Scenario II in order to determine the optimal period.  
In that case, a minimization of the energy per length yields the following
scaling behavior of the optimal period (up to a logarithm)
\begin{equation*}
 w_a\sim (\ell t)^{1/2}Q^{-1/2}(1+Q^{-1}h_\ext)^{1/4}. 
\end{equation*}
As we know from Section \ref{coarsening section} the experimentally
more relevant quantity
is the marginally stable period, i.e., the
largest period (as a function of the external field) for which the
minimal energy is convex. At the cross-over we expect that the marginally stable
period is of the order $\sim t Q^{-1}$, cf.\ Figure
\ref{scenario}. In fact, due to $ (Q^{-1}\ell^{-1}t)^2\sim
1+Q^{-1}h_\ext$ at the cross-over, we have that $w\sim \ell^2 t^{-1} Q (1+Q^{-1} h_\ext)\sim tQ^{-1}$, see \eqref{scalingwall} together with the fact that $w_s\sim \ell {m_2}_a$.
For a period of that order
the minimal energy in Scenario II turns out to be convex. Hence we
expect that the coarsening stops once $(Q^{-1}\ell^{-1}t)^2\ll
1+Q^{-1}h_\ext\ll 1$. Still the transversal 
component of the magnetization grows as $m_2\sim 
(1+Q^{-1}h_\ext)^{1/2}$ so that size and height of the closure
domains decrease.
\begin{figure}[htb]
\includegraphics[width=0.09\textwidth,height=0.075\textwidth]{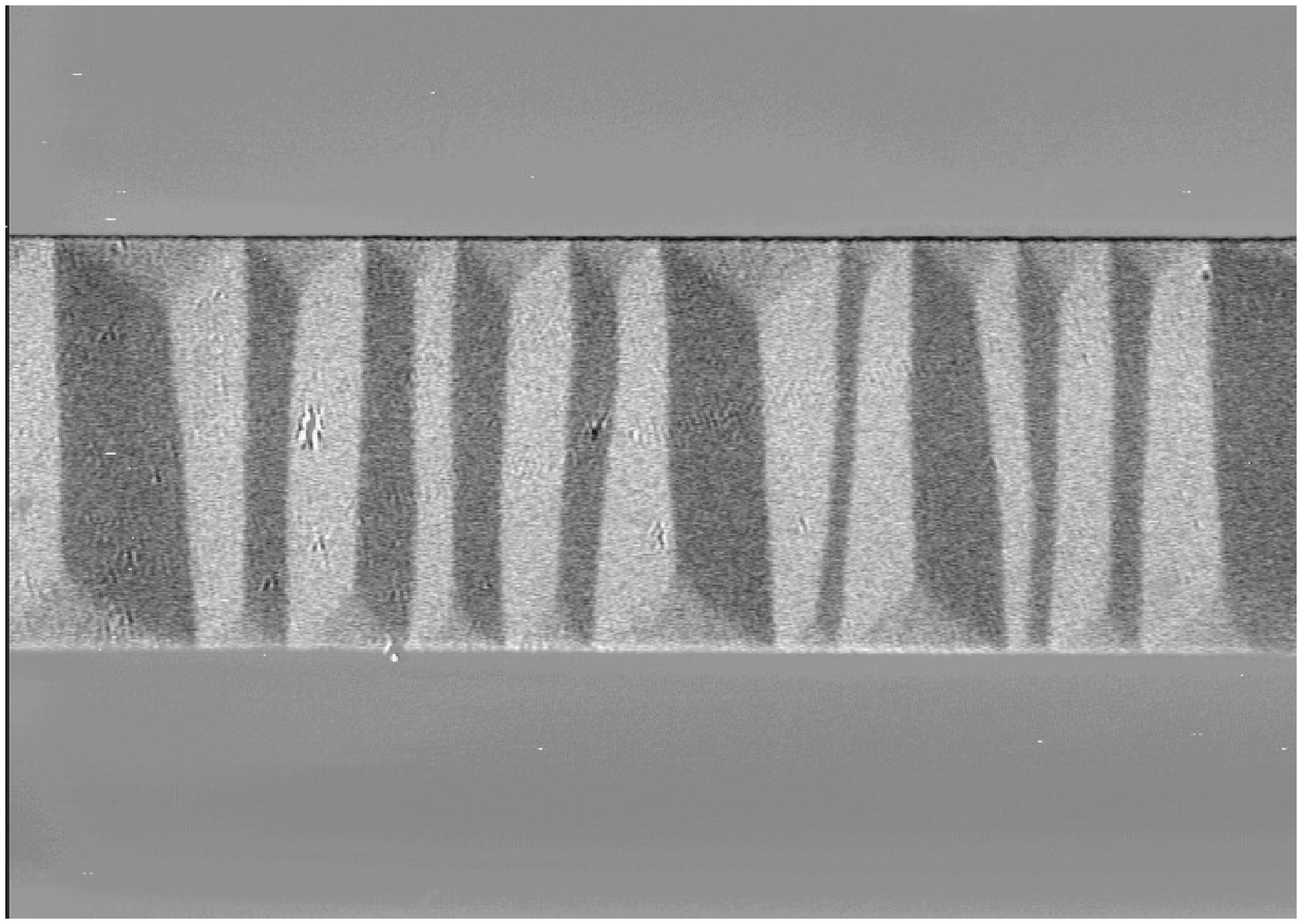}
\includegraphics[width=0.09\textwidth,height=0.075\textwidth]{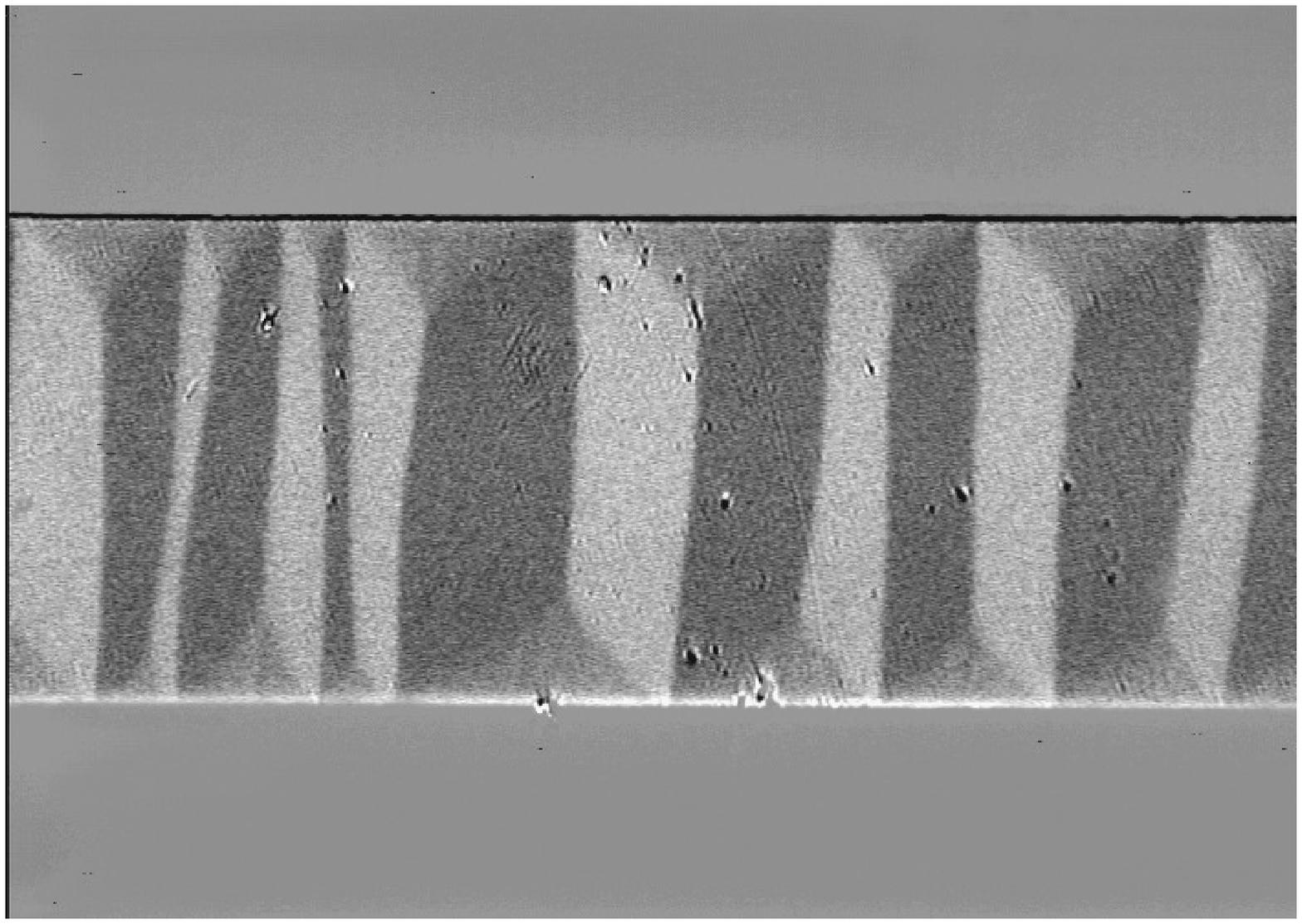}\qquad
\includegraphics[width=0.09\textwidth,height=0.075\textwidth]{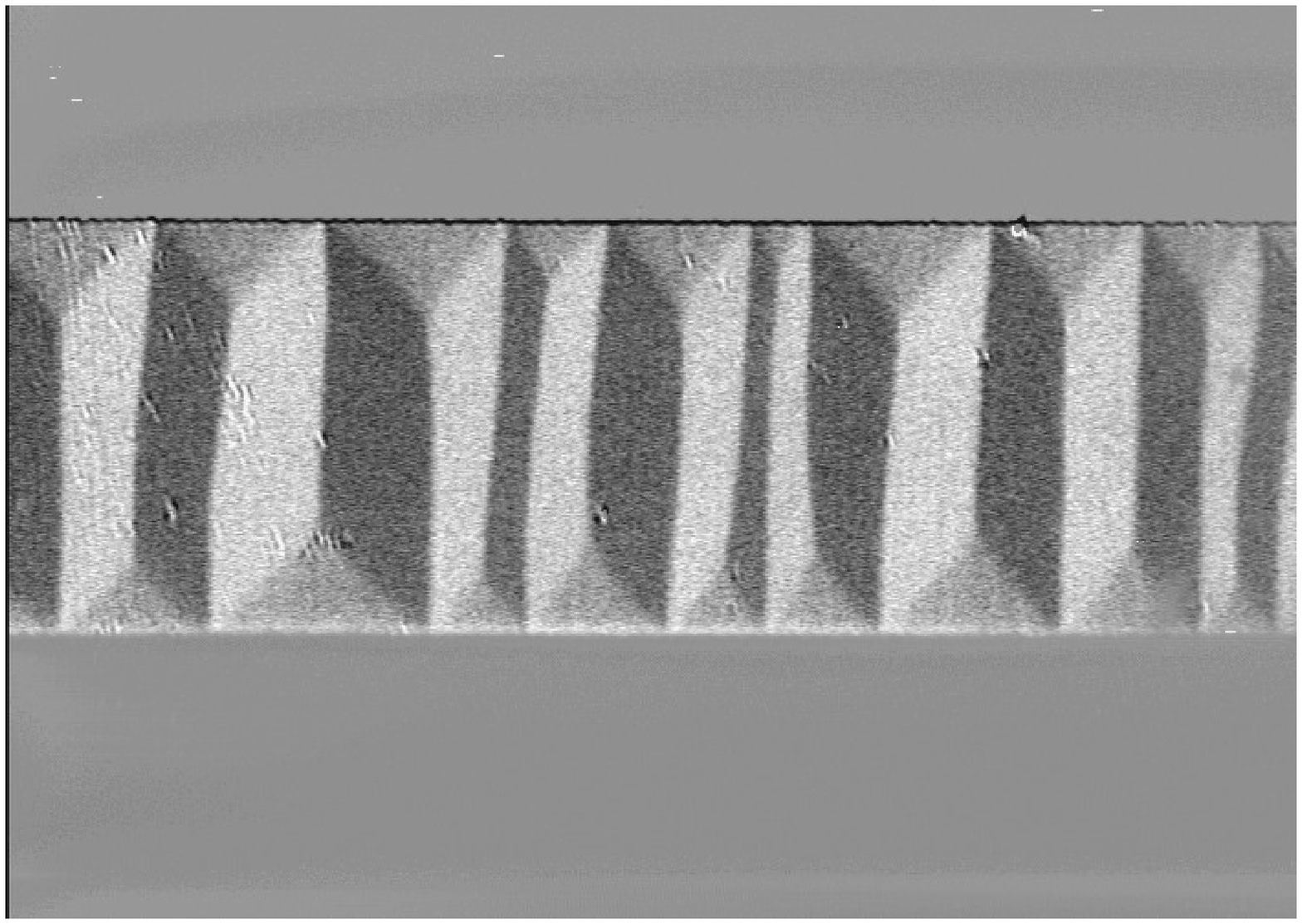}
\includegraphics[width=0.09\textwidth,height=0.075\textwidth]{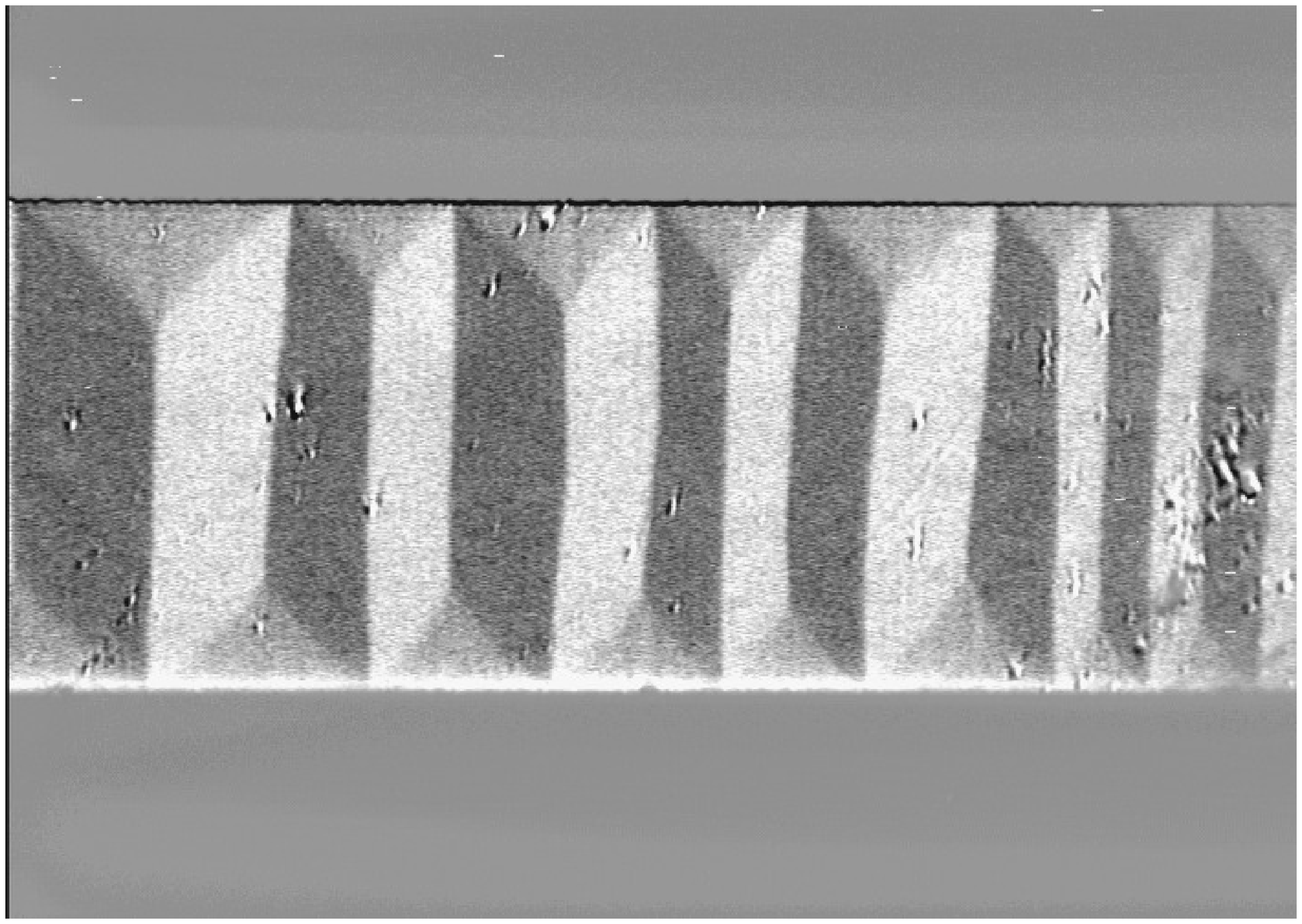}
\\
\includegraphics[width=0.09\textwidth,height=0.075\textwidth]{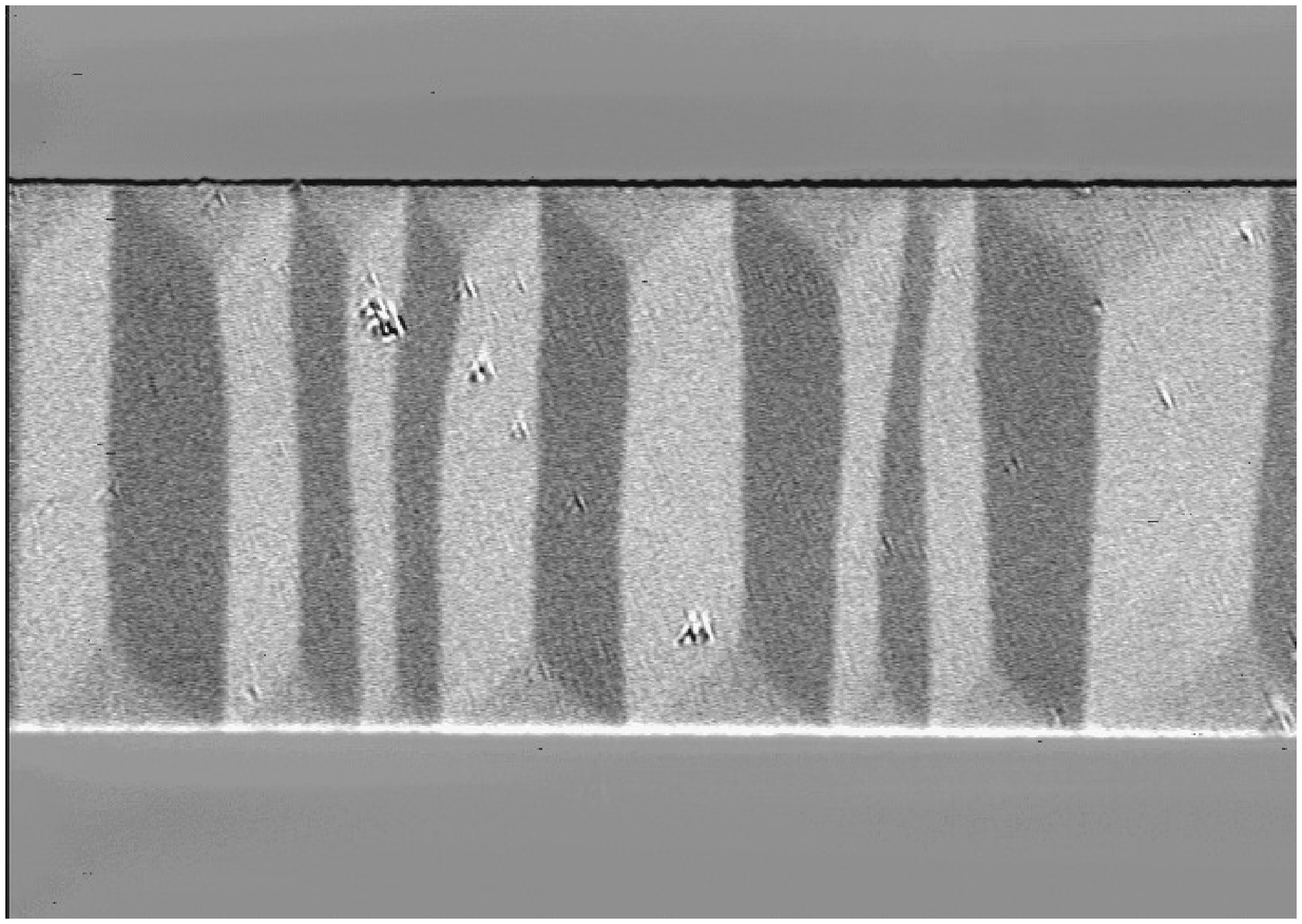}
\includegraphics[width=0.09\textwidth,height=0.075\textwidth]{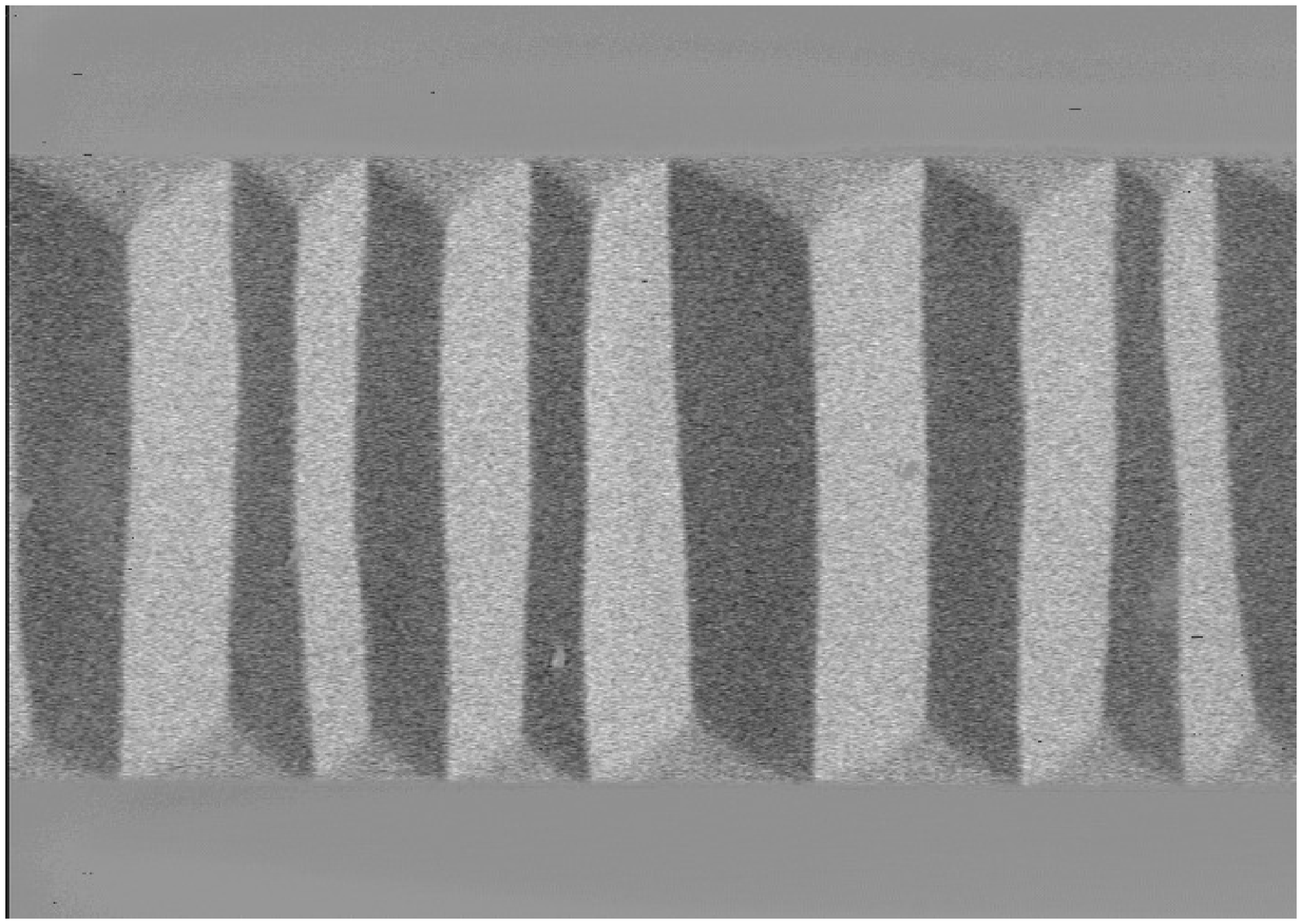}\qquad
\includegraphics[width=0.09\textwidth,height=0.075\textwidth]{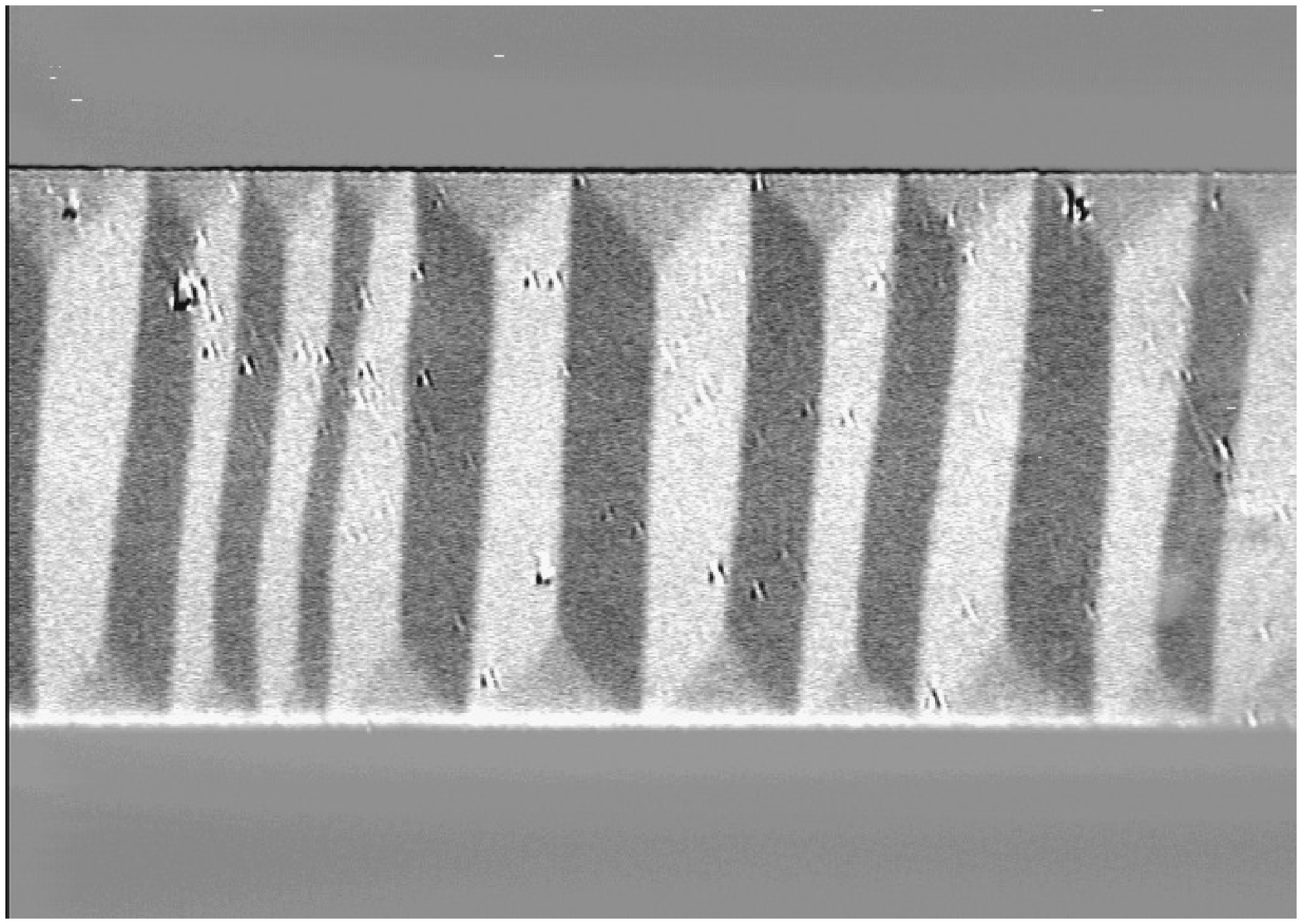}
\includegraphics[width=0.09\textwidth,height=0.075\textwidth]{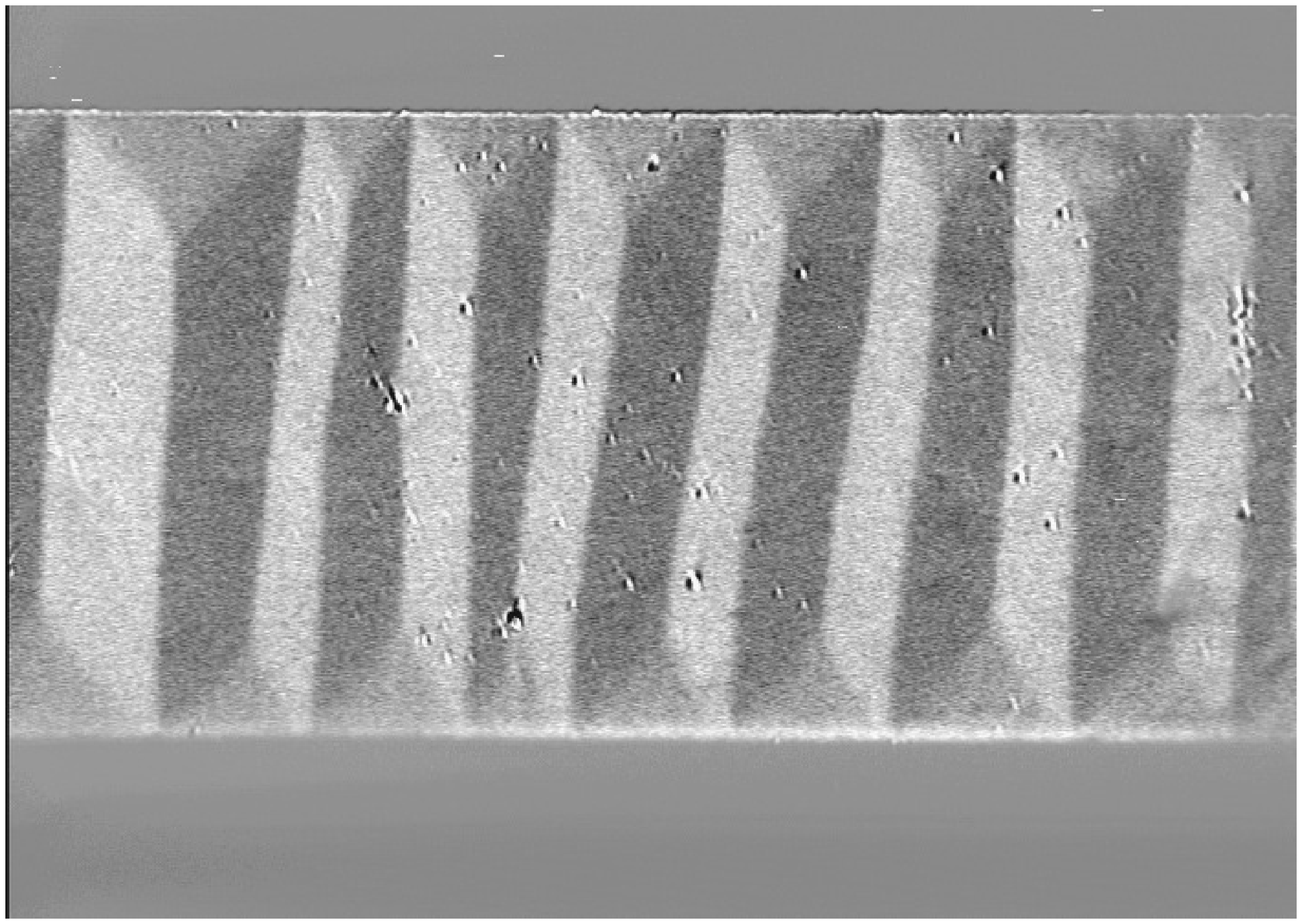}\\
\includegraphics[width=0.09\textwidth,height=0.075\textwidth]{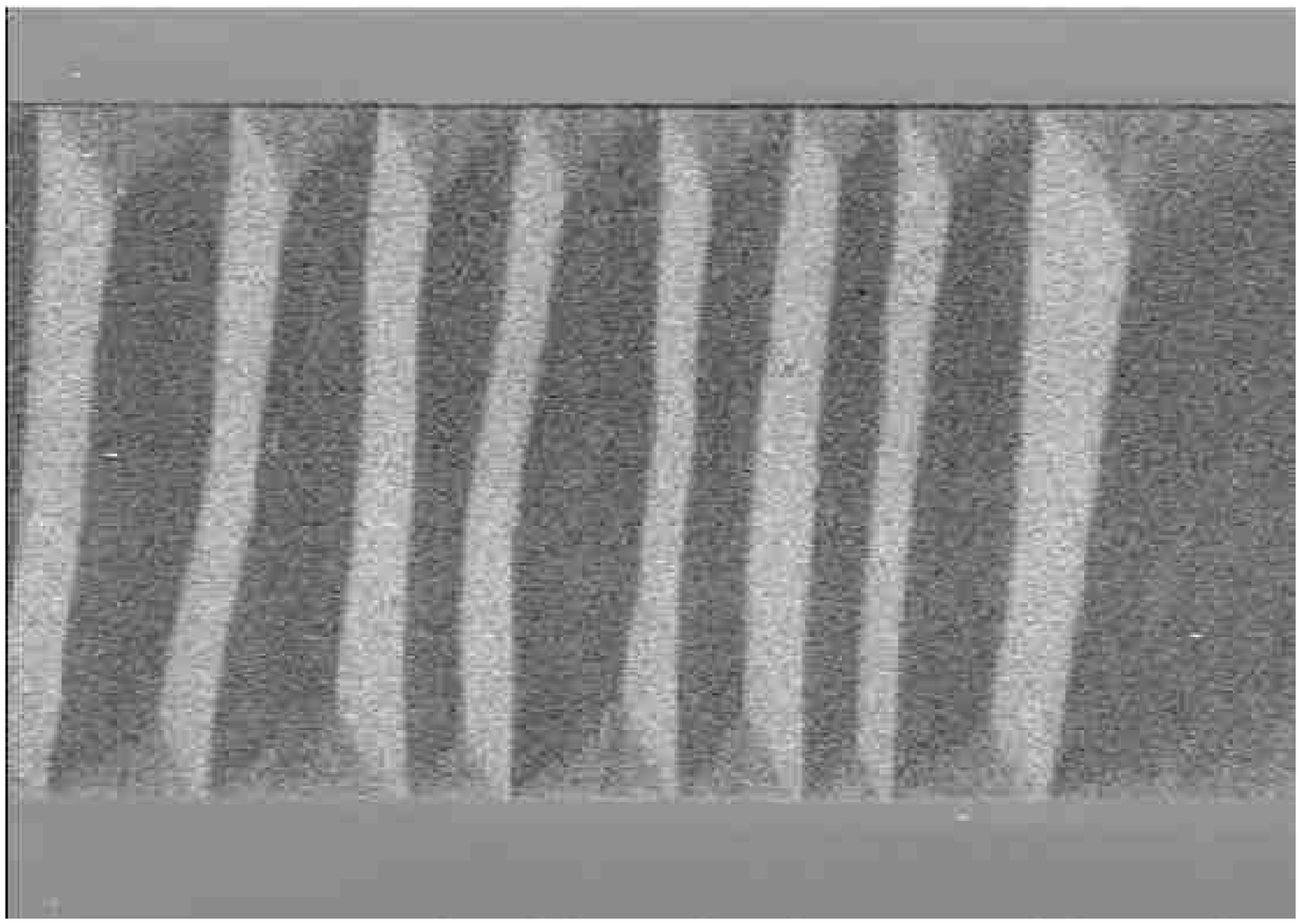}
\includegraphics[width=0.09\textwidth,height=0.075\textwidth]{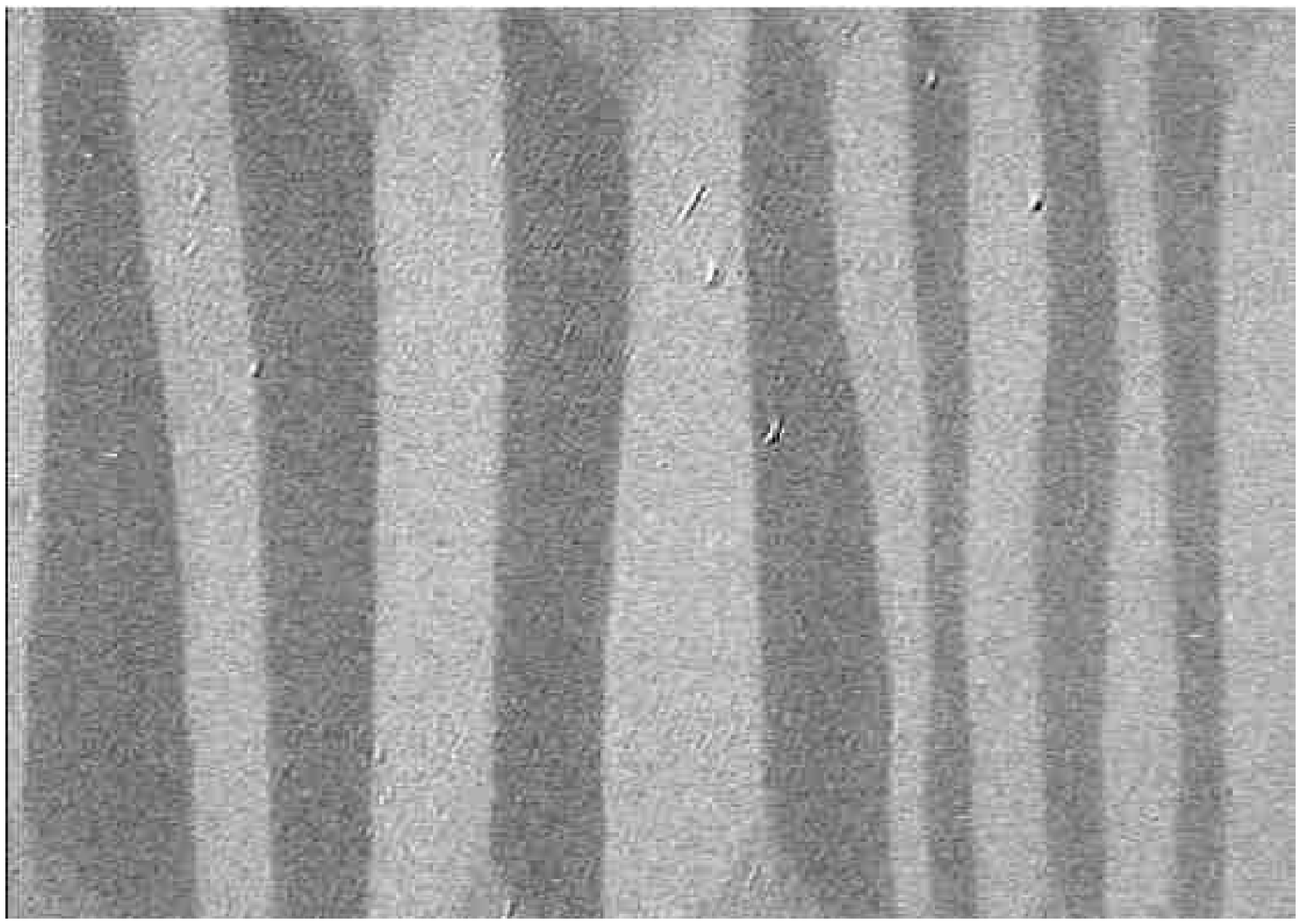}\qquad
\includegraphics[width=0.09\textwidth,height=0.075\textwidth]{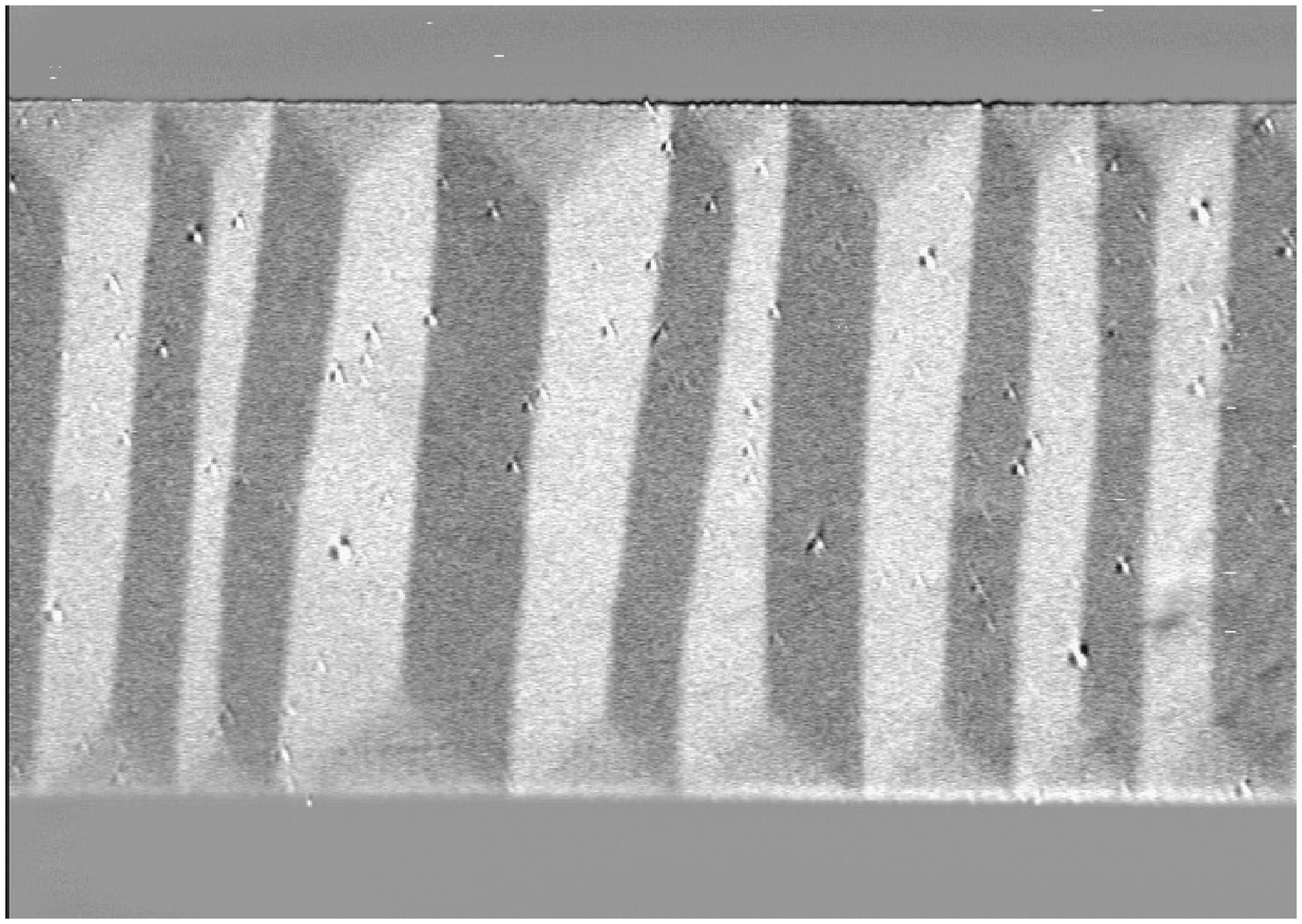}
\includegraphics[width=0.09\textwidth,height=0.075\textwidth]{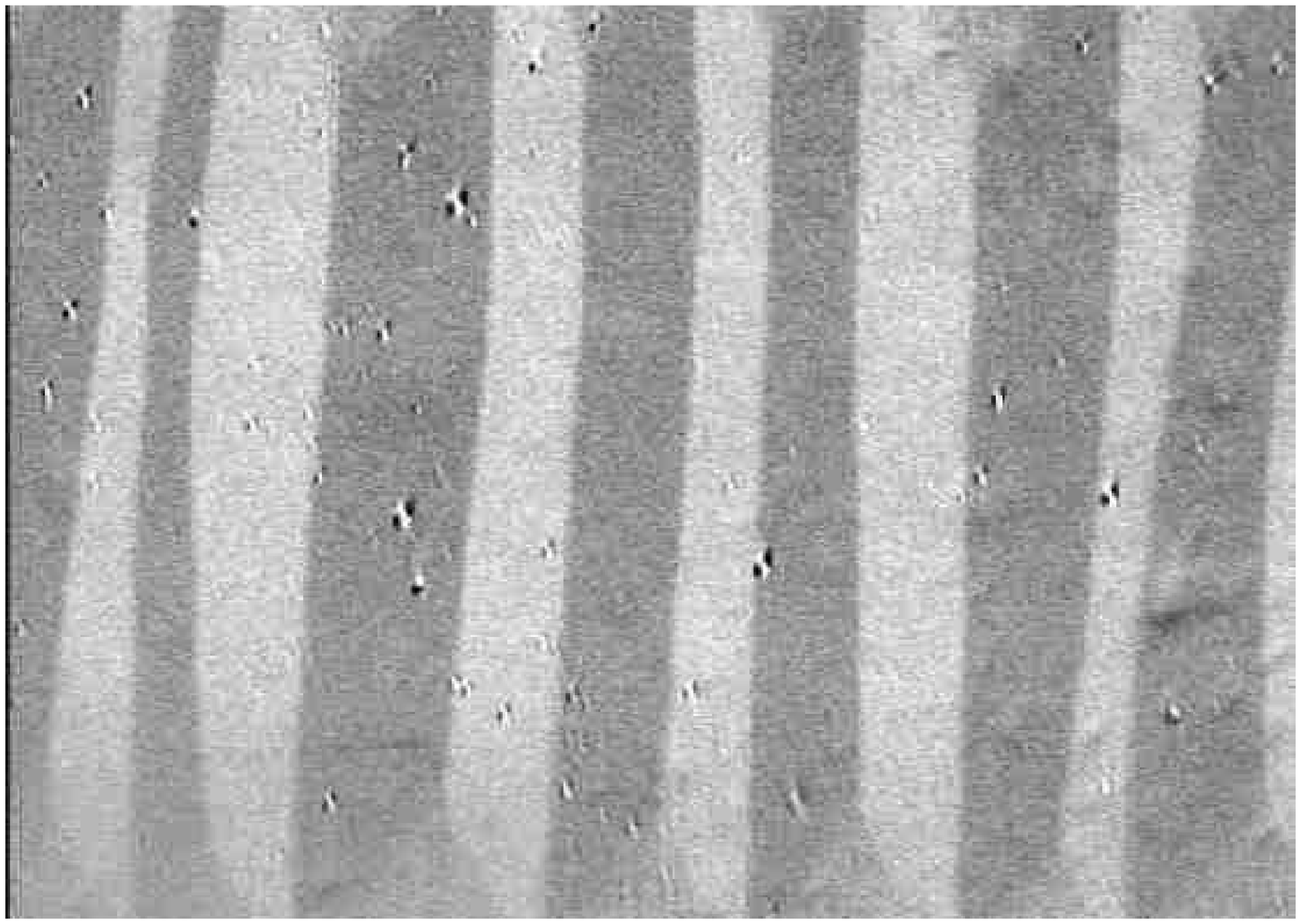}
\caption{Experiment: Permalloy samples of width $60\upmu$m to $150\upmu$m of high
  anisotropy and at the end of the coarsening process. The $6$ samples on the right are of thickness $30$nm,
  the $6$ samples on the right are of thickness $50$nm. The period of
the pattern appears to be
independent of the width of the
samples, in agreement with our theoretical prediction as an effect of anisotropy.}
\end{figure}
\smallskip

Figure \ref{scenario} displays the transition of the scaling
behavior in the optimal period, the marginally stable period, and the
amplitude of the transversal magnetization component. At the cross-over we have that $m_2\sim
t(\ell Q)^{-1}\ll 1$ and $w\sim t Q^{-1} \ll \ell$. This is
consistent with the assumptions of the reduced model, i.e., the
low-angle approximation and the scale separation of the dominant
length scales with respect to $x_1$ and $x_2$. For the same reason and
due to the
observation that $h_\ext+Q\gg d^{2/3}
\ell^{-4/3}t^{2/3}$ implies $\hat h_\ext\gg 1$, also (low-angle)
domain theory is applicable up to the cross-over to Scenario II. 
(As $m_2$ tends towards one in Scenario II solely domain theory is
applicable, the low-angle approximation has to be dropped -- in
particular for the wall energy.)
\smallskip

\begin{figure}[htb]
  \centering
\psset{unit=0.25} 
  \begin{pspicture}(0,-1.5)(9,8)
\psset{linewidth=0.05,linecolor=black,arrows=c-c}
    \pnode(0,0){a}
    \pnode(8,0){b}
    \pnode(0,8){c}
    \pnode(8,8){d}
    \pnode(1,8){e}
    \pnode(1,2){f}
    \pnode(5,8){g}
    \pnode(5,2){h}
    \pnode(3,0){i}
    \pnode(3,6){j}
    \pnode(7,0){k}
    \pnode(7,6){l}
\SpecialCoor
\psset{linewidth=0.05,linecolor=black,arrows=c-c}
    \ncline{a}{b}
    \ncline{c}{d}
\psset{linewidth=0.05,linecolor=gray,arrows=c-c}    
    \ncline{e}{f}
    \ncline{g}{h}
    \ncline{i}{j}
    \ncline{k}{l}
    \ncline{f}{i} 
    \ncline{e}{j}
    \ncline{j}{g}
    \ncline{i}{h}
    \ncline{l}{g}
    \ncline{k}{h}
\psline[linecolor=blue]{<-}(2,4.5)(2,3.5)
\psline[linecolor=blue]{<-}(4,3.5)(4,4.5)
\psline[linecolor=blue]{<-}(6,4.5)(6,3.5)
\psline[linecolor=blue]{->}(4.5,1)(5.5,1)
\psline[linecolor=blue]{->}(2.5,7)(3.5,7)
    \put(8.3,.5){{\psscaleboxto(.3,2){\(\}\)}}}
    \rput[l](8.7,1){$\tfrac{w}{\sqrt 2}$}

\put(3,-0.5) {\rotateright{\psscaleboxto(.5,4){\(\}\)}}} 
\rput[c](5,-1){${w}$}
\psline[linecolor=black]{<->}(8.5,4)(9.5,4)
\end{pspicture}
  \begin{pspicture}(0,-1.5)(9,8)
\psset{linewidth=0.05,linecolor=black,arrows=c-c}
    \pnode(0,0){a}
    \pnode(8,0){b}
    \pnode(0,8){c}
    \pnode(8,8){d}
    \pnode(1.5,8){e}
    \pnode(1,7.5){ee}
    \pnode(1,1.5){f}
    \pnode(4.5,8){gg}
    \pnode(5,7.5){g}
    \pnode(5,1.5){h}
    \pnode(2.5,0){i}
    \pnode(3,.5){ii}
    \pnode(3.5,0){iii}
    \pnode(3,6.5){j}
    \pnode(6.5,0){k}
    \pnode(5.5,8){ll}
    \pnode(7,6.5){l}
\pnode(7,.5){lll}
\SpecialCoor
\psset{linewidth=0.05,linecolor=black,arrows=c-c}
    \ncline{a}{b}
    \ncline{c}{d}
\psset{linewidth=0.05,linecolor=gray,arrows=c-c}    
    \ncline{e}{ee}
    \ncline{ee}{f}
    \ncline{g}{h}
    \ncline{ii}{j}
    \ncline{k}{lll}
    \ncline{f}{i} 
    \ncline{i}{ii} 
    \ncline{e}{j}
    \ncline{j}{gg}
    \ncline{gg}{g}
    \ncline{ii}{iii}
    \ncline{iii}{h}
    \ncline{ll}{g}
    \ncline{l}{lll}
    \ncline{k}{h}
    \ncline{l}{ll}
\psline[linecolor=blue]{<-}(2,4.5)(2,3.5)
\psline[linecolor=blue]{<-}(4,3.5)(4,4.5)
\psline[linecolor=blue]{<-}(6,4.5)(6,3.5)
\psline[linecolor=blue]{->}(4.5,.5)(5.5,.5)
\psline[linecolor=blue]{->}(2.5,7.5)(3.5,7.5)
%
\rput{-45}(8,4){\psline[linecolor=black]{<->}(-.5,0)(.5,0)}
\end{pspicture}
 \begin{pspicture}(0,-1.5)(9,8)
\end{pspicture}\\
\begin{pspicture}(0,-1.5)(9,8)
\psset{linewidth=0.05,linecolor=black,arrows=c-c}
    \pnode(0,0){a}
    \pnode(8,0){b}
    \pnode(0,8){c}
    \pnode(8,8){d}
    \pnode(1,6){e}
    \pnode(1,0){f}
    \pnode(5,6){g}
    \pnode(5,0){h}
    \pnode(3,2){i}
    \pnode(3,8){j}
    \pnode(7,2){k}
    \pnode(7,8){l}
\SpecialCoor
\psset{linewidth=0.05,linecolor=black,arrows=c-c}
    \ncline{a}{b}
    \ncline{c}{d}
\psset{linewidth=0.05,linecolor=gray,arrows=c-c}    
    \ncline{e}{f}
    \ncline{g}{h}
    \ncline{i}{j}
    \ncline{k}{l}
    \ncline{f}{i} 
    \ncline{e}{j}
    \ncline{j}{g}
    \ncline{i}{h}
    \ncline{l}{g}
    \ncline{k}{h}
\psline[linecolor=blue]{<-}(2,4.5)(2,3.5)
\psline[linecolor=blue]{<-}(4,3.5)(4,4.5)
\psline[linecolor=blue]{<-}(6,4.5)(6,3.5)
\psline[linecolor=blue]{<-}(2.5,1)(3.5,1)
\psline[linecolor=blue]{<-}(4.5,7)(5.5,7)
\rput{0}(8,4){\psline[linecolor=black]{<->}(-.5,0)(.5,0)}
\end{pspicture} 
 \begin{pspicture}(0,-1.5)(9,8)
\psset{linewidth=0.05,linecolor=black,arrows=c-c}
    \pnode(0,0){a}
    \pnode(8,0){b}
    \pnode(0,8){c}
    \pnode(8,8){d}
    \pnode(2.5,8){e}
    \pnode(1,6.5){ee}
    \pnode(1,.5){f}
    \pnode(3.5,8){gg}
    \pnode(5,6.5){g}
    \pnode(5,.5){h}
    \pnode(1.5,0){i}
    \pnode(3,1.5){ii}
    \pnode(4.5,0){iii}
    \pnode(3,7.5){j}
    \pnode(5.5,0){k}
    \pnode(6.5,8){ll}
    \pnode(7,7.5){l}
\pnode(7,1.5){lll}
\SpecialCoor
\psset{linewidth=0.05,linecolor=black,arrows=c-c}
    \ncline{a}{b}
    \ncline{c}{d}
\psset{linewidth=0.05,linecolor=gray,arrows=c-c}    
    \ncline{e}{ee}
    \ncline{ee}{f}
    \ncline{g}{h}
    \ncline{ii}{j}
    \ncline{k}{lll}
    \ncline{f}{i} 
    \ncline{i}{ii} 
    \ncline{e}{j}
    \ncline{j}{gg}
    \ncline{gg}{g}
    \ncline{ii}{iii}
    \ncline{iii}{h}
    \ncline{ll}{g}
    \ncline{l}{lll}
    \ncline{k}{h}
    \ncline{l}{ll}
\psline[linecolor=blue]{<-}(2,4.5)(2,3.5)
\psline[linecolor=blue]{<-}(4,3.5)(4,4.5)
\psline[linecolor=blue]{<-}(6,4.5)(6,3.5)
\psline[linecolor=blue]{<-}(2.5,.5)(3.5,.5)
\psline[linecolor=blue]{<-}(4.5,7.5)(5.5,7.5)
\rput{45}(8,4){\psline[linecolor=black]{<->}(-.5,0)(.5,0)}
\end{pspicture}  
  \begin{pspicture}(0,-6.5)(9,3)
\psset{linewidth=0.05,linecolor=black,arrows=c-c}
    \pnode(0,0){a}
    \pnode(8,0){b}
    \pnode(0,8){c}
    \pnode(8,8){d}
    \pnode(2,8){e}
    \pnode(1,7){ee}
    \pnode(1,1){f}
    \pnode(4,8){gg}
    \pnode(5,7){g}
    \pnode(5,1){h}
    \pnode(2,0){i}
    \pnode(3,1){ii}
    \pnode(4,0){iii}
    \pnode(3,7){j}
    \pnode(6,0){k}
    \pnode(6,8){ll}
    \pnode(7,7){l}
\pnode(7,1){lll}
\SpecialCoor
\psset{linewidth=0.05,linecolor=black,arrows=c-c}
    \ncline{a}{b}
    \ncline{c}{d}
\psset{linewidth=0.05,linecolor=gray,arrows=c-c}    
    \ncline{e}{ee}
    \ncline{ee}{f}
    \ncline{g}{h}
    \ncline{ii}{j}
    \ncline{k}{lll}
    \ncline{f}{i} 
    \ncline{i}{ii} 
    \ncline{e}{j}
    \ncline{j}{gg}
    \ncline{gg}{g}
    \ncline{ii}{iii}
    \ncline{iii}{h}
    \ncline{ll}{g}
    \ncline{l}{lll}
    \ncline{k}{h}
    \ncline{l}{ll}
\psline[linecolor=blue]{<-}(2,4.5)(2,3.5)
\psline[linecolor=blue]{<-}(4,3.5)(4,4.5)
\psline[linecolor=blue]{<-}(6,4.5)(6,3.5)
\psline[linecolor=blue]{->}(4.5,.5)(5.5,.5)
\psline[linecolor=blue]{<-}(2.5,.5)(3.5,.5)
\psline[linecolor=blue]{->}(2.5,7.5)(3.5,7.5)
\psline[linecolor=blue]{<-}(4.5,7.5)(5.5,7.5)
\end{pspicture}
 \caption{Continuous transition from the concertina pattern via the
   Landau state to the reversed concertina.  Note that the total
   length of the walls and the
 Zeeman energy do not change while the anisotropy energy is smaller in
 case of the Landau state.}
 \label{Landau ansatz}
 \end{figure}
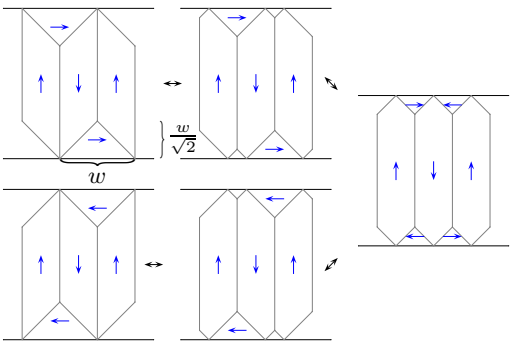

Let us mention another observation supporting the conjecture
that anisotropy effects are most prominent close to the
field strength where the concertina vanishes:
For $Q\gg \ell^{-1} t$,
the ground state for vanishing external field $h_\ext=0$ is no longer
given by the uniform magnetization $m=(\pm 1,0,0)$, but
a Landau or Concertina-type pattern, see Figure \ref{Landau ansatz}, has
lower energy. The optimal periods $w$ of the two latter pattern are determined
by a balance of the wall energy and the anisotropy energy
in the closure domains, and scale as $w\sim Q^{-1/2}(\ell t)^{1/2}$
up to a logarithm.
Hence we expect that in this regime, the concertina does
not switch to $m=(-1,0,0)$, but evolves to the pattern in Figure \ref{Landau ansatz}.
\smallskip

\begin{figure*}[htb]
  \centering
  \input{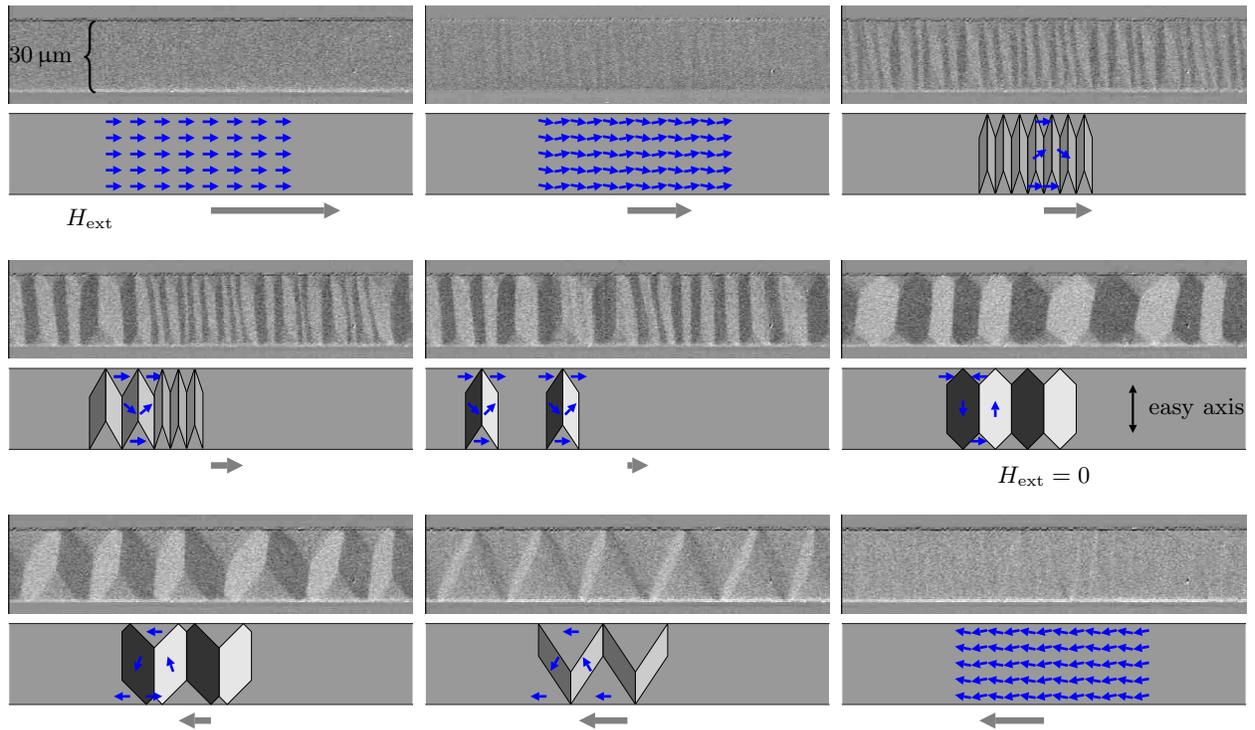}
  \caption{Experiment: Hysteresis of a CoFeB sample of $60$nm thickness 
    and $30\upmu$m width. Following the coarsening we observe a
    transition to a Landau state at $0$ external field which turns
    into a concertina which degenerates and refines, and finally
    disappears.}
\label{CoFeB}
\end{figure*}

In fact, that type of transition of the concertina pattern can be
observed in CoFeB samples -- that posses a stronger (transversal) uniaxial anisotropy, cf.\ Figure \ref{CoFeB}.

\smallskip

\section{Conclusion}
In this work, we addressed the concertina pattern in very elongated
thin-film elements.
We provided an explanation of the formation and
the coarsening of this pattern as the external
field is reduced from saturation. 

\medskip

We identified a parameter regime in which the uniform magnetization 
becomes unstable to an oscillatory buckling mode. In this
parameter regime, we derived a two-dimensional and
thus numerically tractable reduced energy functional from three-dimensional
micromagnetics. 
On the basis of the reduced model, we performed numerical bifurcation analysis:
The bifurcation is slightly subcritical, but has a turning point, 
after which the buckling mode grows into the concertina pattern
with its low-angle N\'eel walls. 
This is an alternative explanation
for the formation of the concertina to the one proposed by van den Berg: 
An outgrow of an unstable mode instead of an in-grow of closure domains.
Over a wide range of sample sizes, there is a good agreement between the 
explicit period of the unstable mode and the measured 
average period of the concertina pattern. In particular, the predicted dependence on film thickness and width is confirmed. However, the 
measured period exceeds the theoretically predicted one by a 
factor up to approximately two.

\medskip

We gave an argument for this initial deviation that at the same time explains the 
coarsening: Domain theory based on the reduced model
-- where low-angle N\'eel walls are replaced by sharp discontinuity lines --
shows that coarsened configurations are energetically favorable. More importantly, 
uncoarsened configurations eventually become unstable because the 
energy per period becomes concave. Based on the reduced model,
we argued by a Bloch-wave Ansatz that this concavity 
indeed translates into a secondary instability of the concertina pattern 
with respect to long wave-length modulations. These secondary
instabilities are confirmed by numerical bifurcation analysis. 
The long wave-length instabilities are further confirmed by an
extended bifurcation analysis that capitalizes on the near-degeneracy
of the primary bifurcation. This extended bifurcation analysis also
showed that that the long wave-length instability of the primary
branch extends all the way down to the turning point.
Hence at the moment of its appearance, the concertina pattern already
has a resulting period larger than the one of the unstable mode.
That qualitatively explains the deviation between the period
of the unstable mode and the measured period of the concertina.
Incidentally, these secondary instabilities are an
asymmetric (with respect to the wave number) version of the Eckhaus instability introduced in the context of convective problems.
\medskip

We gave yet another argument for the deviation of the period
of the unstable mode from the measured period of the concertina
at its formation. Based on the reduced model, we established
a continuous transition from the
magnetization ripple, which is triggered by the polycrystalline
structure of the material, and the concertina pattern. On the
level of the reduced model, the effect of 
an easy axis that varies from grain to grain translates
into a random transversal external field that smears out the subcritical
bifurcation. Hence for a sufficiently strong ripple effect,
as the concertina pattern becomes
discernible from the ripple, it has already coarsened.
\medskip

Finally, we investigated the effects of a weak uniaxial 
material anisotropy on the concertina pattern. We distinguished three effects:
1) a shift of the critical field that changes its sign already
for weak anisotropies, 2) a change in the coarsened concertina
pattern from ``limited by shape anisotropy'' to ``limited by
material anisotropy'' that kicks in for somewhat larger anisotropies,
3) a change from a subcritical to a supercritical bifurcation
for a sufficiently large transversal anisotropy.
\medskip

The various analyses render a fairly complete picture of the energy
landscape that in particular explains the hysteresis of the concertina
pattern. 

\section{Acknowledgment}
JM, RS, and HW thank R. Mattheis and R. Kaltofen for help with the
thin-film deposition. FO and JS thank Alexander Mielke for helpful
discussions on the Eckhaus instability and Martin Zimmermann for
technical support.
\clearpage
\bibliography{lit.bib}
 \end{document}